\documentclass[]{aastex}

\usepackage{emulateapj5}
\usepackage{onecolfloat}
\usepackage{graphicx} 
\usepackage{fancyheadings} 
\usepackage{ulem}
\usepackage{rotating}
\usepackage{lscape}

\newcommand{\ndla}{37}

\newcommand{\lya}{Ly$\alpha$ }

\newcommand{\kms}{km~s$^{-1}$ }
\newcommand{\cm}[1]{\, {\rm cm^{#1}}}
\newcommand{\N}[1]{{N({\rm #1})}}

\newcommand{\sci}[1]{{\rm \; \times \; 10^{#1}}}

\newcommand{\mkms}{{\rm \; km\;s^{-1}}}
\newcommand{\tskip}{\tablevspace{1pt}}

\begin{document}

\twocolumn[%
\submitted{Accepted to the Astrophysical Journal Supplements: April 10, 2003}
\title{The ESI/Keck~II Damped \lya Abundance Database}

\author{ Jason X. Prochaska\altaffilmark{1,2},
Eric Gawiser\altaffilmark{1,3,4,5},  
Arthur M. Wolfe\altaffilmark{1,3}, 
Jeff Cooke\altaffilmark{1,3},
\&
Dawn Gelino\altaffilmark{1,3}}

\begin{abstract} 

This paper presents chemical abundance measurements for \ndla\ damped
\lya systems at $z>2.5$ observed with the Echellette Spectrograph and
Imager on the Keck~II telescope.  We measure the H\,I column densities
of these damped systems through Voigt profile fits to their \lya profiles
and we implement the apparent optical depth method to determine ionic
column densities.  Figures and tables of all relevant data are presented.
A full analysis of the chemical enrichment history described by these
observations will be presented in a future paper.  This dataset is also 
valuable for efficiently planning future echelle observations and for
rough abundance pattern analyses.
We aim to make this entire data set public within three years of this
publication.


\end{abstract}
]

\altaffiltext{1}{Visiting Astronomer, W.M. Keck Telescope.
The Keck Observatory is a joint facility of the University
of California, the California Institute of Technology, and NASA.}
\altaffiltext{2}{UCO/Lick Observatory, University of California, Santa Cruz,
Santa Cruz, CA 95064}
\altaffiltext{3}{Department of Physics, and Center for Astrophysics and 
Space Sciences, University of California, San Diego, C--0424, La Jolla, 
CA 92093-0424}
\altaffiltext{4}{NSF Postdoctoral Fellow, Yale University, New Haven, CT, PO Box 208101, New Haven, CT  06520}
\altaffiltext{5}{Andes Prize Fellow, Universidad de Chile, Casilla 36-D, Santiago, Chile}

\pagestyle{fancyplain}
\lhead[\fancyplain{}{\thepage}]{\fancyplain{}{PROCHASKA ET AL.}}
\rhead[\fancyplain{}{The ESI/Keck~II Damped \lya Abundance Database}]{\fancyplain{}{\thepage}}
\setlength{\headrulewidth=0pt}
\cfoot{}

\section{INTRODUCTION}
\label{sec-intro}

For nearly a decade our group has been pursuing chemical abundance
measurements of the damped \lya systems (DLA) through high resolution spectroscopy
using the HIRES instrument \citep{vogt94} on the Keck~I telescope
\citep{pro99,pro01}.  With over 20 nights of observing, we have built
a sample of $N \approx 40$ damped \lya systems at redshift $z=1.7$ to 4.5.
This dataset has provided a measure of the enrichment history at $z>2$
\citep{pw00}, examined nucleosynthetic processes and dust depletion 
in young protogalaxies \citep{pw02}, and is now enabling measurements
of the star formation rate and history of these galaxies
\citep{pro02,wpg03}.  A number of groups are currently pursuing similar
programs with new echelle instruments on 10-m class telescopes
\citep{molaro00,petit00,ellison01,miro01a,lopez02,pettini02}.
Progress is limited, however, by the expense of echelle observations;
typical high-z quasars require an entire night of observation for 
reliable analysis of their absorption systems.

On August 29, 1999, the Echellette Spectrograph 
and Imager \citep[ESI;][]{sheinis02} saw first light on the Keck~II telescope.
Shortly after, we turned to ESI to
perform a new survey of damped \lya chemical abundances \citep{pgw01}.
Our observing strategy took advantage of ESI's unique characteristics --
high throughput, complete wavelength coverage from 
$\lambda \approx 4000 - 11000$\AA,
both moderate $(R \approx 10000)$ and low resolution modes -- 
to address several key issues related to chemical evolution \citep[e.g.][]{pw00}.
In particular, we focused on $z> 2.5$ damped systems 
where fewer echelle observations had been made.  
In four combined nights of integration, we have acquired abundance 
measurements for \ndla\ damped \lya systems.
In comparison with our HIRES observations, then,
we have observed the same number of damped systems 
in $\approx 1/5$ the observing time.  
Although the lower resolution of ESI limits the scientific impact of
the dataset, 
the results from this survey will significantly improve our understanding
of the chemical evolution history of the early universe (Prochaska et al.\ 2003,
in preparation).
The observations also provide valuable 'first-look' analysis crucial
to efficiently guide follow-up observations with echelle spectroscopy
\citep[e.g.][]{phw03}.
Furthermore, our observations include $\approx 10$
sub-DLA with $\N{HI} \approx 10^{19-20.3} \cm{-2}$ \citep[e.g.][]{peroux01}
and unparalleled 
spectral coverage at $\lambda > 8000$\AA\ valuable for both 
DLA and $z \approx 2$ Mg\,II metal-line studies.

In this paper, we present the chemical abundance measurements for our ESI
damped \lya sample.  The goal of this paper is to present the data and 
ionic column densities for the damped systems with little accompanying
analysis.  Future papers will examine
the implications of this dataset and extend the analysis to the sub-DLA
and $z \approx 2$ Mg\,II systems.  This paper is outlined as follows.
We describe the observations and data reduction routines in $\S$~2.  
In $\S$~3 we detail the methods used to perform the abundance measurements
and detail specific issues related to the ESI instrument.
The individual damped \lya systems are presented in $\S$~4 with full
figures and tables and we present a brief summary in $\S$~5.

\begin{sidewaystable*}\footnotesize
\begin{center}
\caption{
{\sc JOURNAL OF OBSERVATIONS\label{tab:obs}}}
\begin{tabular}{llcccclcrrl}
\tableline
\tableline \tskip
Quasar
& RA (2000) & DEC (2000) & $V/R$ & $z_{em}$ & $z_{abs}$
& Date & Mode & Slit & Exp & Ref \\
\tableline
SDSSp0127-00 & 01 27 00.7 & $-00$ 45 59 & 18.4 & 4.06 & 3.72      & 22jan01 & ECH & 0.75$''$ & 1800 & 1\\
             &            &             &      &      &           & 19jul01 & ECH & 0.50$''$ & 1800 & \\
PSS0133+0400 & 01 33 40.4 & $+04$ 00 59 & 17.9 & 4.13 & 3.69,3.77 & 22jan01 & ECH & 0.75$''$ & 1800 & 2 \\
PSS0134+3317  & 01 34 21.6 & $+33$ 07 56 & 18.9 & 4.52 & 3.76      & 15oct01 & ECH & 0.5$''$  & 6000 & 2,3 \\
PSS0209+0517 & 02 09 44.7 & $+05$ 17 14 & 17.4 & 4.18 & 3.66,3.86 & 22jan01 & ECH & 0.75$''$ & 1200 & 2\\
BRJ0426-2202 & 04 26 10.3 & $-22$ 02 17 & 18.0 & 4.30 & 2.98      & 22jan01 & ECH & 0.5$''$  & 1800 & 3,4\\
FJ0747+2739  & 07 47 11.2 & $+27$ 39 04 & 17.5 & 4.11 & 3.42,3.90 & 15oct01 & ECH & 0.5$''$  & 6000 & 5\\
PSS0808+5215 & 08 08 49.5 & $+52$ 15 16 & 18.8 & 4.45 & 3.11      & 07apr00 & LWD & 1.0$''$  &  400 & 2,6\\
   &&&&&&                                                           07apr00 & ECH & 0.75$''$ & 2400 & \\
FJ0812+32    & 08 12 40.8 & $+32$ 08 08 & 17.6 & 2.70 & 2.07,2.63 & 22jan01 & ECH & 0.5$''$  & 1200 & 5\\
Q0930+28     & 09 33 37.8 & $+28$ 45 35 & 17.5 & 3.42 & 3.23      & 22jan01 & ECH & 0.5$''$  & 1200 & 7\\
PC0953+4749  & 09 56 25.2 & $+47$ 34 44 & 19.5 & 4.46 & 3.40,3.89,4.24 & 22jan01 & ECH & 0.5$''$  &  4800 & 8,9\\
PSS0957+3308 & 09 57 44.5 & $+33$ 08 24 & 17.6 & 4.25 & 3.28,4.18 & 07apr00 & LWD & 1.0$''$  &  400 & 2,6 \\
   &&&&&&                                                           07apr00 & ECH & 0.75$''$ & 1200 & \\
BQ1021+3001     & 10 21 56.8 & $+30$ 01 31 & 17.0 & 3.12 & 2.95      & 22jan01 & ECH & 0.75$''$ & 1800 & 5,10 \\
CTQ460       & 10 39 09.4 & $-23$ 13 26 & 18.0 & 3.13 & 2.77      & 22jan01 & ECH & 0.5$''$  & 1800 & 11,12\\
HS1132+2243  & 11 35 08.0 & $+22$ 27 07 & 17.4 & 2.88 & 2.78      & 22jan01 & ECH & 0.5$''$  & 1200 & 13 \\
Q1209+0919   & 12 11 34.1 & $+09$ 02 17 & 18.5 & 3.30 & 2.58      & 22jan01 & ECH & 0.5$''$  & 1800 & 14 \\
PSS1248+3110 & 12 48 20.2 & $+31$ 10 44 & 18.9 & 4.35 & 3.696     & 07apr00 & LWD & 1.0$''$  &  400 & 2,6 \\
   &&&&&&                                                           07apr00 & ECH & 0.75$''$ & 2400 & \\
PSS1253-0228 & 12 53 36.3 & $-02$ 28 08 & 18.7 & 4.01 & 2.78      & 22jan01 & ECH & 0.5$''$  & 2700 & 2 \\
Q1337+11     & 13 40 02.4 & $+11$ 06 30 & 19.0 & 2.92 & 2.79      & 22jan01 & ECH & 0.5$''$  & 2400 & 14, 15\\
PKS1354-17   & 13 57 05.9 & $-17$ 44 06 & 18.5 & 3.15 & 2.78      & 19jul01 & ECH & 0.5$''$  & 1800 & 16\\
PSS1432+3940 & 14 32 24.9 & $+39$ 40 24 & 18.6 & 4.28 & 3.272     & 07apr00 & LWD & 1.0$''$  &  400 & 2,6 \\
   &&&&&&                                                           07apr00 & ECH & 0.75$''$ & 3600 & \\
Q1502+4837   & 15 02 27.3 & $+48$ 37 09 & 17.9 & 3.20 & 2.56      & 19jul01 & LWD & 1.0$''$  &  300 & 17 \\
   &&&&&&                                                           19jul01 & ECH & 0.50$''$ & 1800 & \\
PSS1506+5220 & 15 06 54.6 & $+52$ 20 05 & 18.1 & 4.18 & 3.22      & 22jan01 & ECH & 0.5$''$  & 1200 & 2\\
PSS1723+2243 & 17 23 23.2 & $+22$ 43 58 & 18.2 & 4.52 & 3.69      & 19jul01 & ECH & 0.5$''$  & 1800 & 2\\
PSS2155+1358 & 21 55 02.1 & $+13$ 58 26 & 17.9 & 4.26 & 3.31      & 31oct00 & ECH & 0.75$''$ & 1800 & 2 \\
Q2223+20     & 22 25 37.0 & $+20$ 40 18 &      & 3.56 & 3.12      & 15oct01 & ECH & 0.5$''$  & 3000 & 9,18\\
PSS2241+1352 & 22 41 47.9 & $+13$ 52 03 & 18.7 & 4.44 & 4.28      & 19jul01 & ECH & 0.50$''$ & 2700 & 2\\
PSS2323+2758 & 23 23 41.0 & $+27$ 58 01 & 18.5 & 4.18 & 3.68      & 19jul01 & ECH & 0.50$''$ & 3600 & 2\\
FJ2334-09    & 23 34 46.4 & $-09$ 08 12 & 17.9 & 3.33 & 3.06      & 15oct01 & ECH & 0.5$''$  & 2400 & 5\\
Q2342+34     & 23 44 51.1 & $+34$ 33 47 & 18.6 & 3.01 & 2.91      & 19jul01 & ECH & 0.50$''$ & 1800 & 18,19\\
PSS2344+0342 & 23 44 03.2 & $+03$ 42 26 & 18.6 & 4.30 & 3.22      & 19jul01 & ECH & 0.50$''$ & 1800 & 2,3 \\
\tableline
\end{tabular}
\end{center}
Key to References -- 1:
\cite{fan01}; 2: \cite{djg98}; 3:
\cite{peroux01}; 4: \cite{storrie01}; 5:
\cite{white00}; 6: \cite{pgw01};
7: \cite{lu98}; 8: \cite{schneider91}; 
9: \cite{storrie00}; 10: \cite{gregg96}; 
11: \cite{maza95}; 12: \cite{lopez01}; 13:
\cite{hagen99}; 14: \cite{wolfe95};
15: \cite{sargent89}; 16: \cite{schilizzi75};
17: \cite{snellen01}; 18: \cite{griffith90};
19: \cite{white93}
\end{sidewaystable*}

\section{OBSERVATIONS AND DATA REDUCTION}
\label{sec:redux}

This observing program totaled 5 scheduled nights which netted $\approx 4$
total nights of scientific observations.  In
Table~\ref{tab:obs} we present a comprehensive journal of observations
for the quasars included in this paper.  Column~1 gives the name,
column~2 and 3 list the RA and DEC (J2000), 
column~4 is the apparent magnitude of the quasar ($R$ for $z>3$, $V$ for $z<3$),
column~5 is the emission redshift,
column~6 lists $z_{abs}$ for the DLA along the sightline, 
column~7 gives the date of the observation,
column~8 details the mode of observation
(LWD is the low-dispersion prism mode; ECH is the echellette mode), 
column~9 shows the slit width (predominantly 0.5$''$), 
column~10 is the exposure time, and
column~11 provides comments.

With this program as the impetus,
Prochaska has built an IDL data reduction package for the ESI instrument.  
Because the ESI modes are nearly static 
(i.e.\ there is only one setup per mode) and flexure is well compensated
by an active control system, we found it advantageous to create a package
tailored to ESI's specific characteristics.  The general approach for
spectroscopy (the package also reduces images taken with ESI) is 
similar to the packages designed by the Sloan Digital Sky Survey (SDSS)
and 
Deep Extragalactic Evolutionary Probe
(DEEP) teams for their spectroscopic observations. 
The ESI package constructs a 2D wavelength map in the original data frame 
from the arc calibration images and 
implements 1D BSPLINE algorithms for sky subtraction and standard approaches for
extraction and coadding.  The package is publically 
available\footnote{http://www2.keck.hawaii.edu/realpublic/inst/esi/ESIRedux/index.html}
and will be intermittently supported for the duration
of ESI.  We now present a more detailed description of the code for
ECH mode reductions.  Future papers will elaborate on the reduction routines
for the LWD and imaging modes.

At the heart of the code is an IDL structure designed to organize all data
from a single night of observing.  This structure contains the vital 
characteristics of a given frame (e.g.\ ESI mode, exposure time, slit size)
and helps guide the reduction process.
In general, one passes this structure to a given IDL routine and the
code automatically coordinates and processes the frames.
Bias subtraction is performed using a bias image constructed from
a series of 0s exposures.  This approach is advantageous to using the overscan
because of a significant bias variation in the sector of the CCD 
containing two chip defects.  
A normalized flat is constructed 
(ideally from dome flats as the internal quartz lamps exhibit significant
scattered light) to remove pixel-to-pixel variations.  In the process, 
the slit-edges of each echellette order are determined from the flat and if
the data were taken in the standard two-amp mode, then the gain 
ratio is calculated.  This gain ratio is applied to all
data taken throughout the night which is effective but not perfect. 
In particular, data near $5570$\AA\ 
(the wavelength affected by the boundary between amplifiers) 
may suffer if the electronics are not perfectly stable throughout the night.

The most important step in the ECH routine is the construction of a 2D 
wavelength image which contains a unique wavelength value for every data
pixel in the ECH image.  To create this wavelength image, the code first
processes and combines all arc frames for a given slit size.  
In sequence, it straightens\footnote{The curvature of the echellette orders
is determined from a sequence of pin-hole observations taken either 
by the observer or from the calibration archive.} 
each echellette order and extracts the central
few columns (the data runs parallel to the columns in ESI)
to provide a 1D arc spectrum.  This spectrum is cross-correlated against 
an archived 1D spectrum to determine any integer offset.  We have found
that ESI is stable to better than 3 pixels for data taken over the
two years since commissioning.
After applying this offset, the program automatically 
identifies arc lines assuming the wavelength solution of
the archived 1D spectrum.  It then centroids these lines and determines
a final 1D solution by fitting a polynomial with order between 5 and 9.
The code traces bright, individual arc lines to determine their curvature
relative to the slit which is typically
$\approx 0.1$~pix from slit-edge to slit-edge.  A 2D polynomial is simultaneously
fit to the centroids of all of the  arc lines of a given echellette order 
and a 2D wavelength map is created with typical RMS $< 2$km/s.
Finally, this wavelength map is inverted to the original data frame
to create a 2D wavelength image (vacuum corrected)
that is used by all data frames with identical slit width.  
For a given exposure, every data pixel now has a unique $\lambda$ , flux pair.

Sky subtraction, extraction and the creation of a final 1D spectrum 
proceed in relatively standard fashion.  The data images are bias
subtracted and flattened.  All significant objects are 
identified within the slit and masked for sky subtraction.  For the
bluest orders where sky lines are negligible, a low order polynomial
is fit row by row to the sky and subtracted from the data.
For the orders with $\lambda > 5000$\AA, the sky subtraction algorithm
utilizes the 2D wavelength image.  The $\lambda$, flux pairs for all
sky pixels in a given order are collapsed to 1D and fit with a 1D BSPLINE to 
create a single, functional representation of the sky at all wavelength.  
This approach is particularly advantageous for fitting bright sky lines
and the user can modify the location of the BSPLINE breakpoints to 
improve the fitting of specific sky lines as desired.
The 1D BSPLINE solution is evaluated at every data pixel and 
subtracted from the image.  The principal science object is extracted
from the sky-subtracted image using a boxcar aperture (determined uniquely
for each echellette order) and the $\lambda$, flux pairs are collapsed
for each order to produce ten 1D spectra with 10 km/s pixels.
Note that during this process the $\lambda$ values are heliocentric corrected.
The error array is derived in a similar fashion and includes uncertainty
due to object, sky, and readnoise.  Cosmic rays are identified during the
extraction process by comparing the collapsed object profile to the 
row-by-row profile.  These are then flagged and all data at that wavelength
are rejected from future analysis.
At this stage, individual exposures are optimally coadded with rejection
and the resulting data is fluxed using a standard star from that night
or the archive.  Finally, the individual orders are coadded to create
a single, continuous 1D spectrum.  

Ideally, the package is fully automated with minimal user interaction
beyond bookkeeping and optional checking.  Several procedures are
computationally expensive (construction of the arc image; sky subtraction)
in part owing to the number of pixels involved and in part due to the
implementation of IDL instead of C or FORTRAN.  The current code will
reduce and extract a single data frame in $\approx 15$min on a 900MHz
Sunblade processor.  
Future generations of the code will include
procedures for optimal extraction and generic binning modes and will
also include updated archival calibration frames.

\section{METHODS}
\label{sec:methods}

The techniques employed in this paper to determine column densities
and errors do not differ significantly from those of our previous
work with echelle data \citep[e.g.][]{pro01}.  In general, we 
implement the apparent optical depth method \citep[AODM,][]{sav91}
to evaluate column densities of unsaturated, unblended metal-line 
transitions.  We stress one difference from our previous
work: we establish a stricter saturation limit to 
prevent ``hidden saturation'' allowed by 
the lower resolution of the ESI instrument. 
Also note that the upper limits reported are formally 
$3 \sigma$ statistical limits and the measurement uncertainties for
detections are reported as $1 \sigma$ errors derived from the AODM.
Unless otherwise noted, these limits were derived by integrating the optical
depth over the velocity range of the spectra covering all of the
the absorption in the detected transitions.
Because these uncertainties do not take into account 
systematic error (i.e.\ continuum placement), we caution the reader
when interpreting them.  For example, a reported error
of 0.01~dex is too optimistic for this dataset.  We recommend adopting
a minimum error of 0.1~dex for the measurements of single transitions. 
In turn, we would recommend assuming that relative 
abundances (e.g.\ Si/Fe) have at least 0.15~dex uncertainty.
Finally, the incidence of contradiction between our reported
limits and the measured value for ions with multiple observed transitions
suggests the limits refer to $\approx 95\%$ c.l.

\begin{figure}[ht]
\begin{center}
\includegraphics[height=3.6in, width=2.8in,angle=-90]{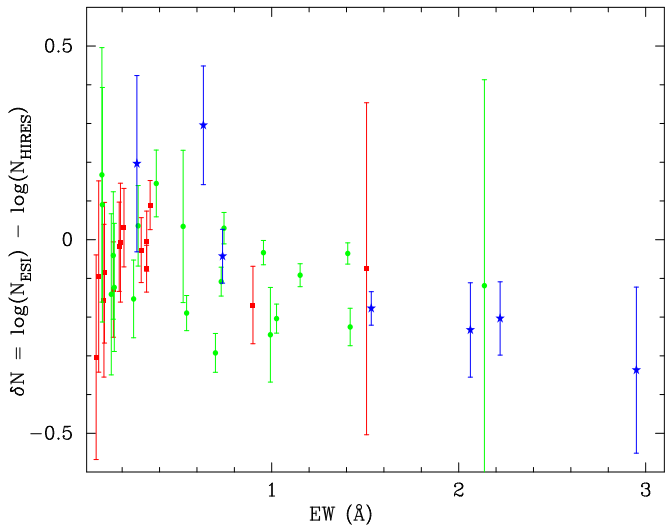}
\figcaption{Observed differences $\delta$ between column density measurements as
a function of observed equivalent width (EW)
for a series of metal-line transitions from three DLA 
(medium gray squares: PH957;
light gray circles: FJ0812+32;
dark gray stars: PSS0957+33, $z=4.18$)
observed with both ESI and HIRES.  Because of the effects of resolution, 
the ESI values are significantly lower when EW~$> 1$\AA.  There is
a milder effect for transitions with $0.6 < \rm{EW} < 1$\AA.  
\label{fig:sat}}
\end{center}
\end{figure}

The accuracy of our ESI measurements and the effects of saturation are
well demonstrated in Figure~\ref{fig:sat}.  In the figure, we plot 
$\delta_N \equiv \log N(X)_{ESI} - \log N(X)_{HIRES}$ against EW$_{ESI}(X)$,
the observed equivalent width of transition X for a set of transitions
in three damped \lya systems with both ESI and HIRES observations.
For transitions with EW~$< 0.6$\AA, the differences are consistent with
measurement error while transitions with EW~$> 1$\AA\ clearly suffer from
saturation in the ESI observations.
For transitions with 0.6~$<$~EW~$< 1$\AA, the ESI observations are
systematically low but the effect is mild.  In the following analysis,
we impose lower limits to the column densities measured from transitions
with EW exceeding $\approx 0.6$\AA.

\begin{figure}[ht]
\begin{center}
\includegraphics[height=3.6in, width=2.8in,angle=-90]{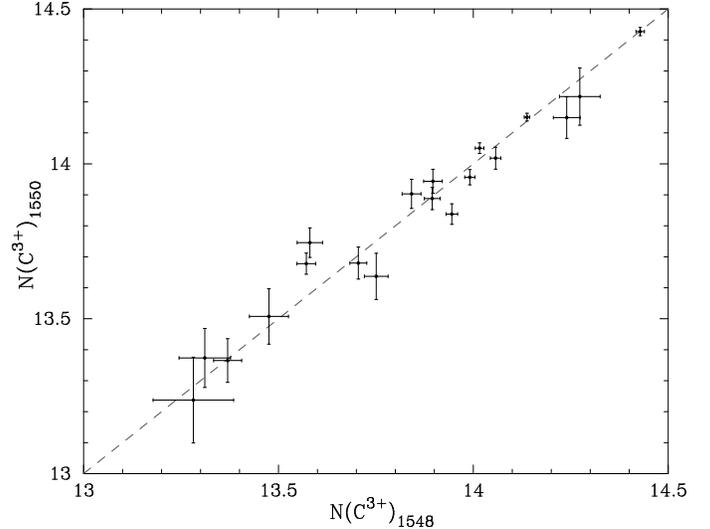}
\figcaption{A comparison of C$^{3+}$ column densities measured independently
from the C~IV 1548 and 1550 transitions in the same DLA using the apparent 
optical depth method (AODM).  The dashed line traces the line of equality.  
The good agreement between the two values over a large range in
$\N{C^{3+}}$ indicates the AODM is an accurate method for measuring
column densities in unblended transitions.\label{fig:civtst}}
\end{center}
\end{figure}

\begin{table*}\footnotesize
\begin{center}
\caption{ 
{\sc ATOMIC DATA \label{tab:fosc}}}
\begin{tabular}{lcccc}
\tableline
\tableline
Transition &$\lambda$ &$f$ & Ref\\
\tableline\tskip
   HI 972  &  972.5368 & 0.02900000 &  1  \\
   OI 988  &  988.7734 & 0.04318000 &  1  \\
 NIII 989  &  989.7990 & 0.10660000 &  1  \\
+   HI 1025 & 1025.7223 & 0.07912000 &  1  \\
  CII 1036 & 1036.3367 & 0.12310000 &  1  \\
   OI 1039 & 1039.2304 & 0.00919700 &  1  \\
  NII 1083 & 1083.9900 & 0.10310000 &  1  \\
FeIII 1122 & 1122.5260 & 0.05390000 & 16  \\
   NI 1134 & 1134.1653 & 0.01342000 &  1  \\
   NI 1134 & 1134.4149 & 0.02683000 &  1  \\
   NI 1134 & 1134.9803 & 0.04023000 &  1  \\
 FeII 1144 & 1144.9379 & 0.10600000 &  2  \\
 SiII 1190 & 1190.4158 & 0.25020000 &  1  \\
 SiII 1193 & 1193.2897 & 0.49910000 &  1  \\
   NI 1199 & 1199.5496 & 0.13000000 &  1  \\
   NI 1200 & 1200.2233 & 0.08620000 &  1  \\
   NI 1200 & 1200.7098 & 0.04300000 &  1  \\
SiIII 1206 & 1206.5000 & 1.66000000 &  1  \\
   HI 1215 & 1215.6701 & 0.41640000 &  1  \\
 MgII 1239 & 1239.9253 & 0.00063300 &  3  \\
 MgII 1240 & 1240.3947 & 0.00035500 &  3  \\
  SII 1250 & 1250.5840 & 0.00545300 &  1  \\
  SII 1253 & 1253.8110 & 0.01088000 &  1  \\
  SII 1259 & 1259.5190 & 0.01624000 &  1  \\
 SiII 1260 & 1260.4221 & 1.00700000 &  1  \\
   CI 1277 & 1277.2450 & 0.09665000 &  1  \\
   OI 1302 & 1302.1685 & 0.04887000 &  1  \\
 SiII 1304 & 1304.3702 & 0.09400000 &  4  \\
 NiII 1317 & 1317.2170 & 0.07786000 &  1  \\
  CII 1334 & 1334.5323 & 0.12780000 &  1  \\
 CII* 1335 & 1335.7077 & 0.11490000 &  1  \\
   OI 1355 & 1355.5977 & 0.00000124 &  1  \\
 NiII 1370 & 1370.1310 & 0.07690000 &  5  \\
 SiIV 1393 & 1393.7550 & 0.52800000 &  1  \\
 SiIV 1402 & 1402.7700 & 0.26200000 &  1  \\
 NiII 1454 & 1454.8420 & 0.03230000 &  7  \\
 SiII 1526 & 1526.7066 & 0.12700000 &  9  \\
  CIV 1548 & 1548.1950 & 0.19080000 &  1  \\
  CIV 1550 & 1550.7700 & 0.09522000 &  1  \\
   CI 1560 & 1560.3092 & 0.08041000 &  1  \\
 FeII 1608 & 1608.4511 & 0.05800000 & 17  \\
 FeII 1611 & 1611.2005 & 0.00136000 & 16  \\
   CI 1656 & 1656.9283 & 0.14050000 &  1  \\
 AlII 1670 & 1670.7874 & 1.88000000 &  1  \\
 NiII 1703 & 1703.4050 & 0.00600000 &  7  \\
 NiII 1709 & 1709.6042 & 0.03240000 &  7  \\
 NiII 1741 & 1741.5531 & 0.04270000 &  7  \\
 NiII 1751 & 1751.9157 & 0.02770000 &  7  \\
 SiII 1808 & 1808.0130 & 0.00218600 & 11  \\
AlIII 1854 & 1854.7164 & 0.53900000 &  1  \\
AlIII 1862 & 1862.7895 & 0.26800000 &  1  \\
TiII 1910a & 1910.6000 & 0.20200000 & 12  \\
TiII 1910b & 1910.9380 & 0.09800000 & 12  \\
 ZnII 2026 & 2026.1360 & 0.48900000 & 13  \\
 CrII 2056 & 2056.2539 & 0.10500000 & 13  \\
 CrII 2062 & 2062.2340 & 0.07800000 & 13  \\
 ZnII 2062 & 2062.6640 & 0.25600000 & 13  \\
 CrII 2066 & 2066.1610 & 0.05150000 & 13  \\
 FeII 2249 & 2249.8768 & 0.00182100 & 14  \\
 FeII 2260 & 2260.7805 & 0.00244000 & 14  \\
 FeII 2344 & 2344.2140 & 0.11400000 &  3  \\
 FeII 2374 & 2374.4612 & 0.03130000 &  3  \\
 FeII 2382 & 2382.7650 & 0.32000000 &  3  \\
 MnII 2576 & 2576.8770 & 0.35080000 &  1  \\
 FeII 2586 & 2586.6500 & 0.06910000 &  3  \\
 MnII 2594 & 2594.4990 & 0.27100000 &  1  \\
 FeII 2600 & 2600.1729 & 0.23900000 &  3  \\
 MgII 2796 & 2796.3520 & 0.61230000 & 15  \\
 MgII 2803 & 2803.5310 & 0.30540000 & 15  \\
  MgI 2852 & 2852.9642 & 1.81000000 &  1  \\
\tableline
\end{tabular}
\end{center}
Key to References -- 1:
\cite{morton91}; 2: \cite{howk00}; 3:
\cite{morton03}; 4: \cite{tripp96}; 5:
\cite{fedchak99}; 
7: \cite{fedchak00}; 9: \cite{schect98}; 11:
\cite{bergs93b}; 12: \cite{wiese01}; 13:
\cite{bergs93}; 14: \cite{bergs94}; 15:
\cite{verner96}; 17: \cite{bergs96}
\end{table*}

In Figure~\ref{fig:civtst} we present a quantitative test of our error analysis
by plotting $\N{C^{3+}}$ measured independently from the C\,IV 1548 and 1550
transitions of $\approx 20$ damped systems.  With the exception of one or
two data points which may suffer from unidentified blending, the measurements
are consistent with a reduced $\chi^2_\nu \approx 1$.  One notes no
obvious effect from line saturation although several obvious cases are
not included in Figure~\ref{fig:civtst}.

To determine the $\N{HI}$ value of the damped \lya systems in our sample,
we have analysed the fluxed \lya profiles from our echellette observations.
We have found this to be a crucial aspect of the \lya profile
fitting process.  In particular, we suspect that by continuum fitting the
data prior to performing the \lya profile analysis that we tended to 
underestimate the $\N{HI}$ values.  It is possible that this systematic
error holds for other studies where the authors continuum fit the quasar
prior to the damped \lya analysis.
Although the absolute fluxing of the ESI data is not accurate because of
uncorrected slit losses, the
relative fluxing is far superior to results obtained for HIRES
and other echelle observations.  In short, we have found that the resolution
and fluxing of the ESI instrument is nearly ideal for this analysis although
we suspect that higher sampling would be advantageous.  We emphasize
that our analysis is not strictly quantitative; the adopted $\N{HI}$ values
do not correspond to a minimization of $\chi^2$ for regions of the spectrum.
Instead, we present our favored value and place conservative error estimates
to these values.  Our experience with line-fitting is that standard $\chi^2$
analysis would yield unrealistically small error estimates, in particular
because the continuum uncertainty is difficult to quantify.  The error
estimates provided in this paper are sufficiently conservative, however,
that the reader should consider them to be roughly $95\%$~c.l.

\begin{table}[ht]\footnotesize
\begin{center}
\caption{ 
{\sc SOLAR ABUNDANCES\label{tab:solar}}}
\begin{tabular}{lc}
\tableline
\tableline
Elm &$\epsilon^a$\\
\tableline
Al  &   6.49  \\
Ar  &   6.52  \\
As  &   2.37  \\
B   &   2.79  \\
C   &   8.59  \\
Ca  &   6.35  \\
Cl  &   5.28  \\
Co  &   4.91  \\
Cr  &   5.67  \\
Cu  &   4.29  \\
Fe  &   7.50  \\
Ga  &   3.13  \\
Ge  &   3.63  \\
H   &  12.00  \\
K   &   5.13  \\
Kr  &   3.23  \\
Li  &   3.31  \\
Mg  &   7.58  \\
Mn  &   5.53  \\
N   &   7.93  \\
Na  &   6.32  \\
Ne  &   8.08  \\
Ni  &   6.25  \\
O   &   8.74  \\
P   &   5.53  \\
Pb  &   2.06  \\
S   &   7.20  \\
Si  &   7.56  \\
Sn  &   2.14  \\
Ti  &   4.94  \\
Xe  &   2.23  \\
Zn  &   4.67  \\
Zr  &   2.61  \\
\tableline
\end{tabular}
\end{center}
$^a$Meteoritic abundances except C,N,O
\end{table}

To determine $\N{HI}$ we tied the
centroid of a Voigt profile to the metal-line 
transitions\footnote{Even when the metal-line transitions show broad 
(up to 200 km/s), complex kinematics we are able to estimate the systematic 
velocity centroid to within $\sim$30 km/s i.e. $\delta z \simeq 0.0001$.}
and varied the $\N{HI}$ value and the quasar continuum simultaneously to
achieve a solution which ideally matched both the core and wings of the
profile.  In all cases, we required that the solution never fell 
significantly below the data values in any portion of the \lya profile.
Nevertheless, because unidentified \lya blending is always present, it 
is plausible that the resulting $\N{HI}$ values are over-estimated in some cases.
On the other hand, one is systematically likely to underestimate the
quasar continuum and therefore underestimate $\N{HI}$.  It would be valuable
to fully assess these effects with numerical simulations.
As is standard practice in DLA searches 
we rejected Broad Absorption Line quasars which make DLA identification 
extremely difficult and do not consider systems found within 
3000 km/s of the quasar emission redshift to be DLAs owing to the chance 
that these systems are physically associated with the quasar
\citep[but see][]{moller98,ellison02}.

Throughout the paper, we adopt the wavelengths
and oscillator strengths presented in Table~\ref{tab:fosc}. 
When possible, we have adopted laboratory values for the oscillator
strengths. 
We also adopt solar meteoritic abundances from \cite{grvss96} and 
\cite{holweger01} for the elemental abundances listed in the paper
(Table~\ref{tab:solar}).

\section{INDIVIDUAL SYSTEMS}

In this section we present the metal-line and \lya profiles of the
DLA and tabulate the measured ionic column densities.  Regarding
the metal-line figures, $v=0$ \kms\ is arbitrarily defined and blends from
coincident absorption lines are indicated by dotted lines.
The normalized continuum is shown by the dotted line and the zero-flux
line is marked by a dash-dot line.
For the low-ion transitions -- 
the dominant ionic transition of a given element in 
an H\,I region -- we tabulate $N_{adopt}$ by taking the weighted
mean (or limit) of all measured transitions.  For these cases, we also
convert $N_{adopt}$ into an elemental abundance [X/H] assuming no
ionization corrections and adopting solar meteoritic abundances.
In the \lya figures, a vertical dashed line indicates the centroid
of the \lya profile while the dotted line traces the quasar continuum.
Furthermore, the best fit solution for the \lya profile is traced by
a solid line and is bounded by two dash-dot lines which indicate our
estimate of the error in the \lya solution.

\begin{figure}[ht]
\begin{center}
\includegraphics[height=3.6in, width=2.8in,angle=90]{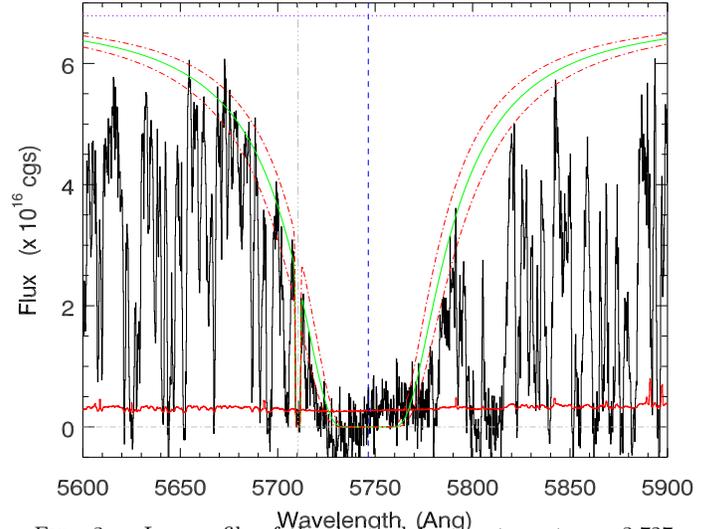}
\figcaption{Lya profile of the damped \lya system at $z=3.727$
toward SDSS0127-00.  
The overplotted solid line and accompanying
dash-dot lines trace the best fit solution and the estimated bounds
corresponding to $\log \N{HI} = 21.15^{+0.10}_{-0.10}$.  
Examining the region at $\lambda \approx 5760$\AA, one notes
the data have flux significantly higher than the overplotted
Voigt profile.  This is due an incorrect matching in the gain
of the two amplifiers for ESI ($\S$~\ref{sec:redux}).
This solution is well constrained
by the left wing of the \lya profile and is simplified by the relatively
well behaved continuum (dotted line) across this region.  \label{fig:sdss0127_lya}}
\end{center}
\end{figure}

\begin{figure}[ht]
\begin{center}
\includegraphics[height=6.1in, width=3.9in]{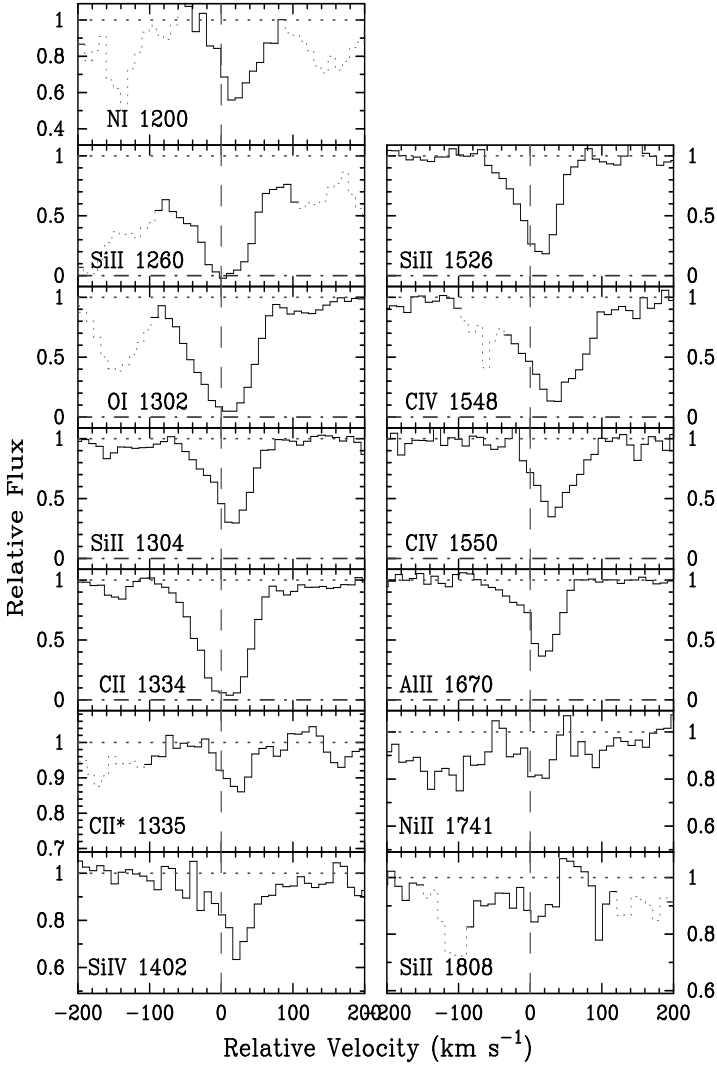}
\figcaption{Velocity plot of the metal-line transitions for the 
damped \lya system at $z = 3.727$ toward SDSS01027-00.
The vertical line at $v=0$ corresponds to $z = 3.72743$.  
\label{fig:sdss0127_mtl}}
\end{center}
\end{figure}

\begin{table}[ht]\footnotesize
\begin{center}
\caption{ {\sc
IONIC COLUMN DENSITIES: SDSS0127-00, $z = 3.727$ \label{tab:SDSS0127-00_3.727}}}
\begin{tabular}{lcccc}
\tableline
\tableline
Ion & $\lambda$ & AODM & $N_{\rm adopt}$ & [X/H] \\
\tableline
C  II &1334.5&$>14.672$&$>14.672$&$>-3.068$\\  
C  II*&1335.7&$13.139 \pm  0.064$\\  
C  IV &1550.8&$14.129 \pm  0.036$\\  
N  I  &1200.2&$14.028 \pm  0.037$&$14.028 \pm  0.037$&$-3.052 \pm  0.107$\\  
O  I  &1302.2&$>15.185$&$>15.185$&$>-2.705$\\  
Al II &1670.8&$>12.743$&$>12.743$&$>-2.897$\\  
Si II &1304.4&$>14.306$&$>14.306$&$>-2.404$\\  
Si II &1526.7&$>14.221$\\  
Si II &1808.0&$<14.760$\\  
Si IV &1402.8&$13.439 \pm  0.049$\\  
Ni II &1741.6&$<13.691$&$<13.691$&$<-1.709$\\  
\tableline
\end{tabular}
\end{center}
\end{table}

\subsection{SDSS0127-00, $z = 3.727$ \label{subsec:SDSS0127-00_3.727}}

The quasar SDSS0127-00 is a member of the Early Data Release of the
Sloan Digital Sky Survey 
\citep{fan01}.  Even in the SDSS
discovery spectrum, a damped \lya system is obvious at $\approx 5770$\AA.
In Figure~\ref{fig:sdss0127_lya} we present the \lya profile and our 
solution which is well constrained by the left wing of the profile. 
One notes a complication with the current IDL package: 
normalizing the gain of the two amplifiers. 
This issue is only manifest
at $\lambda \approx 5770$\AA\ in the spectrum and only arises when
the data frame had a significantly different gain ratio from the
flats.  Of our observations, this is the most severe case.
Future versions of the ESI IDL package will provide a means for correcting
this effect.
The metal-line profiles of this damped system are shown in 
Figure~\ref{fig:sdss0127_mtl} and the ionic column densities 
are listed in Table~\ref{tab:SDSS0127-00_3.727}.


\begin{figure}[ht]
\begin{center}
\includegraphics[height=3.6in, width=2.8in,angle=90]{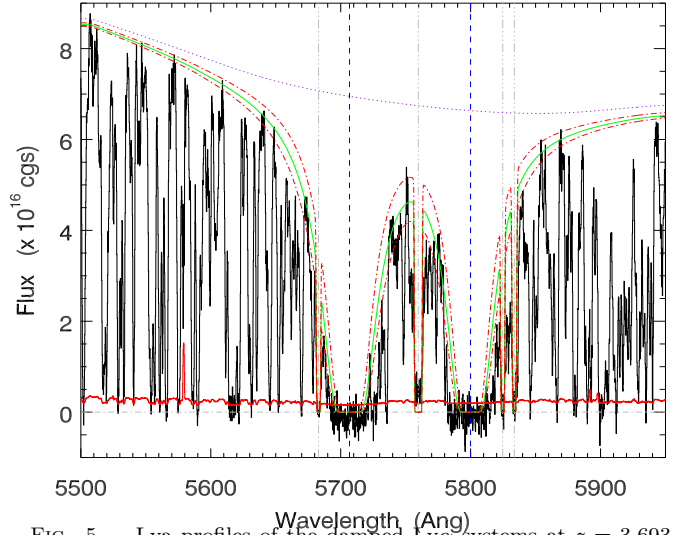}
\figcaption{Lya profiles of the damped \lya systems at $z=3.693$
and 3.774 toward PSS0133+0400.  
The overplotted solid line and accompanying
dash-dot lines trace the best fit solution and the estimated 
bounds corresponding to 
$\log \N{HI} = 20.70^{+0.10}_{-0.15}$ and
$\log \N{HI} = 20.55^{+0.10}_{-0.15}$ respectively.
The vertical dash-dot lines correspond to coincident \lya forest
lines that we included in the analysis.
While the $z=3.693$ DLA is well constrained by the observations,
the core of the $z=3.77$ DLA is significantly wider than the 
\lya wings suggest.  Therefore, this DLA is more sensitive to the
continuum placement (dotted line) and the $\N{HI}$ value is more
uncertain.  Note that we introduced additional components to investigate
their effect on the DLA profiles and found minimal effect.
\label{fig:pss0133_lya}}
\end{center}
\end{figure}

\begin{table}[ht]\footnotesize
\begin{center}
\caption{ {\sc
IONIC COLUMN DENSITIES: PSS0133+0400, $z = 3.693$ \label{tab:PSS0133+0400_3.693}}}
\begin{tabular}{lcccc}
\tableline
\tableline
Ion & $\lambda$ & AODM & $N_{\rm adopt}$ & [X/H] \\
\tableline
C  II &1334.5&$>14.160$&$>14.160$&$>-3.130$\\  
C  II*&1335.7&$<12.563$\\  
C  IV &1548.2&$13.523 \pm  0.028$\\  
C  IV &1550.8&$<13.787$\\  
Al II &1670.8&$12.372 \pm  0.031$&$12.372 \pm  0.031$&$-2.818 \pm  0.129$\\  
Si II &1526.7&$>14.256$&$>14.256$&$>-2.004$\\  
Si II &1808.0&$<14.546$\\  
Si IV &1402.8&$<12.724$\\  
Fe II &1608.5&$13.511 \pm  0.065$&$13.511 \pm  0.065$&$-2.689 \pm  0.141$\\  
\tableline
\end{tabular}
\end{center}
\end{table}

\begin{table}[ht]\footnotesize
\begin{center}
\caption{ {\sc
IONIC COLUMN DENSITIES: PSS0133+0400, $z = 3.774$ \label{tab:PSS0133+0400_3.774}}}
\begin{tabular}{lcccc}
\tableline
\tableline
Ion & $\lambda$ & AODM & $N_{\rm adopt}$ & [X/H] \\
\tableline
C  II &1334.5&$>15.309$&$>15.309$&$>-1.831$\\  
C  II*&1335.7&$<13.949$\\  
C  IV &1548.2&$>14.306$\\  
C  IV &1550.8&$14.355 \pm  0.016$\\  
Al II &1670.8&$>13.792$&$>13.792$&$>-1.248$\\  
Al III&1854.7&$13.164 \pm  0.027$\\  
Al III&1862.8&$13.108 \pm  0.054$\\  
Si II &1526.7&$>15.079$&$15.466 \pm  0.035$&$-0.644 \pm  0.130$\\  
Si II &1808.0&$15.466 \pm  0.035$\\  
Si IV &1393.8&$13.799 \pm  0.011$\\  
Si IV &1402.8&$13.894 \pm  0.015$\\  
Ti II &1910.6&$13.041 \pm  0.066$&$13.041 \pm  0.066$&$-0.449 \pm  0.141$\\  
Cr II &2056.3&$<13.240$&$<13.240$&$<-0.980$\\  
Fe II &1608.5&$>14.868$&$>14.868$&$>-1.182$\\  
Fe II &1611.2&$<15.375$\\  
Ni II &1370.1&$13.999 \pm  0.029$&$13.978 \pm  0.026$&$-0.822 \pm  0.128$\\  
Ni II &1741.6&$13.899 \pm  0.066$\\  
Zn II &2026.1&$<13.099$&$<13.099$&$<-0.121$\\  
\tableline
\end{tabular}
\end{center}
\end{table}

\begin{figure}[ht]
\begin{center}
\includegraphics[height=5.5in, width=3.7in]{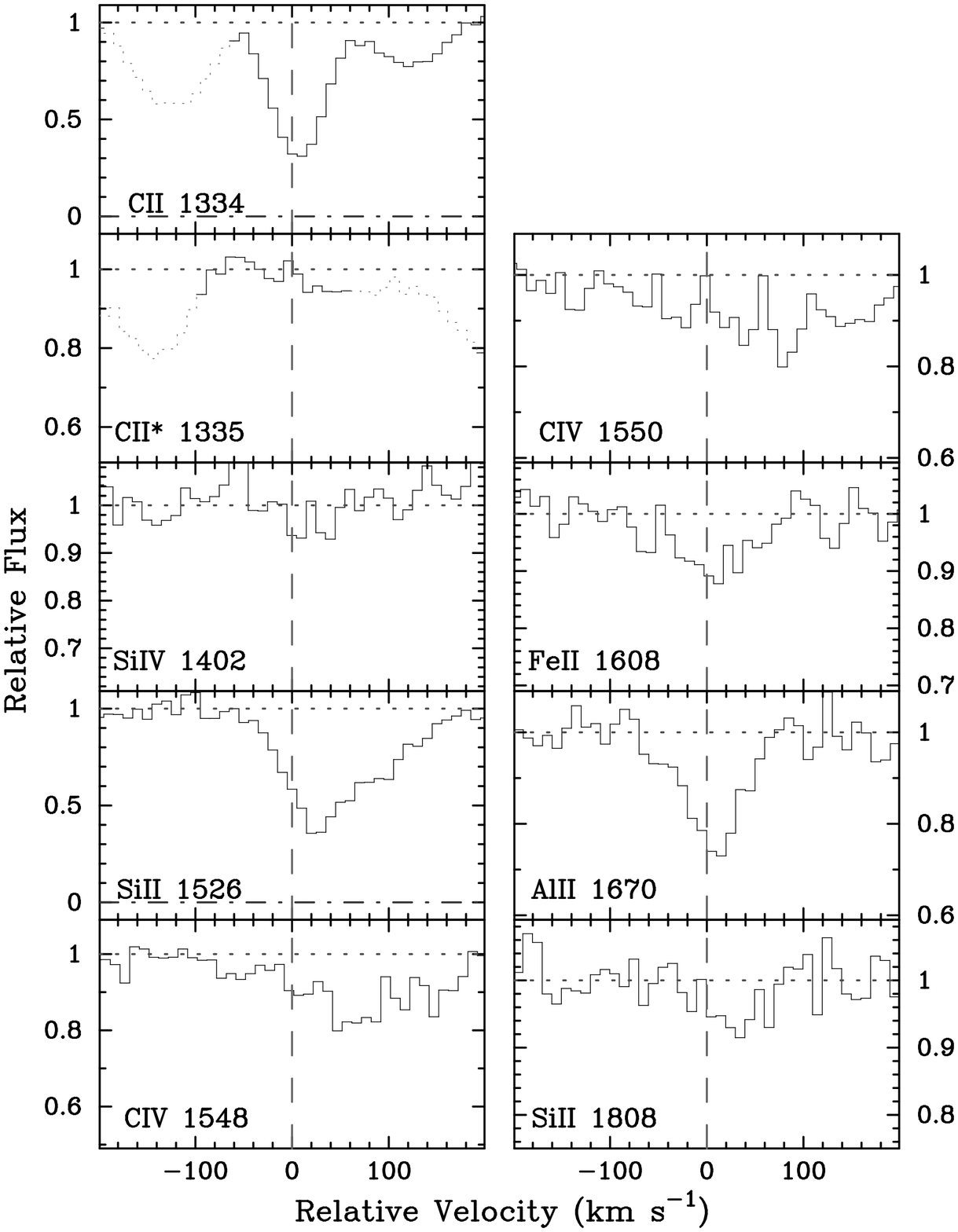}
\figcaption{Velocity plot of the metal-line transitions for the 
damped \lya system at $z = 3.693$ toward PSS0133+0400.
The vertical line at $v=0$ corresponds to $z = 3.6926$. 
\label{fig:pss0133A_mtl}}
\end{center}
\end{figure}

\begin{figure}[ht]
\begin{center}
\includegraphics[height=6.1in, width=3.9in]{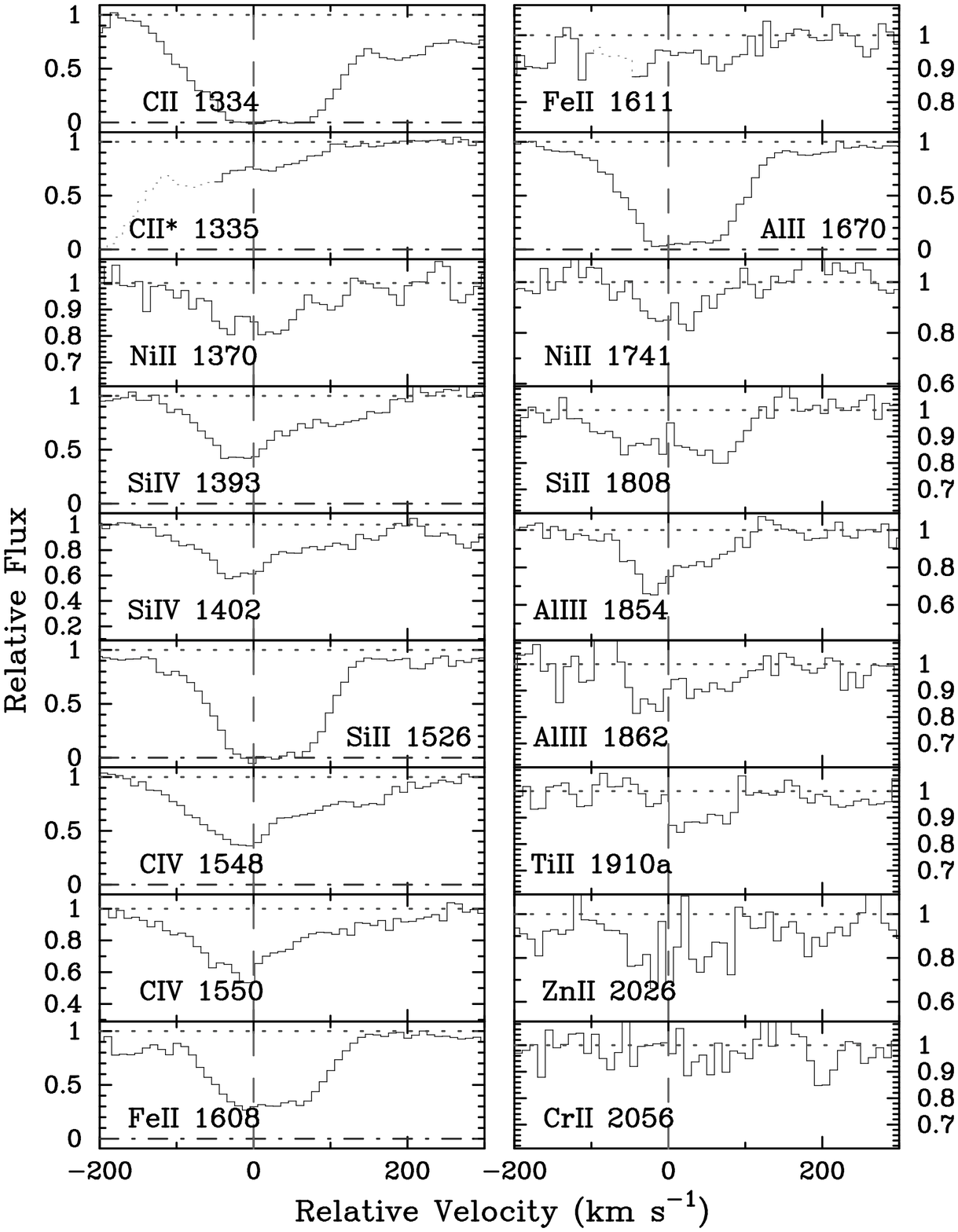}
\figcaption{Velocity plot of the metal-line transitions for the 
damped \lya system at $z = 3.774$ toward PSS0133+0400.
The vertical line at $v=0$ corresponds to $z = 3.77356$.
\label{fig:pss0133B_mtl}}
\end{center}
\end{figure}

\begin{figure}[ht]
\begin{center}
\includegraphics[height=6.1in, width=3.9in]{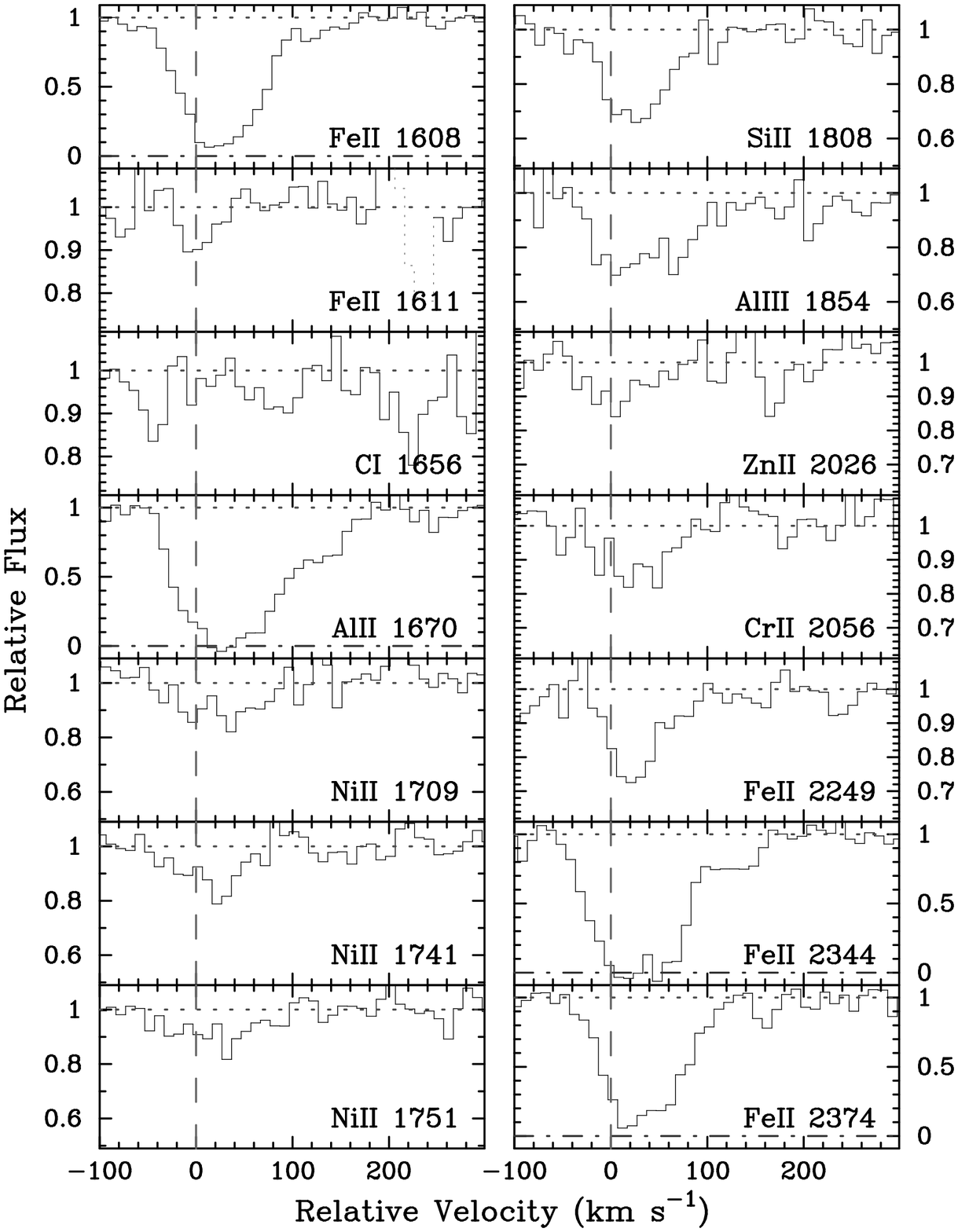}
\end{center}
\end{figure}

\subsection{PSS0133+0400, $z = 3.693,3.774$ \label{subsec:PSS0133+0400_3.693}}

We identified two damped systems toward this quasar from the 
Palomar Sky Survey \citep[PSS][]{djg98}.  
The \lya profiles of the damped systems are 
shown in Figure~\ref{fig:pss0133_lya}.  Because of their proximity, 
we found it necessary to fit the two profiles simultaneously.
While the damped system at $z=3.693$ is well constrained by the observations,
the core of the $z=3.774$ DLA is significantly wider than the damping wings
suggest. This indicates blending from coincident \lya forest lines
with $\N{HI} > 10^{15} \cm{-2}$ and the net result is that the $\N{HI}$ value
is more sensitive to continuum placement.  

Velocity profiles of the metal-line transitions are presented in 
Figures~\ref{fig:pss0133A_mtl} and \ref{fig:pss0133B_mtl} and
the measured ionic column densities are listed in 
Tables~\ref{tab:PSS0133+0400_3.693} and \ref{tab:PSS0133+0400_3.774}.
The system at $z=3.774$ is notable for exhibiting the 
highest DLA metallicity at $z>3$ and also a relatively low level
of differential depletion ([Si/Ni]~$\approx 0.2$).  Follow-up echelle
observations of this sightline may be particularly valuable.


\begin{figure}[ht]
\begin{center}
\includegraphics[height=3.6in, width=2.8in,angle=90]{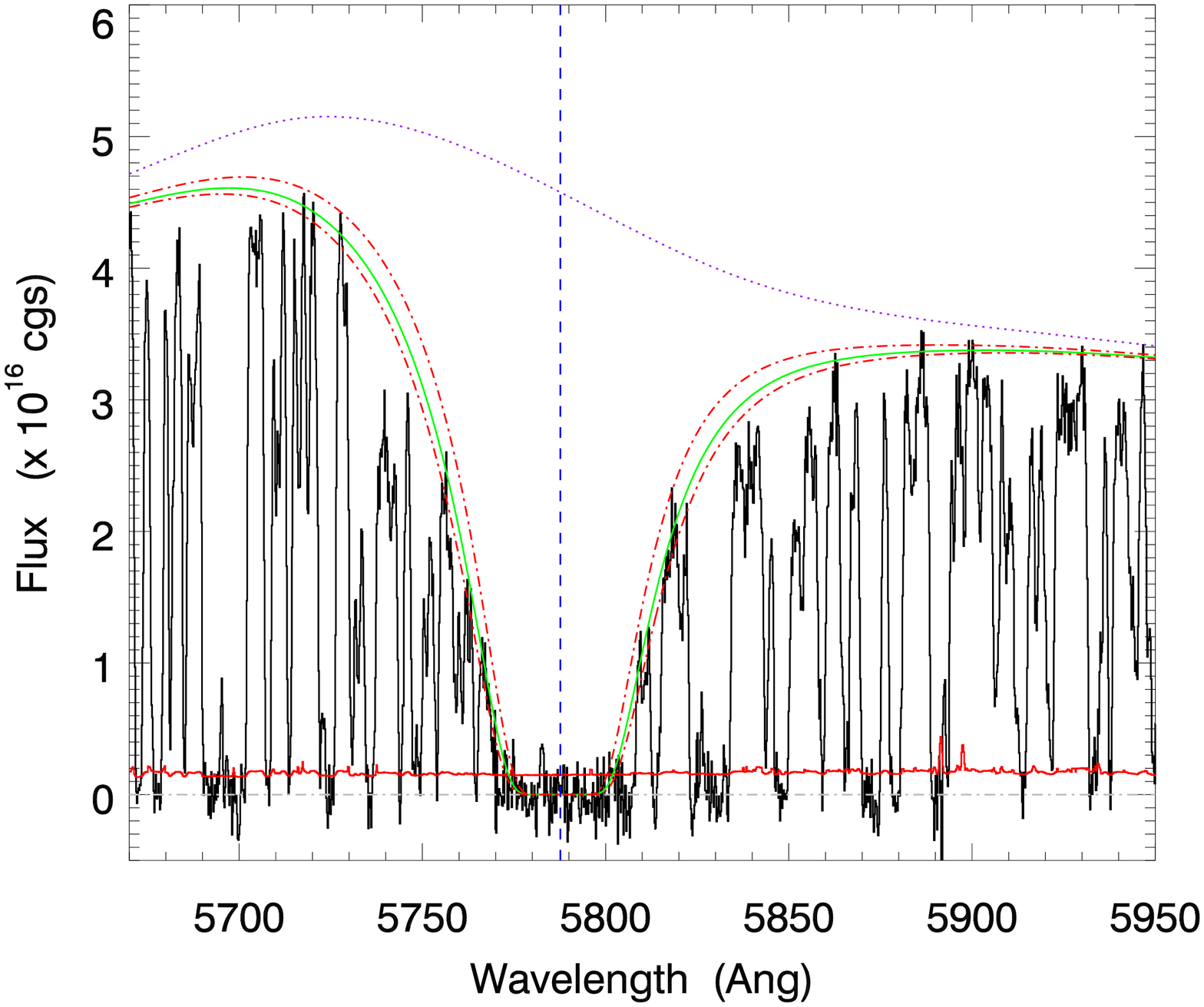}
\figcaption{Lya profile of the damped \lya system at $z=3.761$
toward PSS0134+3317.
The overplotted solid line and accompanying
dash-dot lines trace the best fit solution and the estimated 
bounds corresponding to 
$\log \N{HI} = 20.85^{+0.05}_{-0.10}$.  
Although this DLA is positioned on the O\,VI peak of the quasar,
the core is well resolved and we consider the $\N{HI}$ to be well
constrained.  \label{fig:px0134_lya}}
\end{center}
\end{figure}

\begin{table}[ht]\footnotesize
\begin{center}
\caption{ {\sc
IONIC COLUMN DENSITIES: PSS0134+3317, $z = 3.761$ \label{tab:PSS0134+3317_3.761}}}
\begin{tabular}{lcccc}
\tableline
\tableline
Ion & $\lambda$ & AODM & $N_{\rm adopt}$ & [X/H] \\
\tableline
C  IV &1548.2&$13.282 \pm  0.103$\\  
C  IV &1550.8&$13.237 \pm  0.138$\\  
Al II &1670.8&$12.612 \pm  0.027$&$12.612 \pm  0.027$&$-2.728 \pm  0.080$\\  
Si II &1193.3&$>13.889$&$>13.889$&$>-2.521$\\  
Si II &1808.0&$<14.681$\\  
Si III&1206.5&$>13.815$\\  
\tableline
\end{tabular}
\end{center}
\end{table}

\begin{figure}[ht]
\begin{center}
\includegraphics[height=6.1in, width=3.9in]{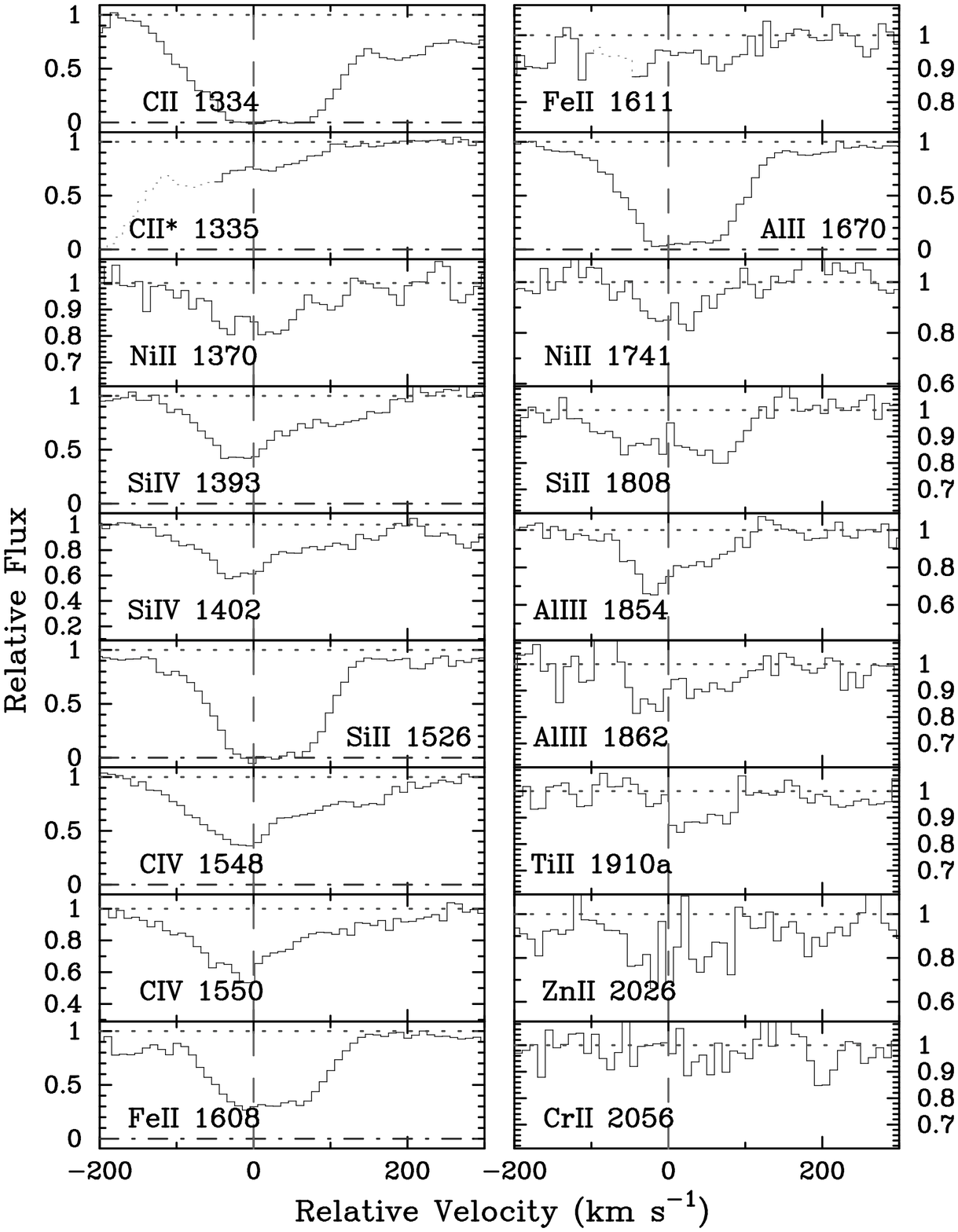}
\figcaption{Velocity plot of the metal-line transitions for the 
damped \lya system at $z = 3.761$ toward PSS0134+3317.
The vertical line at $v=0$ corresponds to $z = 3.76087$.  
\label{fig:px0134_mtl}}
\end{center}
\end{figure}

\subsection{PSS0134+3317, $z = 3.761$ \label{subsec:PSS0134+3317_3.761}}

This quasar is a member of the PSS sample and the damped \lya systems
was discovered by \cite{peroux01}.
The \lya profile corresponding to this DLA is given in 
Figure~\ref{fig:px0134_lya}.  Although the profile is coincident with
the O\,VI peak of PSS0134+3307, the H\,I column density is well constrained
by the observations.  This DLA has a low metallicity
and the majority of the strongest transitions are lost in the \lya forest.
The few valuable transitions are shown in Figure~\ref{fig:px0134_mtl}
and provide column density limits for only Si$^+$ and Al$^+$ 
(Table~\ref{tab:PSS0134+3317_3.761}).


\begin{table}[ht]\footnotesize
\begin{center}
\caption{ {\sc
IONIC COLUMN DENSITIES: PSS0209+0517, $z = 3.667$ \label{tab:PSS0209+0517_3.667}}}
\begin{tabular}{lcccc}
\tableline
\tableline
Ion & $\lambda$ & AODM & $N_{\rm adopt}$ & [X/H] \\
\tableline
C  I  &1560.3&$<12.910$\\  
C  IV &1548.2&$13.181 \pm  0.049$\\  
O  I  &1302.2&$>14.774$&$>14.774$&$>-2.416$\\  
Al II &1670.8&$12.521 \pm  0.037$&$12.520 \pm  0.037$&$-2.420 \pm  0.107$\\  
Al III&1854.7&$<12.217$\\  
Si II &1260.4&$>14.107$&$>14.107$&$>-1.903$\\  
Si II &1526.7&$>13.885$\\  
Si II &1808.0&$<14.455$\\  
Si III&1206.5&$>13.846$\\  
Si IV &1402.8&$13.161 \pm  0.051$\\  
Cr II &2056.3&$<13.079$&$<13.079$&$<-1.041$\\  
Fe II &1608.5&$13.635 \pm  0.053$&$13.635 \pm  0.053$&$-2.315 \pm  0.113$\\  
Ni II &1709.6&$<13.249$&$<13.250$&$<-1.450$\\  
Ni II &1751.9&$<13.621$\\  
\tableline
\end{tabular}
\end{center}
\end{table}

\begin{table}[ht]\footnotesize
\begin{center}
\caption{ {\sc
IONIC COLUMN DENSITIES: PSS0209+0517, $z = 3.864$ \label{tab:PSS0209+0517_3.864}}}
\begin{tabular}{lcccc}
\tableline
\tableline
Ion & $\lambda$ & AODM & $N_{\rm adopt}$ & [X/H] \\
\tableline
C  II &1334.5&$>14.020$&$>14.020$&$>-3.120$\\  
C  II*&1335.7&$<12.946$\\  
C  IV &1548.2&$13.311 \pm  0.066$\\  
C  IV &1550.8&$13.373 \pm  0.095$\\  
O  I  &1302.2&$>14.368$&$>14.368$&$>-2.922$\\  
Al II &1670.8&$12.146 \pm  0.048$&$12.146 \pm  0.048$&$-2.894 \pm  0.111$\\  
Al III&1854.7&$<12.150$\\  
Si II &1260.4&$>13.819$&$13.463 \pm  0.034$&$-2.647 \pm  0.106$\\  
Si II &1304.4&$13.542 \pm  0.044$\\  
Si II &1526.7&$13.396 \pm  0.052$\\  
Si IV &1393.8&$12.922 \pm  0.052$\\  
Si IV &1402.8&$12.934 \pm  0.099$\\  
Fe II &1608.5&$<13.340$&$<13.340$&$<-2.710$\\  
Ni II &1370.1&$<13.123$&$<13.123$&$<-1.677$\\  
Ni II &1741.6&$<13.236$\\  
\tableline
\end{tabular}
\end{center}
\end{table}

\begin{figure}[ht]
\begin{center}
\includegraphics[height=3.6in, width=2.8in,angle=90]{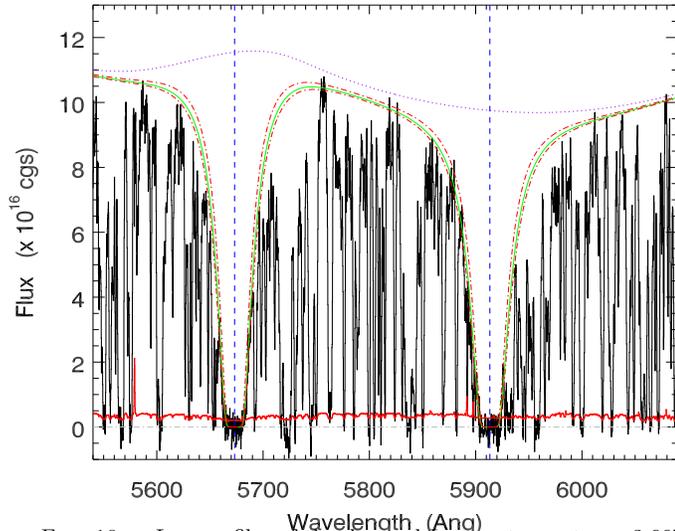}
\figcaption{Lya profiles of the damped \lya systems at $z=3.667$
and 3.864 toward PSS0209+0517.  
The overplotted solid line and accompanying
dash-dot lines trace the best fit solution and the estimated 
bounds corresponding to 
$\log \N{HI} = 20.45^{+0.1}_{-0.1}$ and
$\log \N{HI} = 20.55^{+0.1}_{-0.1}$ respectively.
In both cases, the core of the \lya profile is relatively unblended
with coincident \lya lines and provides a good determination of $\N{HI}$
even though the continuum (dotted line) shows modest variation across
the profiles.  Note that the enhancement in the continuum at 
$\lambda \approx 5680$\AA\ is due to O\,VI emission by the QSO.
\label{fig:pss0209_lya}}
\end{center}
\end{figure}

\begin{figure}[ht]
\begin{center}
\includegraphics[height=6.1in, width=3.9in]{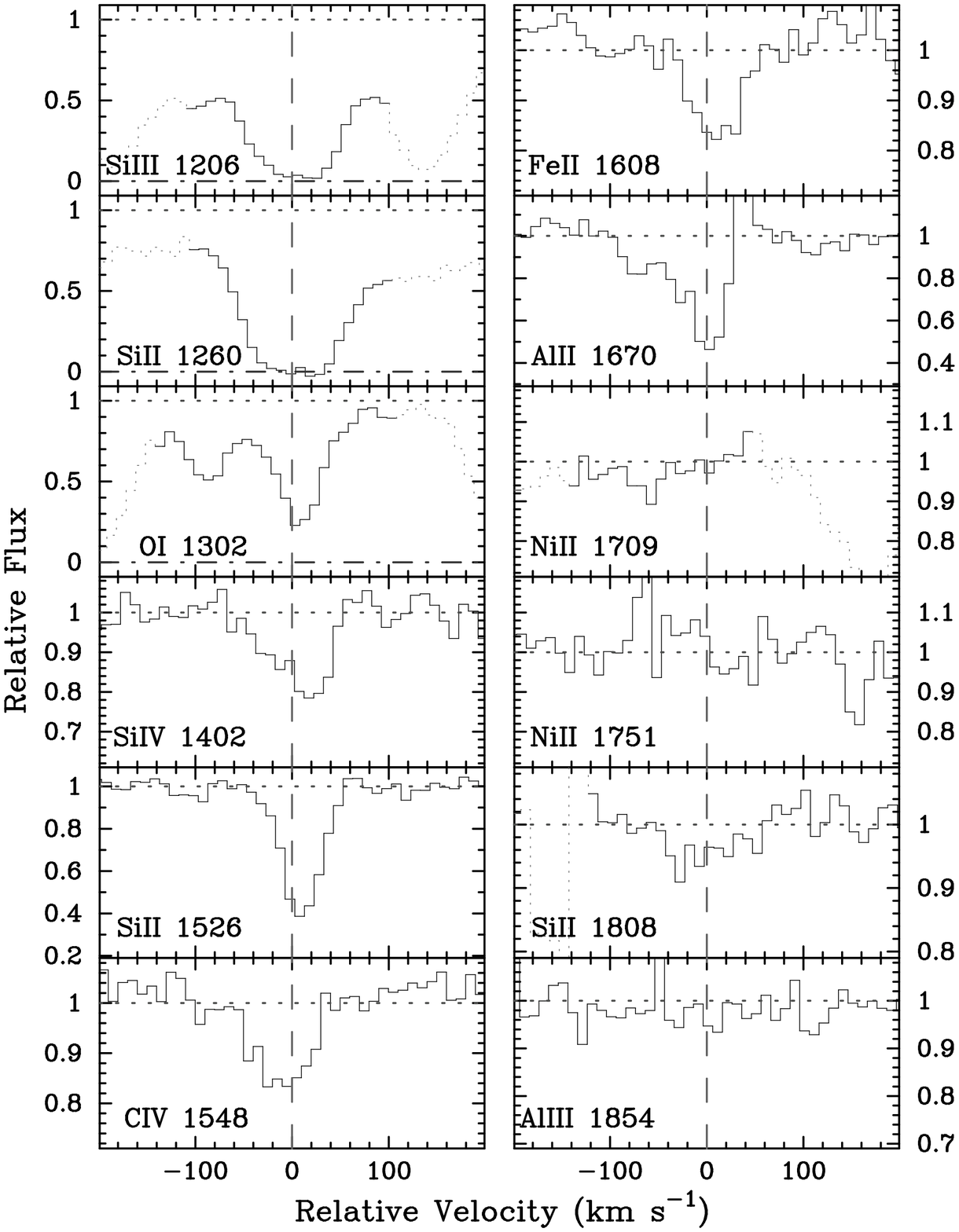}
\figcaption{Velocity plot of the metal-line transitions for the 
damped \lya system at $z = 3.667$ toward PSS0209+0517.
The vertical line at $v=0$ corresponds to $z = 3.66674$.  
\label{fig:pss0209A_mtl}}
\end{center}
\end{figure}

\begin{figure}[ht]
\begin{center}
\includegraphics[height=6.1in, width=3.9in]{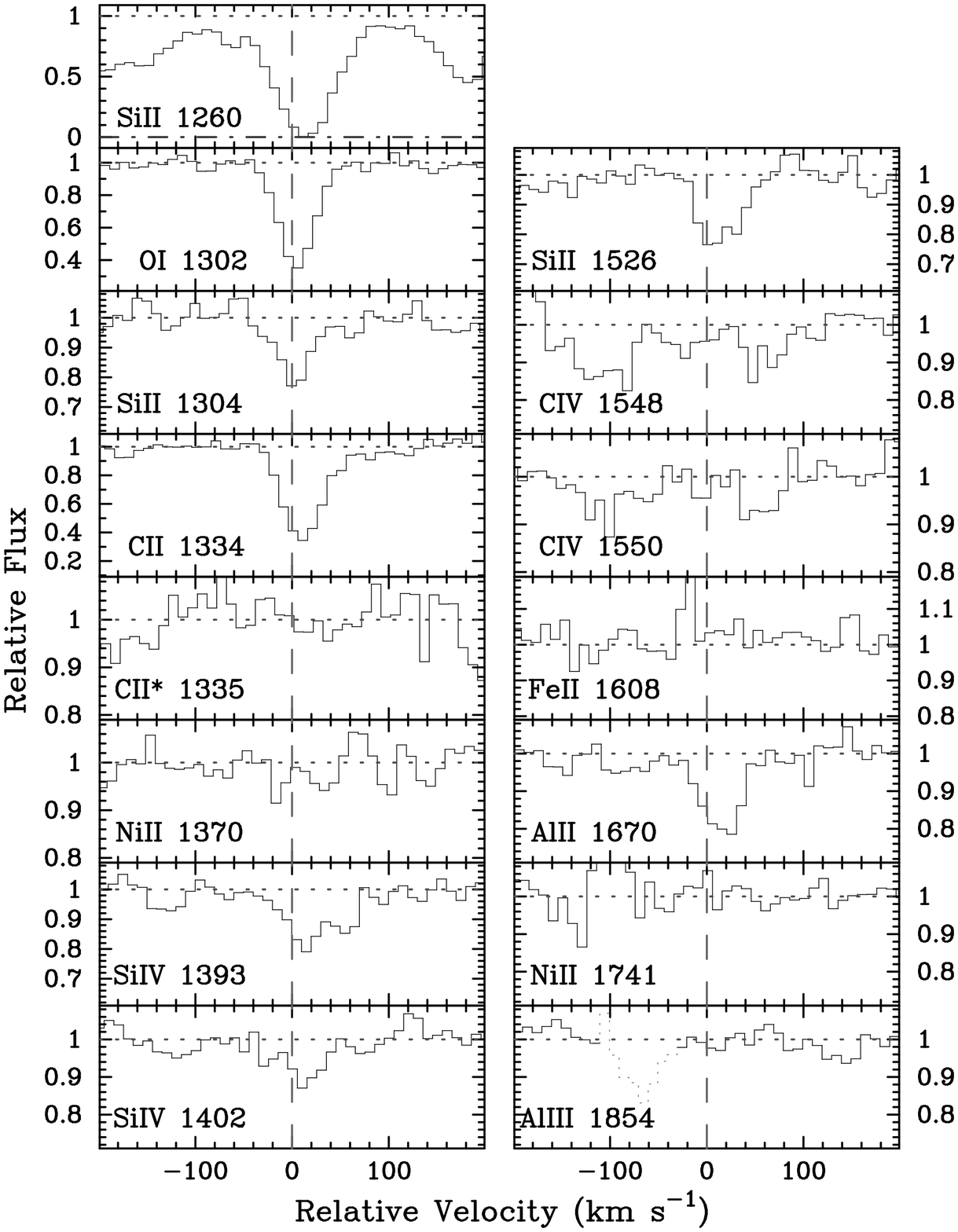}
\figcaption{Velocity plot of the metal-line transitions for the 
damped \lya system at $z = 3.864$ toward PSS0209+0517.
The vertical line at $v=0$ corresponds to $z = 3.86427$.  
\label{fig:pss0209B_mtl}}
\end{center}
\end{figure}

\subsection{PSS0209+0517, $z = 3.667,3.864$ \label{subsec:PSS0209+0517}}

This quasar exhibits two damped \lya systems at $z=3.667$ and $3.864$
(discovered by Peroux et al.\ 2001) 
whose \lya profiles are presented in Figure~\ref{fig:pss0209_lya}.
The cores of both \lya profiles are well determined and provide a
precise determination of $\N{HI}$.  The metal-line profiles of
the two DLA are presented in Figures~\ref{fig:pss0209A_mtl} and
\ref{fig:pss0209B_mtl} and the ionic column densities are listed
in Tables~\ref{tab:PSS0209+0517_3.667}, \ref{tab:PSS0209+0517_3.864}.


\begin{figure}[ht]
\begin{center}
\includegraphics[height=3.6in, width=2.8in,angle=90]{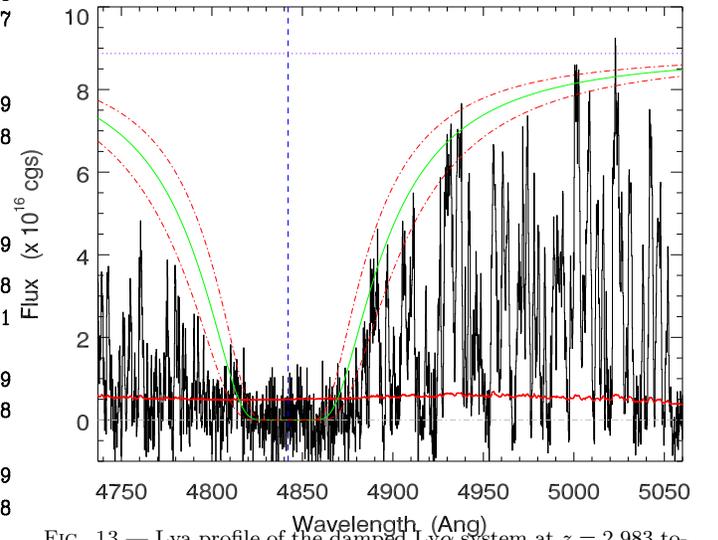}
\figcaption{Lya profile of the damped \lya system at $z=2.983$
toward BRJ0426-2202.
The overplotted solid line and accompanying
dash-dot lines trace the best fit solution and the estimated 
bounds corresponding to 
$\log \N{HI} = 21.50^{+0.15}_{-0.15}$.  
The blue wing of the \lya profile is severely blended with the
higher order Lyman series of a higher redshift Lyman limit system.
The red wing is reasonably free from contamination, although the
resulting fit is sensitive to the continuum placement.  This results
in a more uncertain determination of $\N{HI}$.
\label{fig:brj0426_lya}}
\end{center}
\end{figure}

\begin{figure}[ht]
\begin{center}
\includegraphics[height=6.1in, width=3.9in]{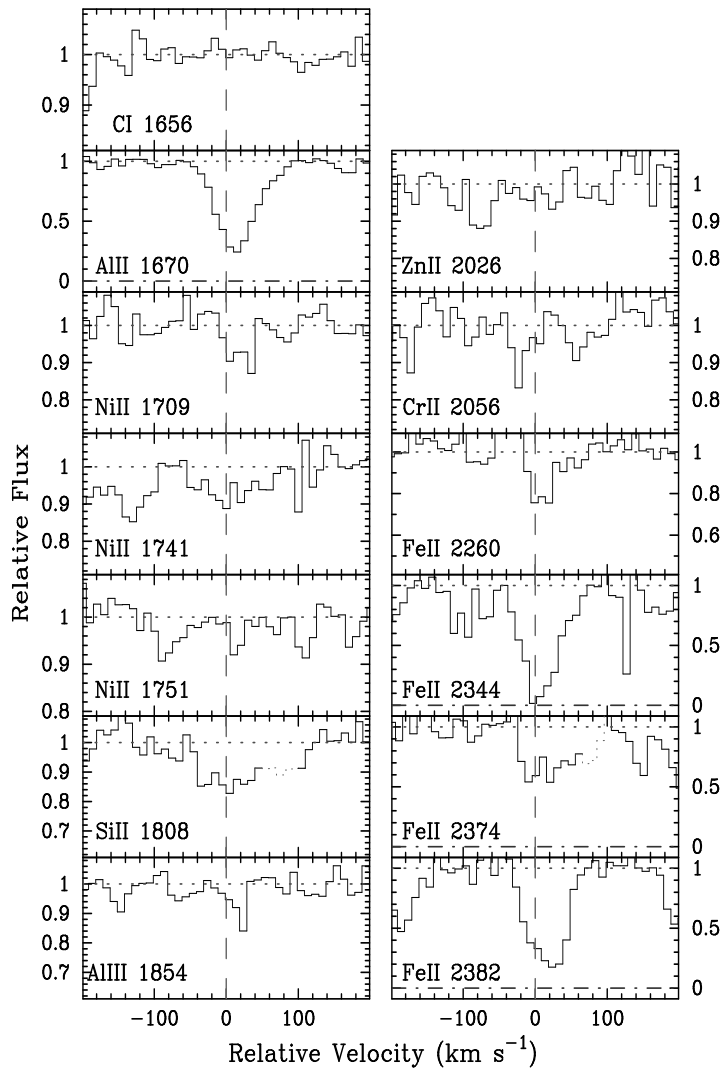}
\figcaption{Velocity plot of the metal-line transitions for the 
damped \lya system at $z = 2.983$ toward BRJ0426-2202.
The vertical line at $v=0$ corresponds to $z = 2.98308$.  
\label{fig:brj0426_mtl}}
\end{center}
\end{figure}

\begin{table}[ht]\footnotesize
\begin{center}
\caption{ {\sc
IONIC COLUMN DENSITIES: BRJ0426-2202, $z = 2.983$ \label{tab:BRJ0426-2202_2.983}}}
\begin{tabular}{lcccc}
\tableline
\tableline
Ion & $\lambda$ & AODM & $N_{\rm adopt}$ & [X/H] \\
\tableline
C  I  &1656.3&$<13.020$\\  
Al II &1670.8&$>12.931$&$>12.931$&$>-3.059$\\  
Al III&1854.7&$<12.196$\\  
Si II &1808.0&$<14.996$&$<14.996$&$<-2.064$\\  
Cr II &2056.3&$<12.901$&$<12.901$&$<-2.269$\\  
Fe II &2260.8&$<14.887$&$14.152 \pm  0.067$&$-2.848 \pm  0.164$\\  
Fe II &2344.2&$>14.240$\\  
Fe II &2374.5&$14.152 \pm  0.067$\\  
Fe II &2382.8&$>13.625$\\  
Ni II &1709.6&$<13.373$&$<13.373$&$<-2.377$\\  
Ni II &1741.6&$<13.512$\\  
Ni II &1751.9&$<13.403$\\  
Zn II &2026.1&$<12.165$&$<12.165$&$<-2.005$\\  
\tableline
\end{tabular}
\end{center}
\end{table}

\subsection{BRJ0426-2202, $z = 2.983$ \label{subsec:BRJ0426-2202_2.983}}

This quasar was discovered by \cite{storrie01} and \cite{peroux01}
identified the damped \lya system.
The \lya profile of this large H\,I DLA is difficult
to constrain because of a higher redshift Lyman limit system which is
significantly suppressing the quasar flux below 4800\AA.  Nevertheless,
the red wing of the DLA is relatively free from contamination and a
reasonable fit to the $\N{HI}$ value was achieved (Figure~\ref{fig:brj0426_lya}).
We note that our value is 0.4~dex larger than the value reported
by \cite{peroux01} which may suggest an uncertainty in $\N{HI}$ even larger
than the 0.15~dex we adopt.

Because the DLA has a substantially lower redshift than the quasar and
a low metallicity, it exhibits only a few metal-line transitions
(Figure~\ref{fig:brj0426_mtl}).  The resulting ionic column densities
are given in Table~\ref{tab:BRJ0426-2202_2.983}.

\begin{table}[ht]\footnotesize
\begin{center}
\caption{ {\sc
IONIC COLUMN DENSITIES: FJ0747+2739, $z = 3.423$ \label{tab:FJ0747+2739_3.423}}}
\begin{tabular}{lcccc}
\tableline
\tableline
Ion & $\lambda$ & AODM & $N_{\rm adopt}$ & [X/H] \\
\tableline
C  II &1334.5&$>15.096$&$>15.096$&$>-2.344$\\  
C  IV &1548.2&$>14.308$\\  
C  IV &1550.8&$>14.403$\\  
Al II &1670.8&$>13.204$&$>13.204$&$>-2.136$\\  
Al III&1854.7&$13.060 \pm  0.031$\\  
Si II &1193.3&$>14.481$&$>14.539$&$>-1.871$\\  
Si II &1526.7&$>14.539$\\  
Si II &1808.0&$<14.932$\\  
Fe II &1608.5&$>14.430$&$>14.430$&$>-1.920$\\  
Fe II &1611.2&$<14.832$\\  
Fe II &2260.8&$<14.701$\\  
Ni II &1751.9&$<13.271$&$<13.272$&$<-1.828$\\  
\tableline
\end{tabular}
\end{center}
\end{table}

\begin{table}[ht]\footnotesize
\begin{center}
\caption{ {\sc
IONIC COLUMN DENSITIES: FJ0747+2739, $z = 3.900$ \label{tab:FJ0747+2739_3.900}}}
\begin{tabular}{lcccc}
\tableline
\tableline
Ion & $\lambda$ & AODM & $N_{\rm adopt}$ & [X/H] \\
\tableline
C  II &1036.3&$>14.380$&$>14.503$&$>-2.587$\\  
C  II &1334.5&$>14.503$\\  
C  II*&1335.7&$13.259 \pm  0.075$\\  
C  IV &1548.2&$13.715 \pm  0.014$\\  
O  I  &1302.2&$>14.806$&$>14.806$&$>-2.434$\\  
Al II &1670.8&$12.461 \pm  0.025$&$12.461 \pm  0.025$&$-2.529 \pm  0.103$\\  
Si II &1304.4&$14.087 \pm  0.015$&$14.031 \pm  0.011$&$-2.029 \pm  0.101$\\  
Si II &1526.7&$13.983 \pm  0.016$\\  
Si II &1808.0&$<14.423$\\  
S  II &1253.8&$<14.361$&$<14.361$&$<-1.339$\\  
Fe II &1608.5&$<13.799$&$<13.799$&$<-2.201$\\  
Ni II &1317.2&$<13.106$&$<13.106$&$<-1.644$\\  
Zn II &2026.1&$<12.398$&$<12.398$&$<-0.772$\\  
\tableline
\end{tabular}
\end{center}
\end{table}

\begin{figure}[ht]
\begin{center}
\includegraphics[height=3.6in, width=2.8in,angle=90]{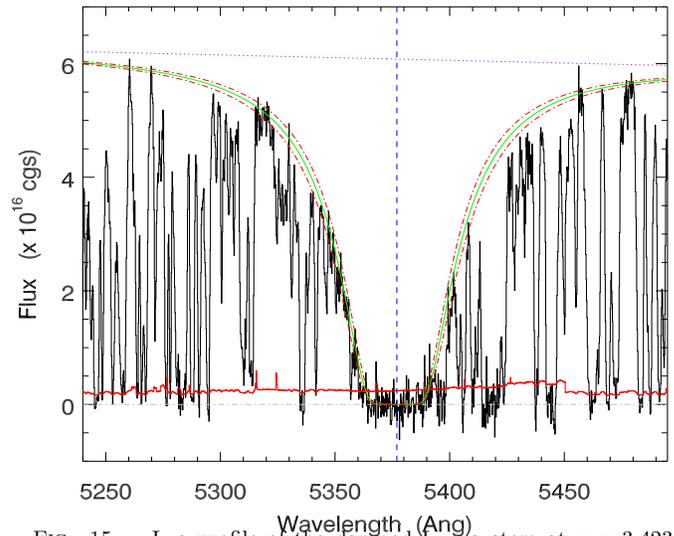}
\figcaption{Lya profile of the damped \lya system at $z=3.423$
toward FJ0747+2739.
The overplotted solid line and accompanying
dash-dot lines trace the best fit solution and the estimated 
bounds corresponding to 
$\log \N{HI} = 20.85^{+0.05}_{-0.05}$.  
Its H\,I value is particularly well determined because of the simplicity
of the continuum and the absence of blending with the core of the profile.\label{fig:fj0747A_lya}}
\end{center}
\end{figure}

\begin{figure}[ht]
\begin{center}
\includegraphics[height=3.6in, width=2.8in,angle=90]{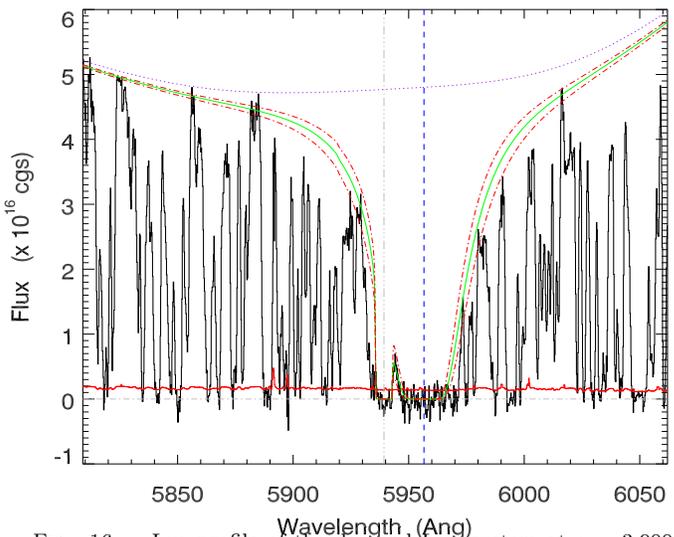}
\figcaption{Lya profile of the damped \lya system at $z=3.900$
toward FJ0747+2739.
The overplotted solid line and accompanying
dash-dot lines trace the best fit solution and the estimated 
bounds corresponding to 
$\log \N{HI} = 20.50^{+0.10}_{-0.10}$.  
This DLA is located near the \lya peak of the quasar and the red
wing is more difficult to determine.  Nevertheless, the blue wing
and therefore the H\,I column density is reasonably well constrained, 
despite the presence of a lower column-density system in the middle 
of the blue wing which was considered as part of the fit.
\label{fig:fj0747B_lya}}
\end{center}
\end{figure}

\begin{figure}[ht]
\begin{center}
\includegraphics[height=6.1in, width=3.9in]{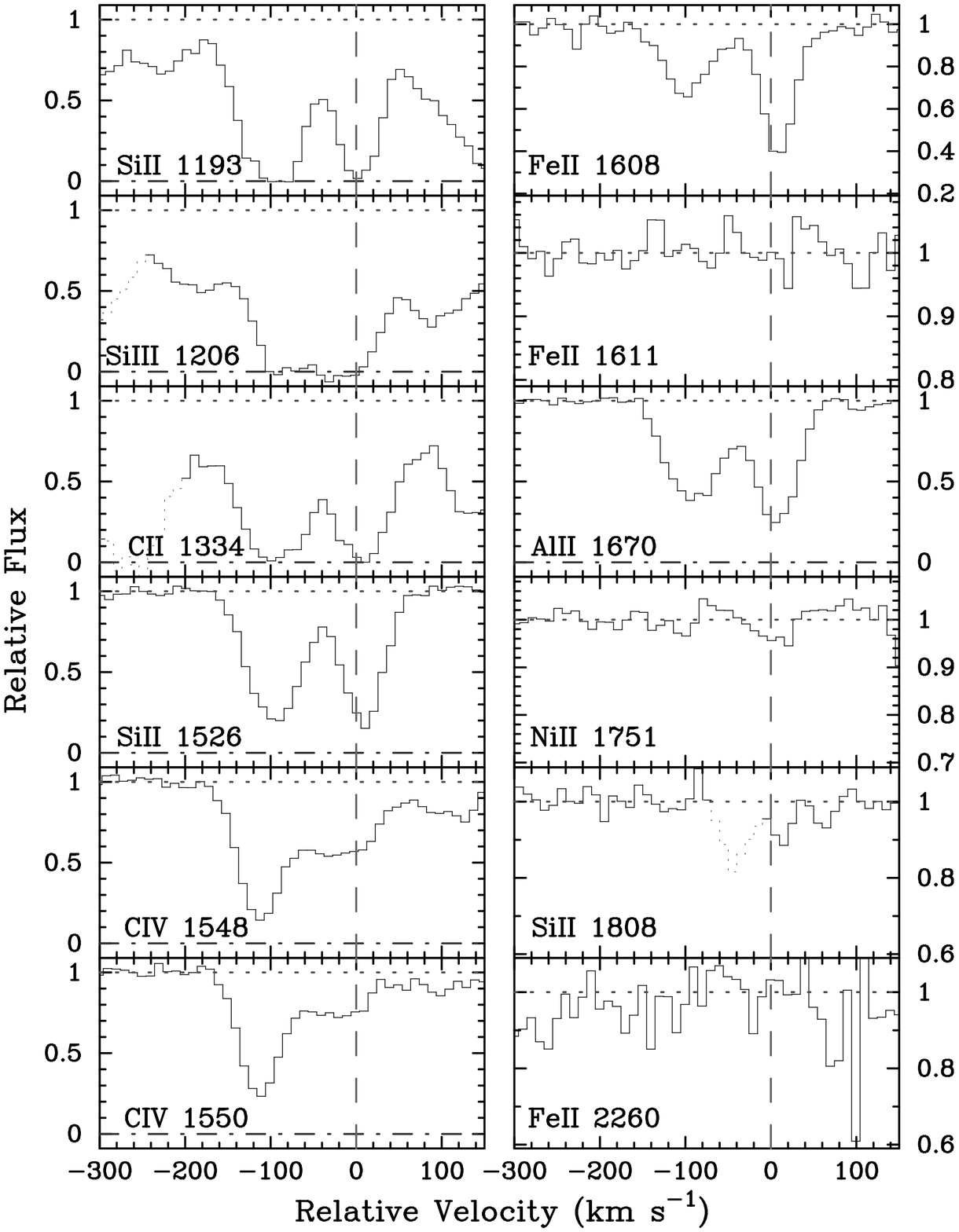}
\figcaption{Velocity plot of the metal-line transitions for the 
damped \lya system at $z = 3.423$ toward FJ0747+2739.
The vertical line at $v=0$ corresponds to $z = 3.4233$.  
\label{fig:fj0747A_mtl}}
\end{center}
\end{figure}

\begin{figure}[ht]
\begin{center}
\includegraphics[height=6.1in, width=3.9in]{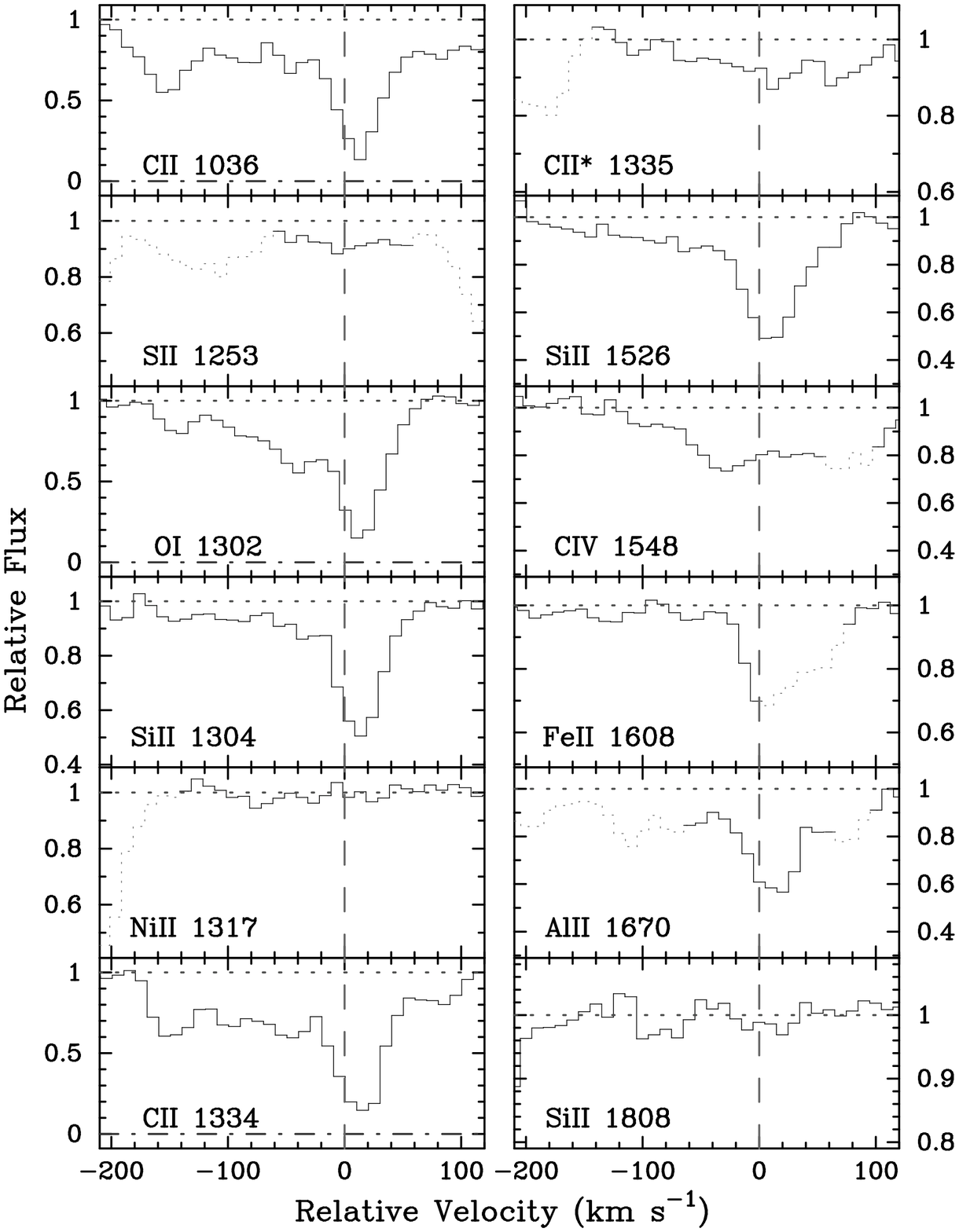}
\figcaption{Velocity plot of the metal-line transitions for the 
damped \lya system at $z = 3.900$ toward FJ0747+2739.
The vertical line at $v=0$ corresponds to $z = 3.9000$.  
\label{fig:fj0747B_mtl}}
\end{center}
\end{figure}

\subsection{FJ0747+2739, $z = 3.423, 3.900$ \label{subsec:FJ0747+2739}}

This bright FIRST quasar \citep{white00} has two, well separated 
damped \lya systems along its sightline.  The lower redshift DLA 
(Figure~\ref{fig:fj0747A_lya}) has a very well determined $\N{HI}$ value
owing to a simple continuum and unblended core.  The DLA at $z=3.900$,
however, has a more uncertain continuum owing to its proximity to the
\lya emission peak (Figure~\ref{fig:fj0747B_lya}).  
 
Figures~\ref{fig:fj0747A_mtl} and \ref{fig:fj0747B_mtl} present the 
metal-line profiles for the $z=3.423$ and $z=3.900$ DLA respectively.
Finally, the ionic column densities of the two DLA are provided in 
Tables~\ref{tab:FJ0747+2739_3.423} and \ref{tab:FJ0747+2739_3.900}.

\begin{table}[ht]\footnotesize
\begin{center}
\caption{ {\sc
IONIC COLUMN DENSITIES: PSS0808+52, $z = 3.113$ \label{tab:PSS0808+52_3.113}}}
\begin{tabular}{lcccc}
\tableline
\tableline
Ion & $\lambda$ & AODM & $N_{\rm adopt}$ & [X/H] \\
\tableline
O  I  &1302.2&$>14.945$&$>14.945$&$>-2.445$\\  
Al III&1862.8&$13.425 \pm  0.014$\\  
Si II &1304.4&$>14.477$&$14.599 \pm  0.120$&$-1.611 \pm  0.139$\\  
Si II &1808.0&$14.599 \pm  0.120$\\  
Cr II &2056.3&$<12.735$&$<12.735$&$<-1.585$\\  
Fe II &2249.9&$<14.626$&$14.170 \pm  0.040$&$-1.980 \pm  0.081$\\  
Fe II &2344.2&$>13.966$\\  
Fe II &2374.5&$14.170 \pm  0.040$\\  
Ni II &1741.6&$<13.238$&$<13.238$&$<-1.662$\\  
Zn II &2026.1&$<12.134$&$<12.134$&$<-1.186$\\  
\tableline
\end{tabular}
\end{center}
\end{table}

\begin{figure}[ht]
\begin{center}
\includegraphics[height=3.6in, width=2.8in,angle=90]{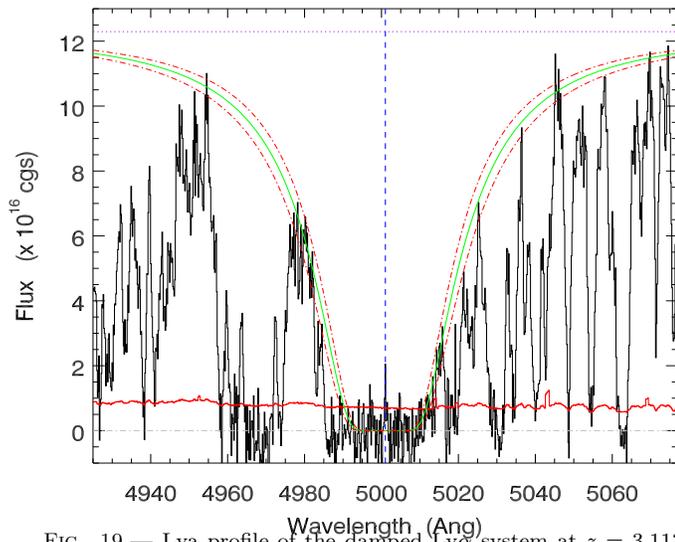}
\figcaption{Lya profile of the damped \lya system at $z=3.113$
toward PSS0808+52.
The overplotted solid line and accompanying
dash-dot lines trace the best fit solution and the estimated 
bounds corresponding to 
$\log \N{HI} = 20.65^{+0.07}_{-0.07}$.  
This DLA has a well sampled core and therefore a very precise
measurement of $\N{HI}$.\label{fig:pss0808_lya}}
\end{center}
\end{figure}

\begin{figure}[ht]
\begin{center}
\includegraphics[height=6.1in, width=3.9in]{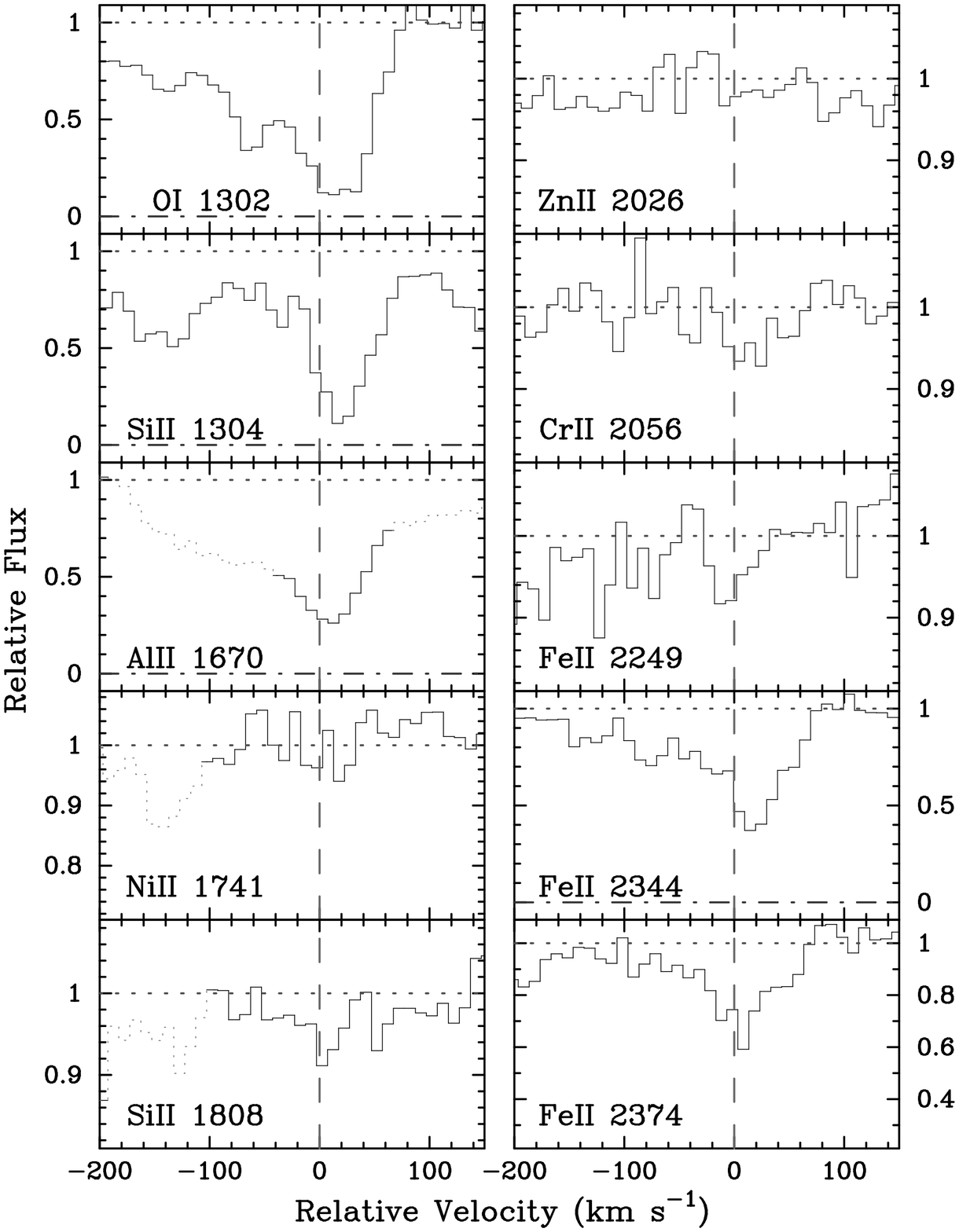}
\figcaption{Velocity plot of the metal-line transitions for the 
damped \lya system at $z = 3.113$ toward PSS0808+52.
The vertical line at $v=0$ corresponds to $z = 3.11318$.  
\label{fig:pss0808_mtl}}
\end{center}
\end{figure}

\subsection{PSS0808+52, $z = 3.113$ \label{subsec:PSS0808+52_3.113}}

We presented a first analysis of this damped \lya system in 
\cite{pgw01}.  Our re-analysis yields very similar results.  The
\lya profile is shown in Figure~\ref{fig:pss0808_lya} and gives a
very precise measure of the H\,I column density.
The complete set of metal-line profiles and ionic column densities
are presented in Figure~\ref{fig:pss0808_mtl} 
and Table~\ref{tab:PSS0808+52_3.113}.

\clearpage

\begin{table}[ht]\footnotesize
\begin{center}
\caption{ {\sc
IONIC COLUMN DENSITIES: FJ0812+32, $z = 2.626$ \label{tab:FJ0812+32_2.626}}}
\begin{tabular}{lcccc}
\tableline
\tableline
Ion & $\lambda$ & AODM & $N_{\rm adopt}$ & [X/H] \\
\tableline
C  I  &1656.9&$12.890 \pm  0.089$\\  
C  II &1334.5&$>15.140$&$>15.140$&$>-2.800$\\  
C  II*&1335.7&$>14.089$\\  
C  IV &1548.2&$>14.496$\\  
C  IV &1550.8&$14.534 \pm  0.008$\\  
N  I  &1159.8&$<17.606$&$<17.606$&$< 0.326$\\  
O  I  &1302.2&$>15.488$&$>15.488$&$>-2.602$\\  
O  I  &1355.6&$<17.981$\\  
Mg II &1239.9&$16.313 \pm  0.080$&$16.324 \pm  0.065$&$-0.606 \pm  0.119$\\  
Mg II &1240.4&$16.349 \pm  0.111$\\  
Mg II &2796.4&$>13.839$\\  
Mg II &2803.5&$>14.030$\\  
Al II &1670.8&$>13.659$&$>13.659$&$>-2.181$\\  
Al III&1854.7&$13.175 \pm  0.016$\\  
Al III&1862.8&$13.243 \pm  0.024$\\  
Si II &1304.4&$>15.052$&$>15.782$&$>-1.128$\\  
Si II &1526.7&$>14.884$\\  
Si II &1808.0&$>15.782$\\  
Si IV &1393.8&$>14.025$\\  
Si IV &1402.8&$14.059 \pm  0.011$\\  
S  II &1250.6&$>15.438$&$>15.438$&$>-1.112$\\  
S  II &1253.8&$>15.284$\\  
S  II &1259.5&$>15.263$\\  
Cr II &2056.3&$13.364 \pm  0.034$&$13.320 \pm  0.032$&$-1.700 \pm  0.105$\\  
Cr II &2062.2&$13.330 \pm  0.041$\\  
Cr II &2066.2&$13.187 \pm  0.081$\\  
Mn II &2594.5&$13.319 \pm  0.037$&$13.319 \pm  0.037$&$-1.561 \pm  0.107$\\  
Fe II &1125.4&$15.121 \pm  0.079$&$15.097 \pm  0.025$&$-1.753 \pm  0.103$\\  
Fe II &1144.9&$>14.790$\\  
Fe II &1608.5&$>14.743$\\  
Fe II &1611.2&$15.132 \pm  0.058$\\  
Fe II &2249.9&$15.153 \pm  0.035$\\  
Fe II &2260.8&$15.019 \pm  0.046$\\  
Fe II &2344.2&$>14.528$\\  
Fe II &2374.5&$>14.764$\\  
Fe II &2382.8&$>14.303$\\  
Ni II &1317.2&$13.825 \pm  0.038$&$13.846 \pm  0.019$&$-1.754 \pm  0.102$\\  
Ni II &1370.1&$13.930 \pm  0.034$\\  
Ni II &1454.8&$13.754 \pm  0.070$\\  
Ni II &1703.4&$<14.003$\\  
Ni II &1741.5&$13.850 \pm  0.037$\\  
Ni II &1751.9&$13.796 \pm  0.063$\\  
Cu II &1358.8&$<12.793$&$<12.793$&$<-0.847$\\  
Zn II &2026.1&$13.036 \pm  0.022$&$13.035 \pm  0.022$&$-0.985 \pm  0.102$\\  
Zn II &2062.7&$<13.100$\\  
\tableline
\end{tabular}
\end{center}
\end{table}

\begin{figure}[ht]
\begin{center}
\includegraphics[height=3.6in, width=2.8in,angle=90]{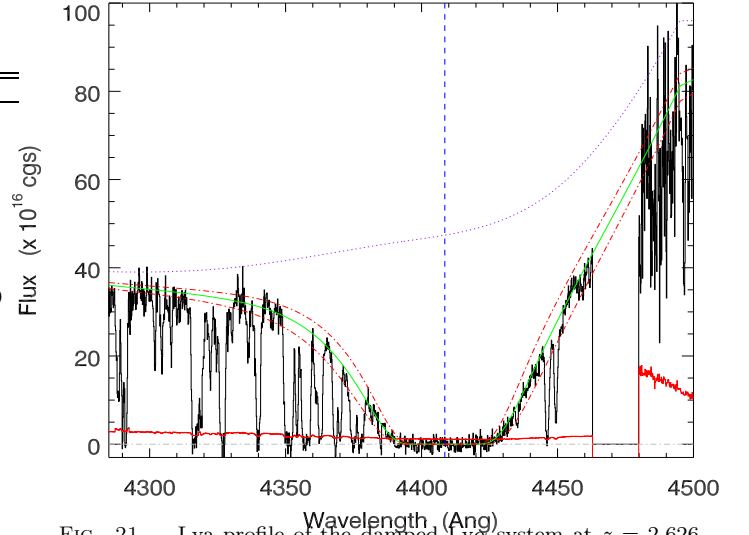}
\figcaption{Lya profile of the damped \lya system at $z=2.626$
toward FJ0812+32.
The overplotted solid line and accompanying
dash-dot lines trace the best fit solution and the estimated 
bounds corresponding to 
$\log \N{HI} = 21.35^{+0.10}_{-0.10}$.  
This \lya profile is complicated by the chip defect at 
$\lambda \approx 4470$\AA\ and the presence of the \lya emission peak.
The core, however, is well sampled and the fit is precise.
\label{fig:fj0812_lya}}
\end{center}
\end{figure}

\begin{figure*}
\begin{center}
\includegraphics[height=8.5in, width=6.0in]{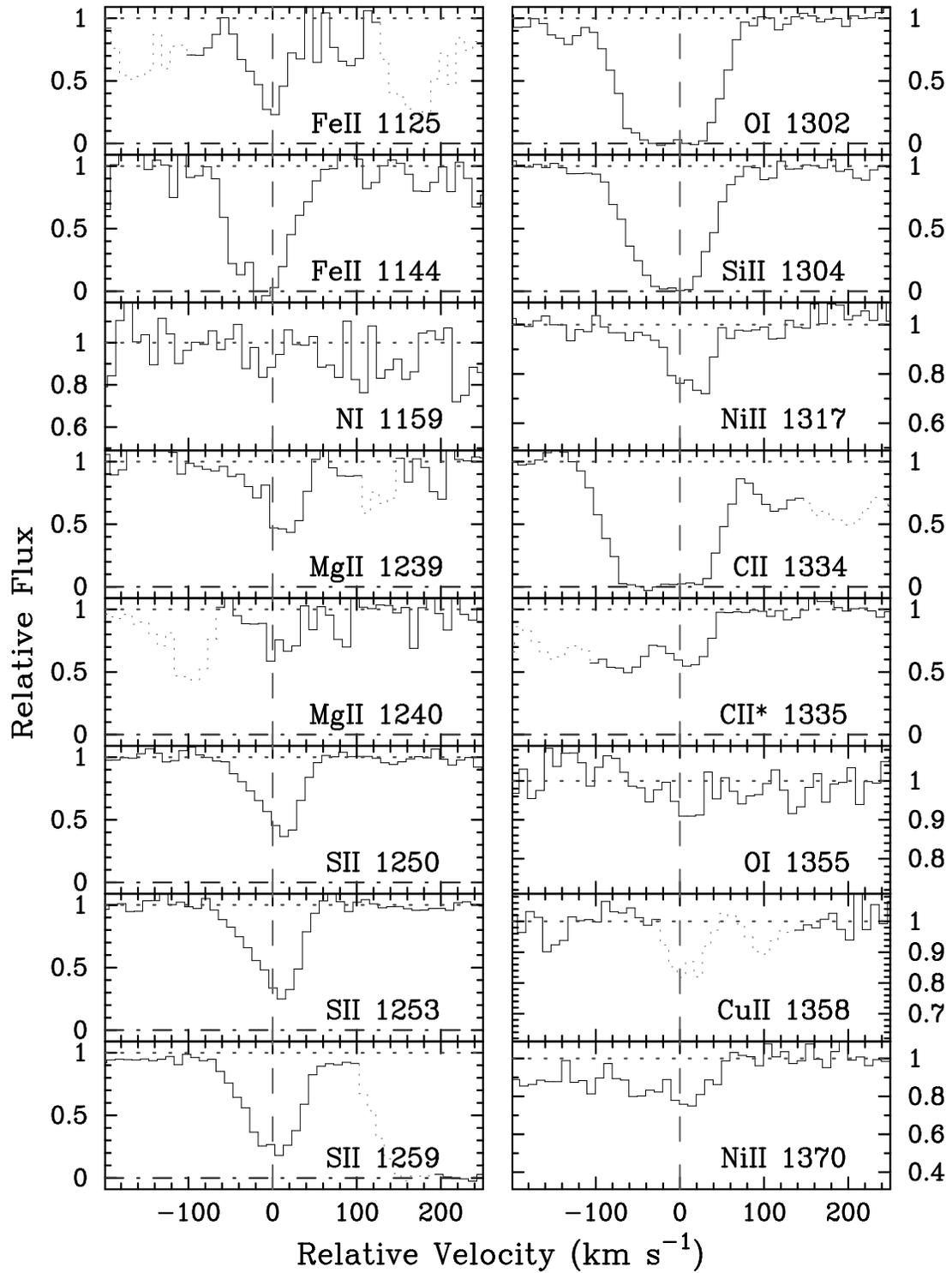}
\figcaption{Velocity plot of the metal-line transitions for the 
damped \lya system at $z = 2.626$ toward FJ0812+32.
The vertical line at $v=0$ corresponds to $z = 2.62644$.  
\label{fig:fj0812_mtl}}
\end{center}
\end{figure*}

\begin{figure*}
\begin{center}
\includegraphics[height=8.5in, width=6.0in]{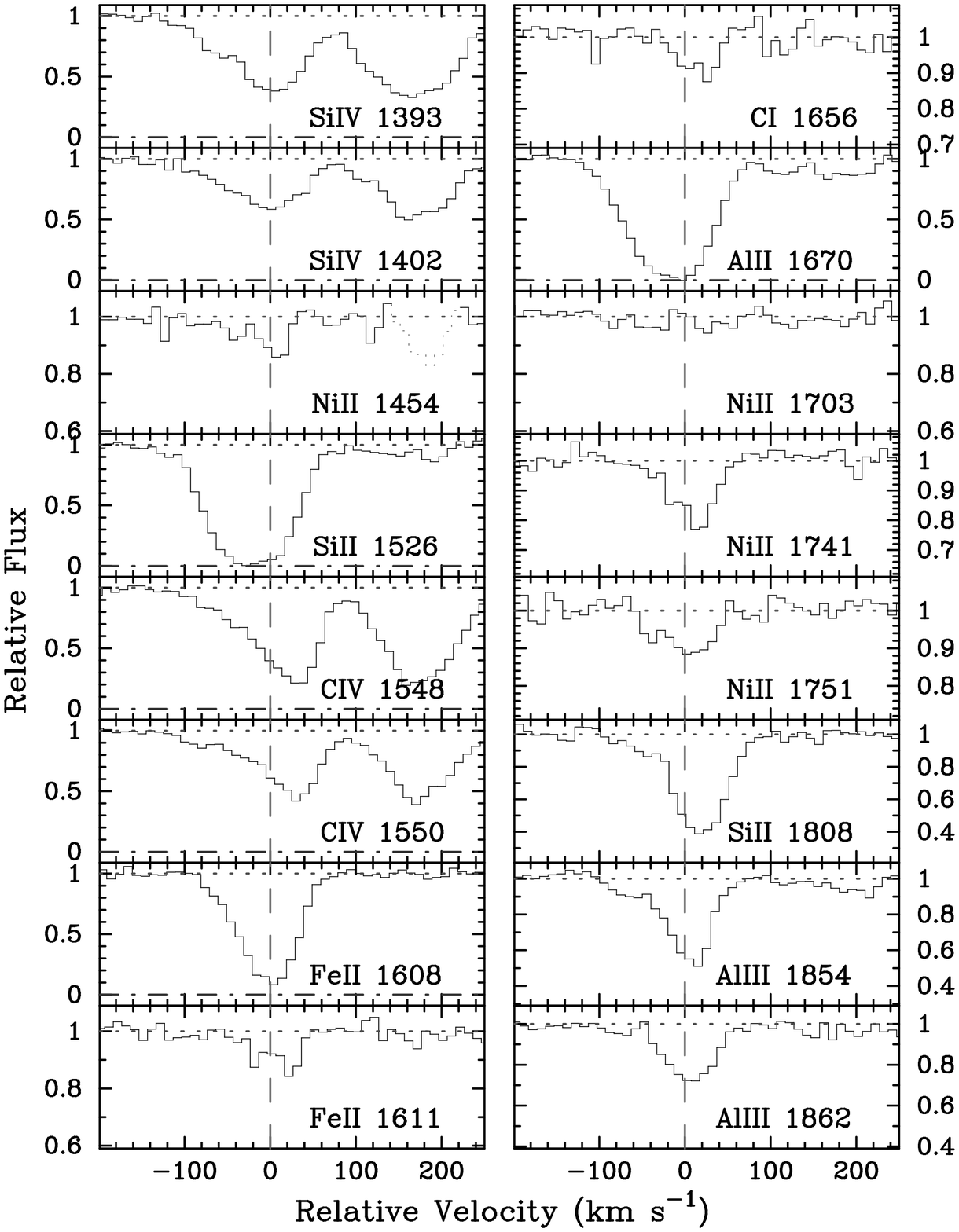}
Fig 22 -- cont
\end{center}
\end{figure*}

\begin{figure*}
\begin{center}
\includegraphics[height=8.5in, width=6.0in]{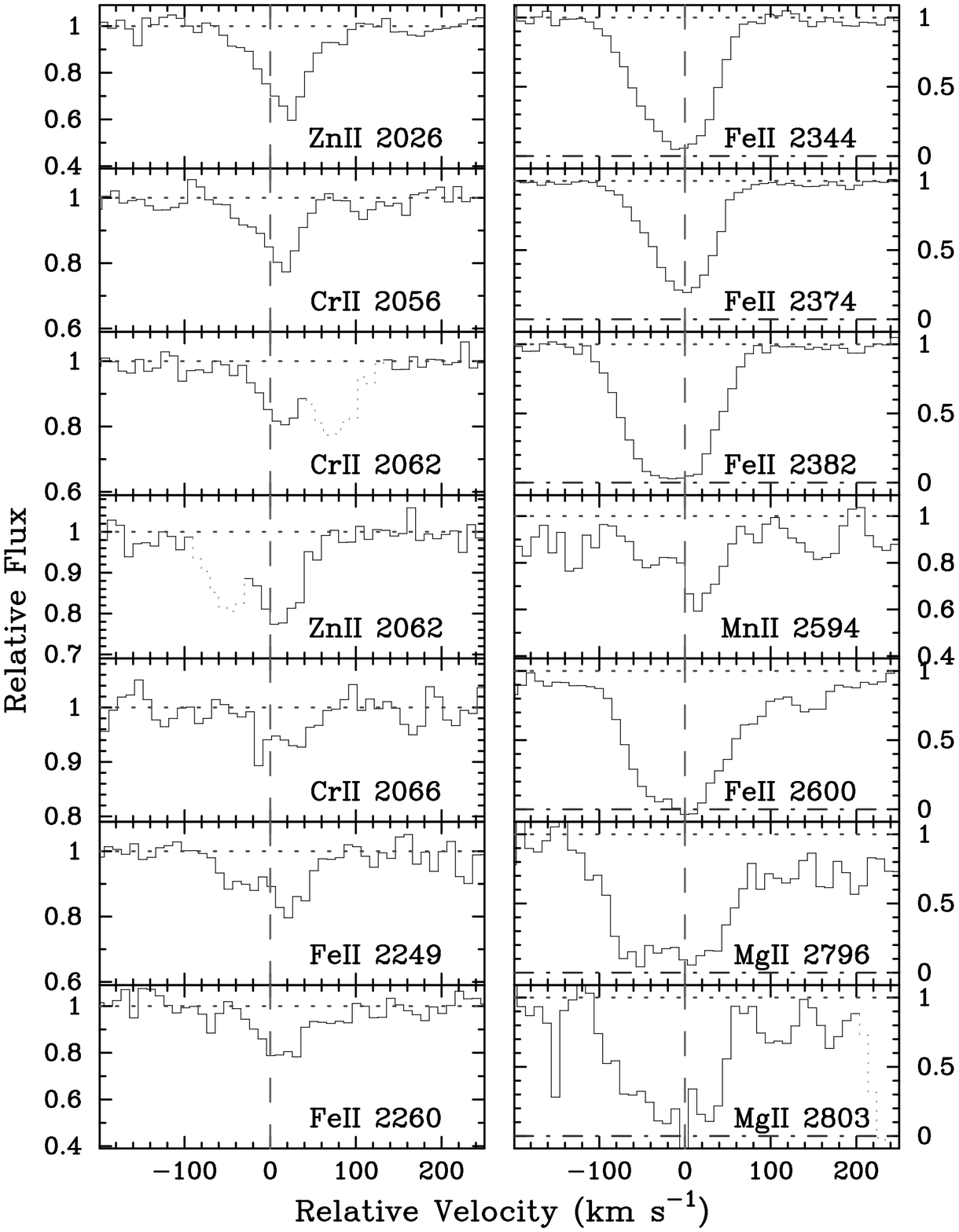}
Fig 22 -- cont
\end{center}
\end{figure*}

\subsection{FJ0812+32, $z = 2.626$ \label{subsec:FJ0812+32}}

This FIRST quasar shows the remarkable damped \lya system at $z=2.626$
which exhibits among the strongest metal-line transitions of any 
damped \lya system.  In fact, our follow-up HIRES 
observations revealed the
the first discovery of Ge$^+$, Cl$^0$, B$^+$, and 
several other ions \citep{phw03}.  The \lya profile from our ESI spectrum
is presented in Figure~\ref{fig:fj0812_lya}.  The fit is somewhat challenging
because of the proximity to the \lya peak and the gap in data quality at
$\lambda \approx 4475$\AA\ owing to a defect on the CCD.  Nevertheless, the
core is well sampled and the blue wing provides complimentary constraints.

An extensive list of metal-line transitions are plotted in 
Figure~\ref{fig:fj0812_mtl} and the corresponding ionic column densities
are provided in Table~\ref{tab:FJ0812+32_2.626}.  A comprehensive analysis
of this sightline including this ESI data will be presented in a future paper.

\begin{table}[ht]\footnotesize
\begin{center}
\caption{ {\sc
IONIC COLUMN DENSITIES: Q0930+28, $z = 3.235$ \label{tab:Q0930+28_3.235}}}
\begin{tabular}{lcccc}
\tableline
\tableline
Ion & $\lambda$ & AODM & $N_{\rm adopt}$ & [X/H] \\
\tableline
C  II &1036.3&$>14.173$&$>14.173$&$>-2.767$\\  
C  II &1334.5&$>13.887$\\  
C  II*&1335.7&$<12.854$\\  
C  IV &1548.2&$13.210 \pm  0.059$\\  
C  IV &1550.8&$<13.152$\\  
O  I  &1302.2&$>14.196$&$>14.196$&$>-2.894$\\  
Al II &1670.8&$12.369 \pm  0.026$&$12.369 \pm  0.026$&$-2.471 \pm  0.103$\\  
Si II &1304.4&$13.793 \pm  0.022$&$13.793 \pm  0.022$&$-2.117 \pm  0.102$\\  
Si II &1526.7&$>13.713$\\  
Si IV &1393.8&$12.749 \pm  0.053$\\  
Si IV &1402.8&$12.749 \pm  0.101$\\  
Fe II &1144.9&$13.389 \pm  0.079$&$13.490 \pm  0.031$&$-2.360 \pm  0.105$\\  
Fe II &1608.5&$13.540 \pm  0.040$\\  
Fe II &2344.2&$13.479 \pm  0.062$\\  
Fe II &2374.5&$<13.713$\\  
\tableline
\end{tabular}
\end{center}
\end{table}

\begin{figure}[ht]
\begin{center}
\includegraphics[height=3.6in, width=2.8in,angle=90]{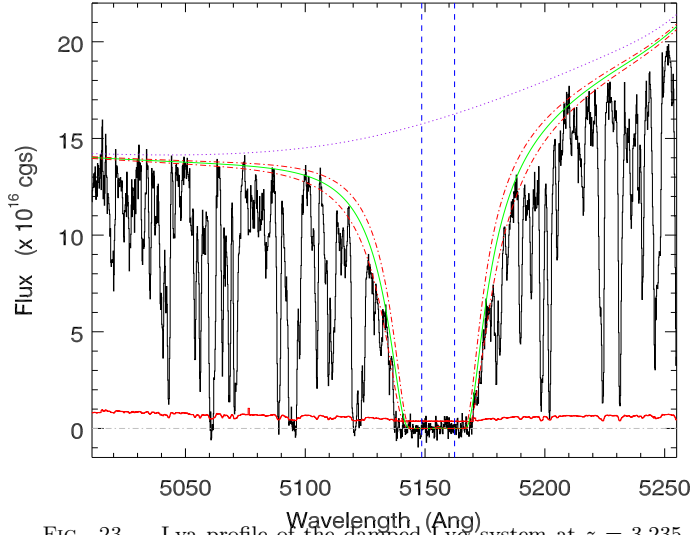}
\figcaption{Lya profile of the damped \lya system at $z=3.235$
toward Q0930+28.
The overplotted solid line and accompanying
dash-dot lines trace the best fit solution and the estimated 
bounds corresponding to 
$\log \N{HI} = 20.35^{+0.10}_{-0.10}$.  
This solution is complicated by the presence of a neighboring
sub-DLA ($z=3.246$, $\N{HI} \approx 10^{20.2} \cm{-2}$) although we consider
the final value to be relatively accurate.
\label{fig:q0930_lya}}
\end{center}
\end{figure}

\begin{figure}[ht]
\begin{center}
\includegraphics[height=6.1in, width=3.9in]{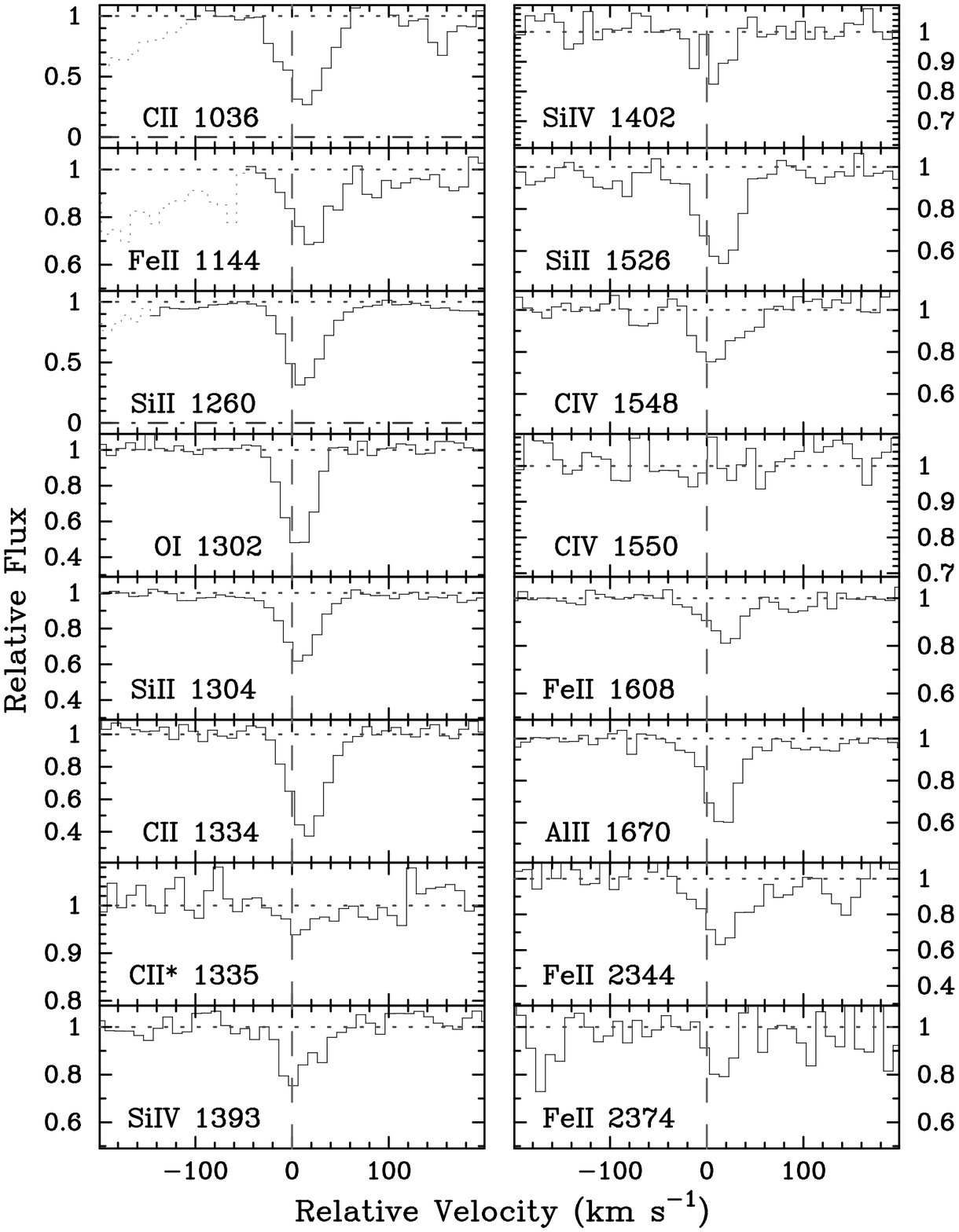}
\figcaption{Velocity plot of the metal-line transitions for the 
damped \lya system at $z = 3.235$ toward Q0930+28.
The vertical line at $v=0$ corresponds to $z = 3.2352$.  
\label{fig:q0930_mtl}}
\end{center}
\end{figure}

\subsection{Q0930+28, $z = 3.235$ \label{subsec:Q0930+28_3.235}}

This damped \lya system and its neighboring sub-DLA ($z=3.246$)
have been studied previously in the literature \citep{lu98,pro02}.  
Our observations provide greater wavelength coverage
and are generally consistent with the echelle observations.   The \lya
profiles of the DLA and sub-DLA are shown in Figure~\ref{fig:q0930_lya}
while the metal-line profiles and ionic column densities for the DLA are
given by Figure~\ref{fig:q0930_mtl} and Table~\ref{tab:Q0930+28_3.235}.

\begin{table}[ht]\footnotesize
\begin{center}
\caption{ {\sc
IONIC COLUMN DENSITIES: PC0953+47, $z = 3.404$ \label{tab:PC0953+47_3.404}}}
\begin{tabular}{lcccc}
\tableline
\tableline
Ion & $\lambda$ & AODM & $N_{\rm adopt}$ & [X/H] \\
\tableline
C  IV &1550.8&$>14.858$\\  
Al II &1670.8&$>13.360$&$>13.360$&$>-2.280$\\  
Si II &1526.7&$>14.617$&$>14.617$&$>-2.093$\\  
Si II &1808.0&$<15.166$\\  
Fe II &1608.5&$>14.472$&$>14.472$&$>-2.178$\\  
Fe II &1611.2&$<15.303$\\  
Ni II &1709.6&$<13.896$&$<13.896$&$<-1.504$\\  
\tableline
\end{tabular}
\end{center}
\end{table}

\begin{table}[ht]\footnotesize
\begin{center}
\caption{ {\sc
IONIC COLUMN DENSITIES: PC0953+47, $z = 3.891$ \label{tab:PC0953+47_3.891}}}
\begin{tabular}{lcccc}
\tableline
\tableline
Ion & $\lambda$ & AODM & $N_{\rm adopt}$ & [X/H] \\
\tableline
C  IV &1548.2&$>14.755$\\  
C  IV &1550.8&$>14.687$\\  
Al II &1670.8&$>13.831$&$>13.831$&$>-1.859$\\  
Si II &1526.7&$>15.161$&$>15.161$&$>-1.599$\\  
Si II &1808.0&$<15.364$\\  
Si IV &1393.8&$>14.138$\\  
Si IV &1402.8&$>13.988$\\  
Fe II &1608.5&$>14.988$&$>14.988$&$>-1.712$\\  
Fe II &1611.2&$<15.530$\\  
Ni II &1370.1&$13.732 \pm  0.062$&$13.752 \pm  0.055$&$-1.698 \pm  0.114$\\  
Ni II &1741.5&$13.859 \pm  0.115$\\  
Ni II &1751.9&$<14.055$\\  
\tableline
\end{tabular}
\end{center}
\end{table}
\begin{table}[ht]\footnotesize
\begin{center}
\caption{ {\sc
IONIC COLUMN DENSITIES: PC0953+47, $z = 4.244$ \label{tab:PC0953+47_4.244}}}
\begin{tabular}{lcccc}
\tableline
\tableline
Ion & $\lambda$ & AODM & $N_{\rm adopt}$ & [X/H] \\
\tableline
C  II &1334.5&$>14.719$&$>14.719$&$>-2.771$\\  
C  II*&1335.7&$13.599 \pm  0.099$\\  
C  IV &1548.2&$14.273 \pm  0.052$\\  
C  IV &1550.8&$14.217 \pm  0.092$\\  
O  I  &1302.2&$>15.162$&$>15.162$&$>-2.478$\\  
Al II &1670.8&$<12.873$&$<12.873$&$<-2.517$\\  
Si II &1304.4&$14.235 \pm  0.027$&$14.235 \pm  0.027$&$-2.225 \pm  0.152$\\  
Si II &1526.7&$>14.104$\\  
Fe II &1608.5&$13.900 \pm  0.068$&$13.900 \pm  0.068$&$-2.500 \pm  0.165$\\  
Fe II &1611.2&$<15.080$\\  
Ni II &1370.1&$<13.613$&$<13.613$&$<-1.537$\\  
Ni II &1741.5&$<13.821$\\  
Ni II &1751.9&$<14.050$\\  
\tableline
\end{tabular}
\end{center}
\end{table}

\begin{figure}[ht]
\begin{center}
\includegraphics[height=3.6in, width=2.8in,angle=90]{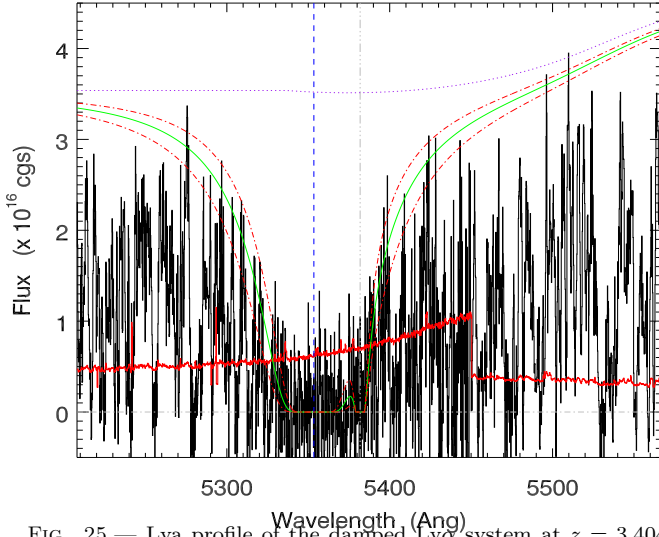}
\figcaption{Lya profile of the damped \lya system at $z=3.404$
toward PC0953+47.
The overplotted solid line and accompanying
dash-dot lines trace the best fit solution and the estimated 
bounds corresponding to 
$\log \N{HI} = 21.15^{+0.15}_{-0.15}$.  
The continuum (dotted line) is not well constrained in this region of the
spectra and the signal-to-noise ratio is poor.  Therefore, we 
ascribe a higher uncertainty to the $\N{HI}$ value of this DLA.
Also note that we were forced to include an additional absorber at
$\lambda \approx 5390$\AA\ to explain the complete core of this
\lya\ profile.
\label{fig:pc0953A_lya}}
\end{center}
\end{figure}

\begin{figure}[ht]
\begin{center}
\includegraphics[height=3.6in, width=2.8in,angle=90]{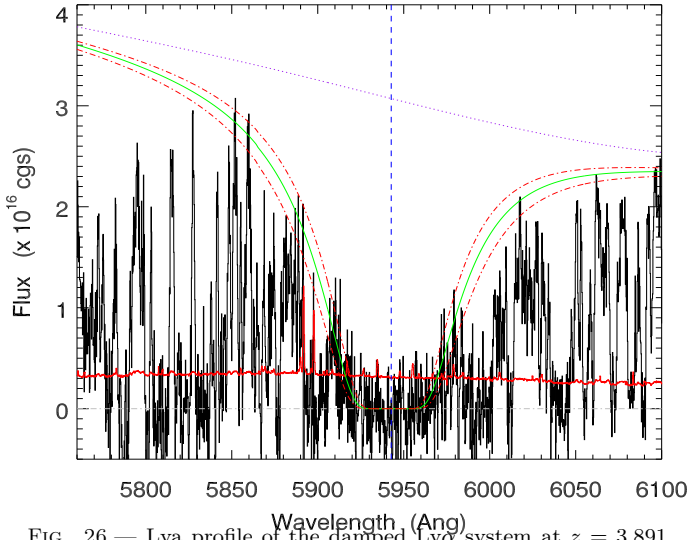}
\figcaption{Lya profile of the damped \lya system at $z=3.891$
toward PC0953+47.
The overplotted solid line and accompanying
dash-dot lines trace the best fit solution and the estimated 
bounds corresponding to 
$\log \N{HI} = 21.20^{+0.10}_{-0.10}$.  
In comparison with the DLA at $z=3.404$, the continuum is better 
determined across this DLA and the resulting solution is nicely
constrained.  
\label{fig:pc0953B_lya}}
\end{center}
\end{figure}

\begin{figure}[ht]
\begin{center}
\includegraphics[height=3.6in, width=2.8in,angle=90]{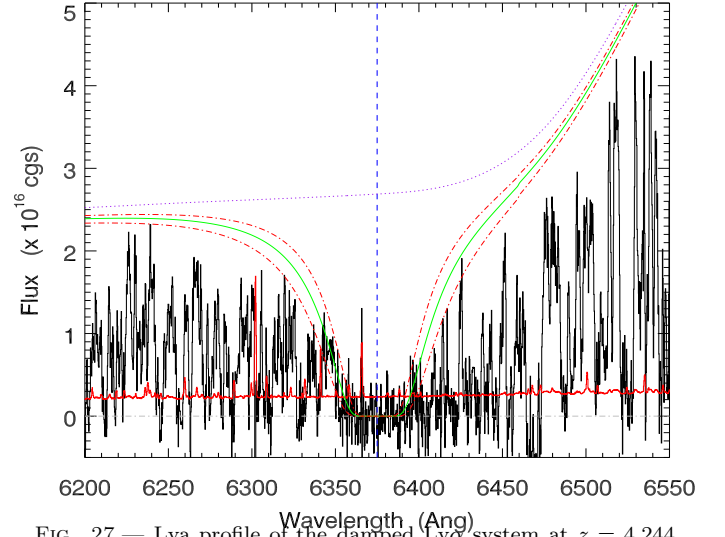}
\figcaption{Lya profile of the damped \lya system at $z=4.244$
toward PC0953+47.
The overplotted solid line and accompanying
dash-dot lines trace the best fit solution and the estimated 
bounds corresponding to 
$\log \N{HI} = 20.90^{+0.15}_{-0.15}$.  There is significant
contamination from the \lya forest in the region surrounding
this DLA and the continuum is complicated at the red end by
the \lya emission peak.  The resulting fit to the \lya profile
is quite uncertain. 
Note the feature at $6362$\AA\ is a skyline as denoted by the
spike in the sigma spectrum.
\label{fig:pc0953C_lya}}
\end{center}
\end{figure}

\begin{figure}[ht]
\begin{center}
\includegraphics[height=5.1in, width=3.5in]{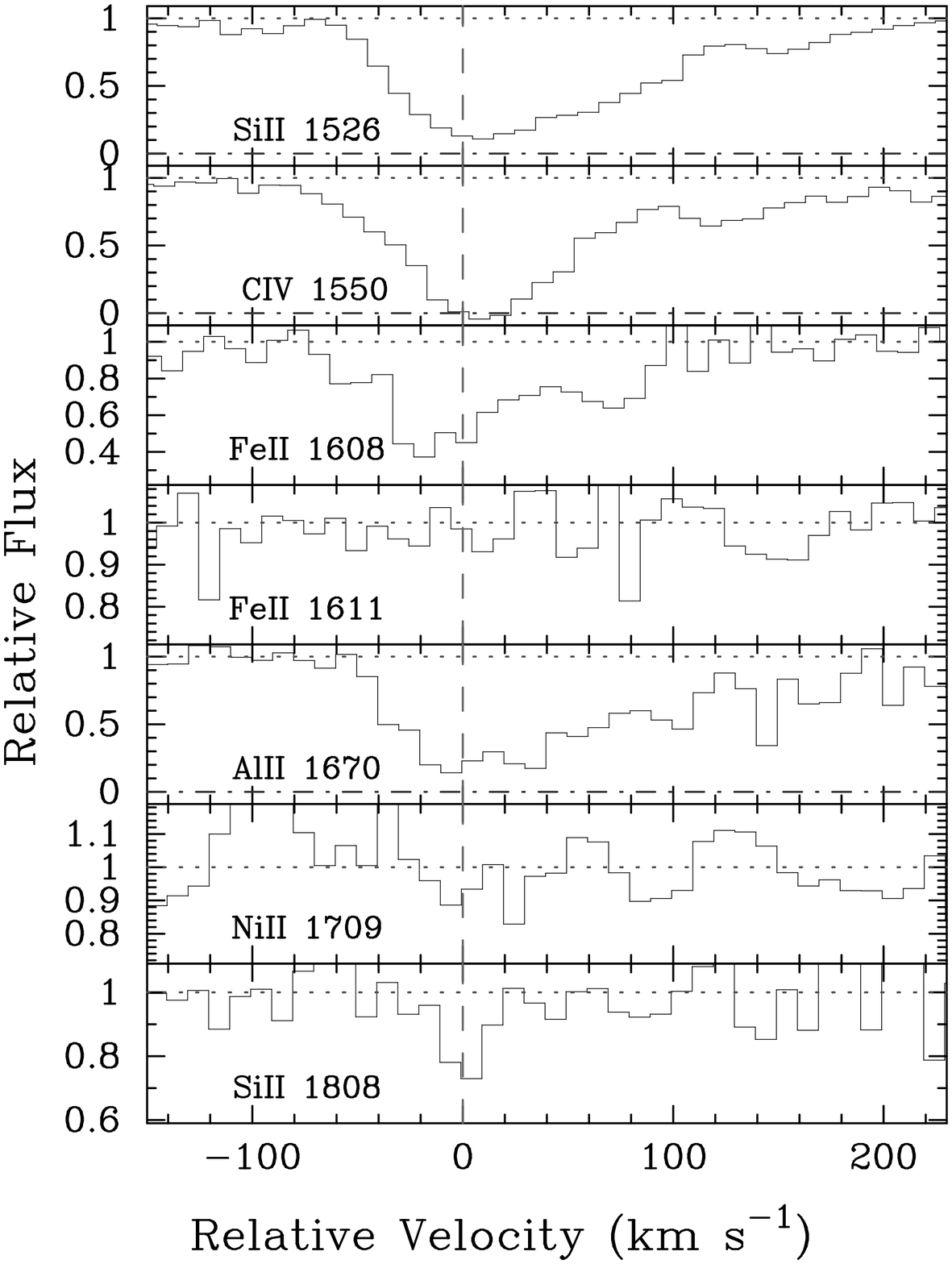}
\figcaption{Velocity plot of the metal-line transitions for the 
damped \lya system at $z = 3.404$ toward PC0953+47.
The vertical line at $v=0$ corresponds to $z = 3.4036$.  
\label{fig:pc0953A_mtl}}
\end{center}
\end{figure}

\begin{figure}[ht]
\begin{center}
\includegraphics[height=6.1in, width=3.9in]{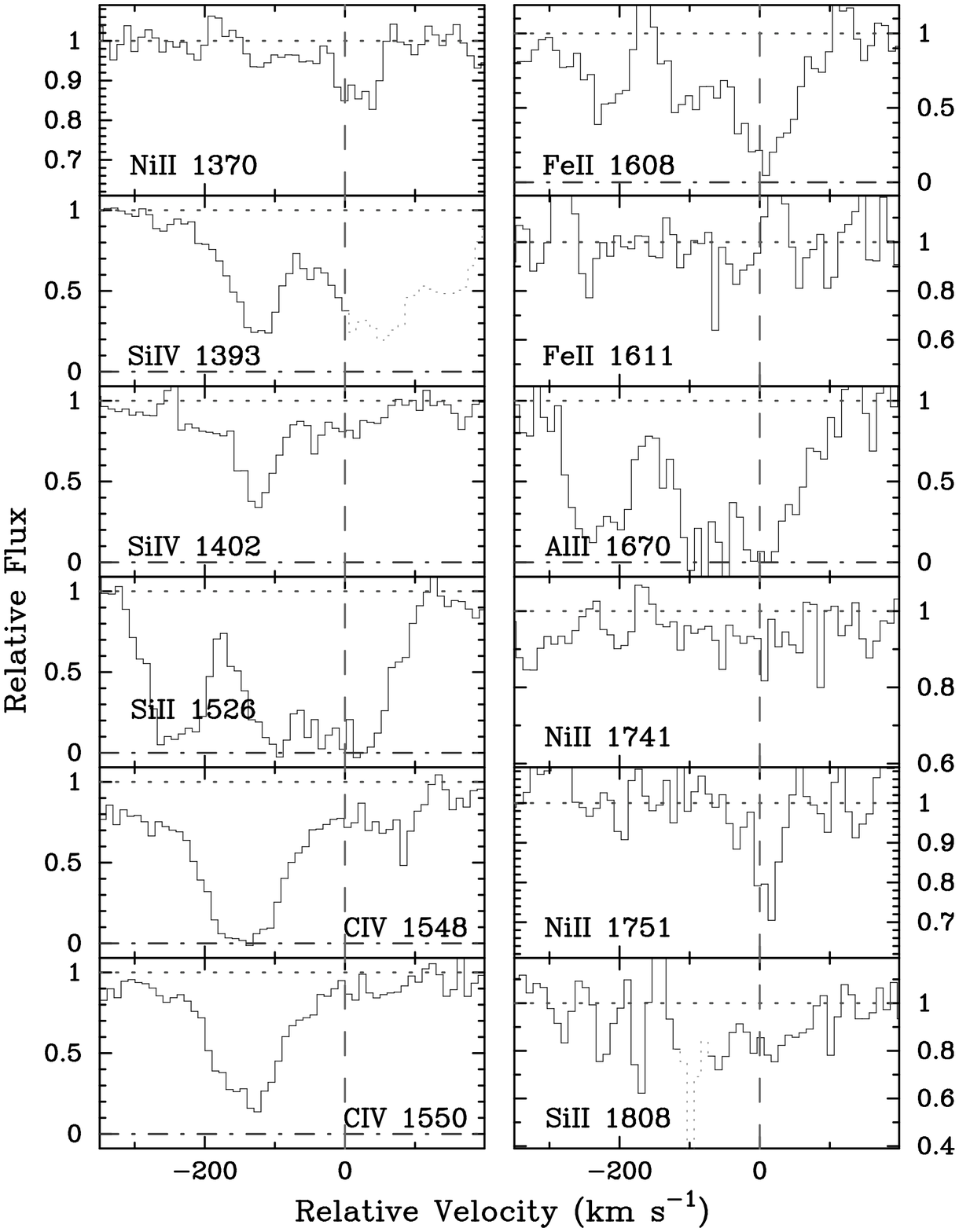}
\figcaption{Velocity plot of the metal-line transitions for the 
damped \lya system at $z = 3.891$ toward PC0953+47.
The vertical line at $v=0$ corresponds to $z = 3.8910$.  
\label{fig:pc0953B_mtl}}
\end{center}
\end{figure}

\begin{figure}[ht]
\begin{center}
\includegraphics[height=6.1in, width=3.9in]{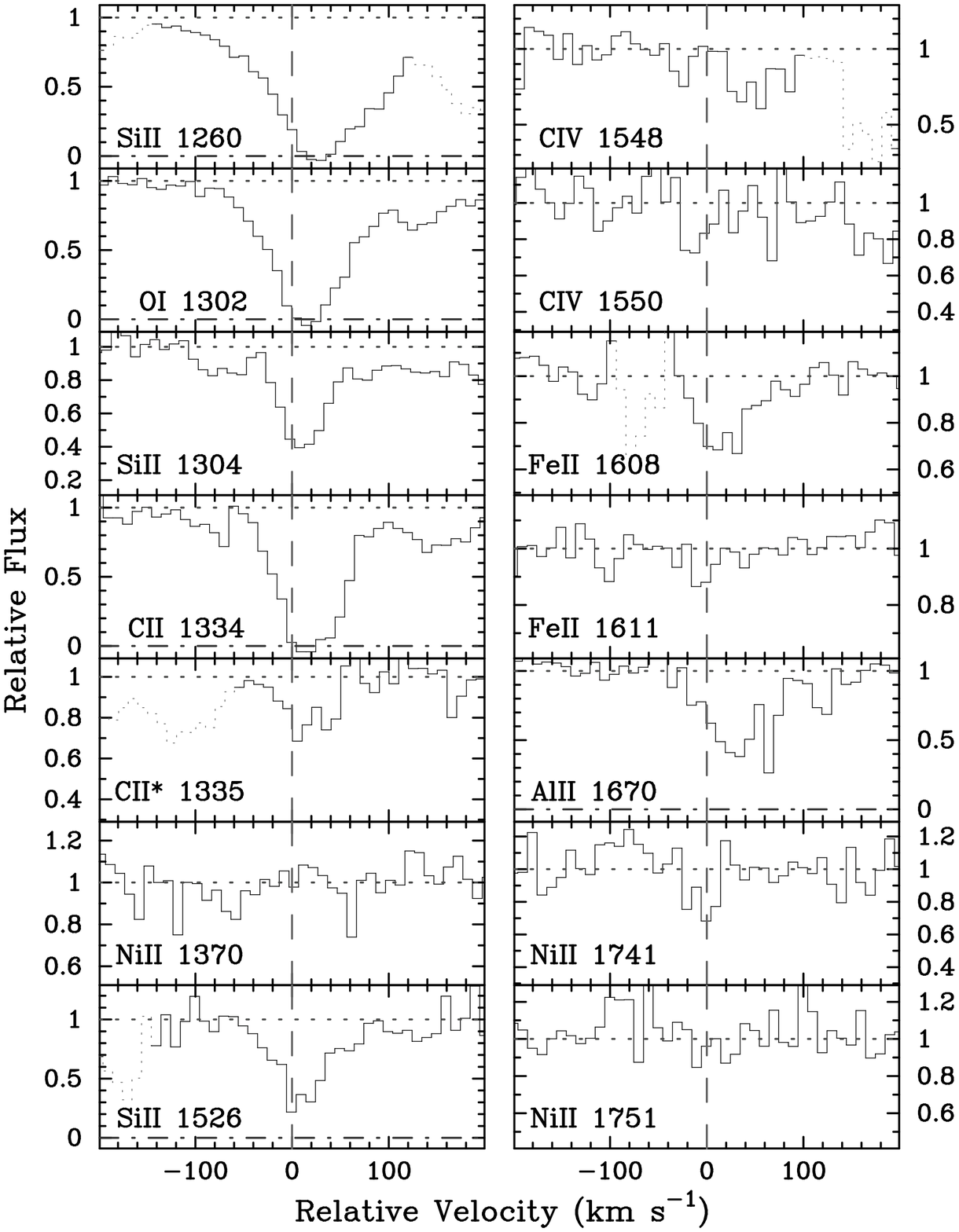}
\figcaption{Velocity plot of the metal-line transitions for the 
damped \lya system at $z = 4.244$ toward PC0953+47.
The vertical line at $v=0$ corresponds to $z = 4.24419$.  
\label{fig:pc0953C_mtl}}
\end{center}
\end{figure}

\subsection{PC0953+47, $z = 3.404, 3.891, 4.244$ \label{subsec:PC0953+47}}

This quasar discovered by \cite{schneider91} and one could identify
three $z>3$ DLA candidates in their discovery spectrum.  
The systems are particularly unusual because all three DLA have among the largest
H\,I column densities at these redshifts.  The \lya profiles are shown
individually in Figures~\ref{fig:pc0953A_lya}-\ref{fig:pc0953C_lya}.  The
S/N of this spectrum is among the lowest in our dataset, primarily because
of poor seeing during this particular observation.  For
the $z=3.404$ and $z=4.244$ DLA, the continuum is poorly constrained and
we place a larger uncertainty on their H\,I column densities.

The metal-line profiles for the three DLA are presented in 
Figures~\ref{fig:pc0953A_mtl}-\ref{fig:pc0953C_mtl} and the ionic column
densities are listed in Tables~\ref{tab:PC0953+47_3.404}-\ref{tab:PC0953+47_4.244}.
Unfortunately, the DLA at $z=3.404$ exhibits transitions which are either
too weak or too strong for an accurate measure of the ionic column density for
the few transitions which lie outside the \lya forest.  This is nearly
the case for the DLA at $z=3.891$ as well.  Because of the remarkable nature of
this sightline, we hope to achieve higher S/N with future observations.

\begin{table}[ht]\footnotesize
\begin{center}
\caption{ {\sc
IONIC COLUMN DENSITIES: PSS0957+33, $z = 3.280$ \label{tab:PSS0957+33_3.280}}}
\begin{tabular}{lcccc}
\tableline
\tableline
Ion & $\lambda$ & AODM & $N_{\rm adopt}$ & [X/H] \\
\tableline
C  IV &1548.2&$13.945 \pm  0.015$\\  
C  IV &1550.8&$13.838 \pm  0.033$\\  
N  I  &1200.7&$<14.444$&$<14.444$&$<-1.936$\\  
Al II &1670.8&$>13.128$&$>13.128$&$>-1.812$\\  
Al III&1854.7&$12.544 \pm  0.042$\\  
Al III&1862.8&$12.396 \pm  0.099$\\  
Si II &1526.7&$>14.461$&$14.676 \pm  0.078$&$-1.334 \pm  0.127$\\  
Si II &1808.0&$14.676 \pm  0.078$\\  
S  II &1259.5&$<14.577$&$<14.577$&$<-1.073$\\  
Cr II &2056.3&$<12.891$&$<12.891$&$<-1.229$\\  
Fe II &1608.5&$<14.261$&$>14.088$&$>-1.862$\\  
Fe II &2249.9&$<14.821$\\  
Fe II &2260.8&$<14.575$\\  
Fe II &2344.2&$>14.088$\\  
Fe II &2374.5&$<14.102$\\  
Ni II &1741.5&$<13.152$&$<13.152$&$<-1.548$\\  
Zn II &2026.1&$<12.168$&$<12.168$&$<-0.952$\\  
\tableline
\end{tabular}
\end{center}
\end{table}

\begin{table}[ht]\footnotesize
\begin{center}
\caption{ {\sc
IONIC COLUMN DENSITIES: PSS0957+33, $z = 4.180$ \label{tab:PSS0957+33_4.180}}}
\begin{tabular}{lcccc}
\tableline
\tableline
Ion & $\lambda$ & AODM & $N_{\rm adopt}$ & [X/H] \\
\tableline
C  II &1334.5&$>14.799$&$>14.799$&$>-2.441$\\  
C  IV &1548.2&$<14.081$\\  
C  IV &1550.8&$13.910 \pm  0.031$\\  
O  I  &1302.2&$>14.972$&$>14.972$&$>-2.418$\\  
Al II &1670.8&$13.038 \pm  0.015$&$13.038 \pm  0.015$&$-2.102 \pm  0.151$\\  
Al III&1854.7&$12.775 \pm  0.103$\\  
Al III&1862.8&$<12.878$\\  
Si II &1304.4&$>14.376$&$>14.376$&$>-1.834$\\  
Si II &1526.7&$>14.250$\\  
Si IV &1393.8&$12.931 \pm  0.052$\\  
Si IV &1402.8&$13.401 \pm  0.035$\\  
Fe II &1608.5&$>13.897$&$>13.897$&$>-2.253$\\  
Fe II &1611.2&$<15.038$\\  
Ni II &1317.2&$<12.948$&$<12.948$&$<-1.952$\\  
\tableline
\end{tabular}
\end{center}
\end{table}

\begin{figure}[ht]
\begin{center}
\includegraphics[height=3.6in, width=2.8in,angle=90]{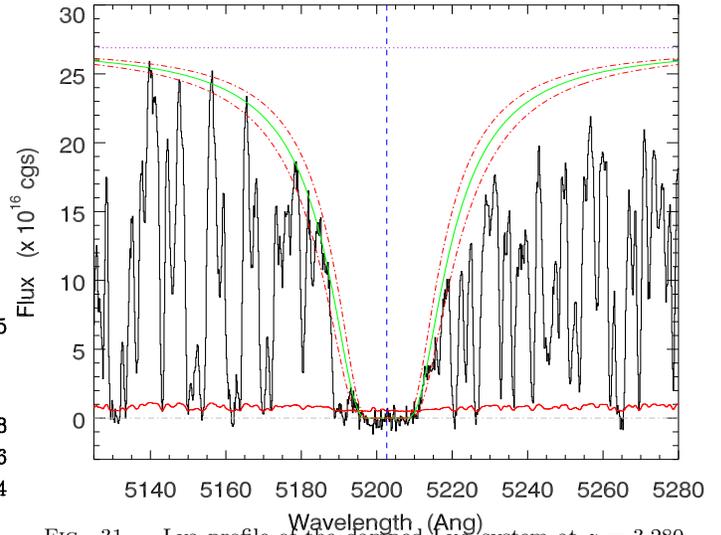}
\figcaption{Lya profile of the damped \lya system at $z=3.280$
toward PSS0957+33.
The overplotted solid line and accompanying
dash-dot lines trace the best fit solution and the estimated 
bounds corresponding to 
$\log \N{HI} = 20.45^{+0.10}_{-0.10}$.  
Aside from the somewhat significant amount of contamination
from coincident \lya clouds, this profile offered a straightforward
solution.
\label{fig:pss0957A_lya}}
\end{center}
\end{figure}

\begin{figure}[ht]
\begin{center}
\includegraphics[height=3.6in, width=2.8in,angle=90]{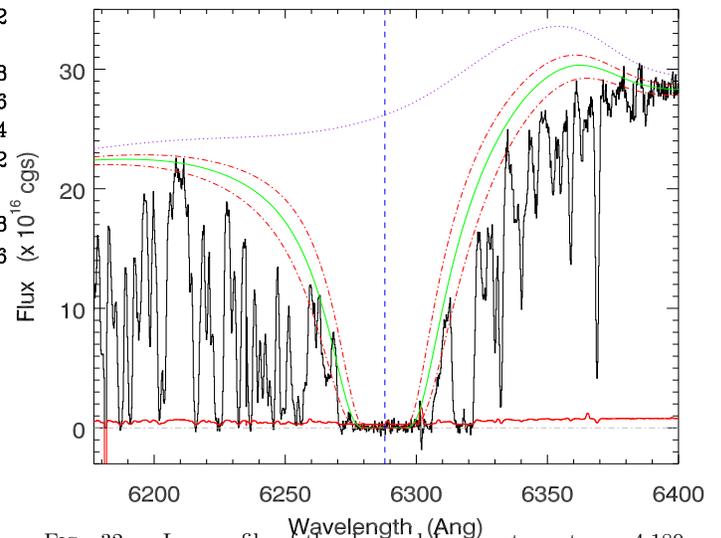}
\figcaption{Lya profile of the damped \lya system at $z=4.180$
toward PSS0957+33.
The overplotted solid line and accompanying
dash-dot lines trace the best fit solution and the estimated 
bounds corresponding to 
$\log \N{HI} = 20.65^{+0.15}_{-0.15}$.  
Both the presence of the nearby \lya emission peak and the absence
of detailed information in the core of the profile lead to a more 
uncertain H\,I solution.
\label{fig:pss0957B_lya}}
\end{center}
\end{figure}

\begin{figure}[ht]
\begin{center}
\includegraphics[height=6.1in, width=3.9in]{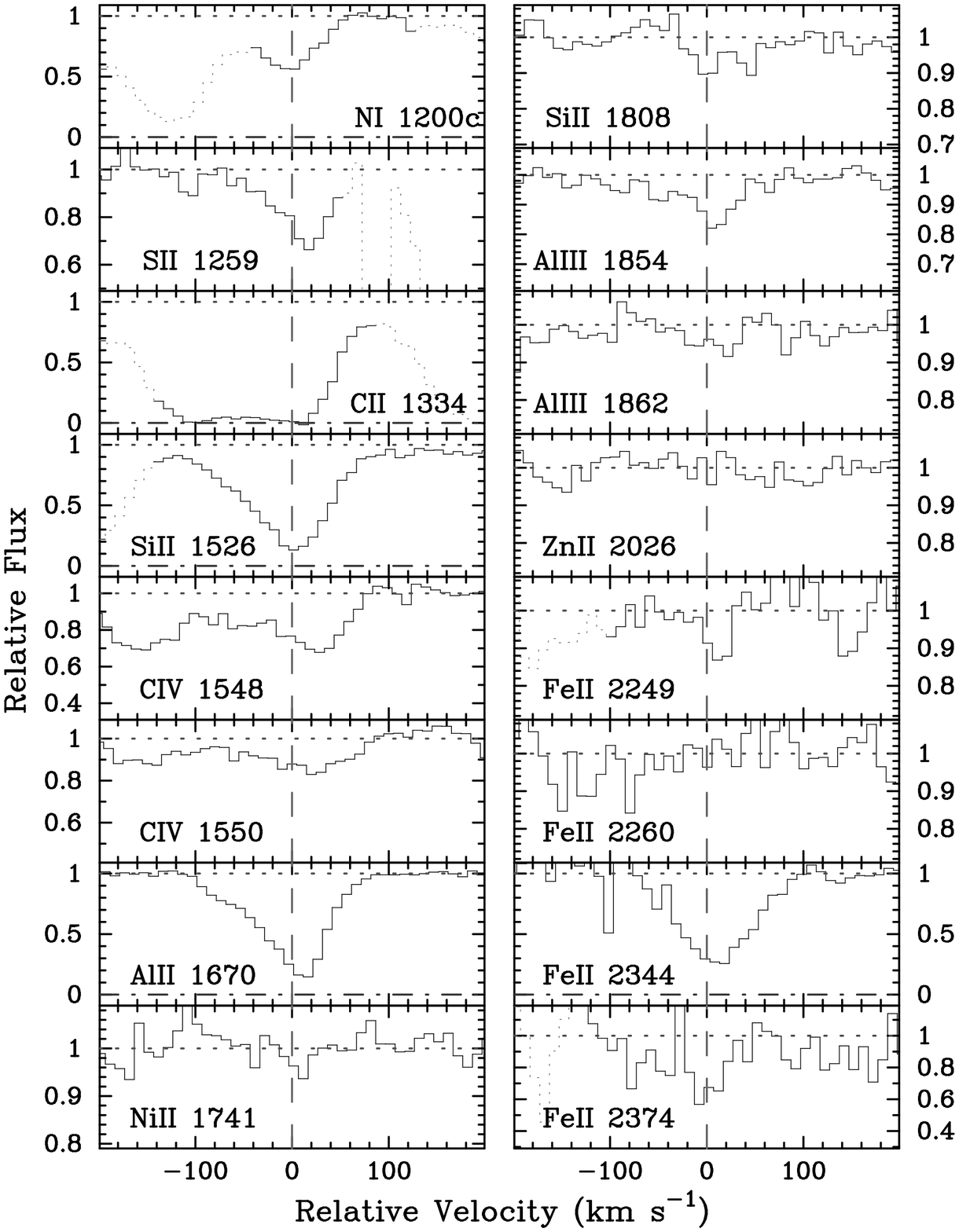}
\figcaption{Velocity plot of the metal-line transitions for the 
damped \lya system at $z = 3.280$ toward PSS0957+33.
The vertical line at $v=0$ corresponds to $z = 3.2797$.  
\label{fig:pss0957A_mtl}}
\end{center}
\end{figure}

\begin{figure}[ht]
\begin{center}
\includegraphics[height=6.1in, width=3.9in]{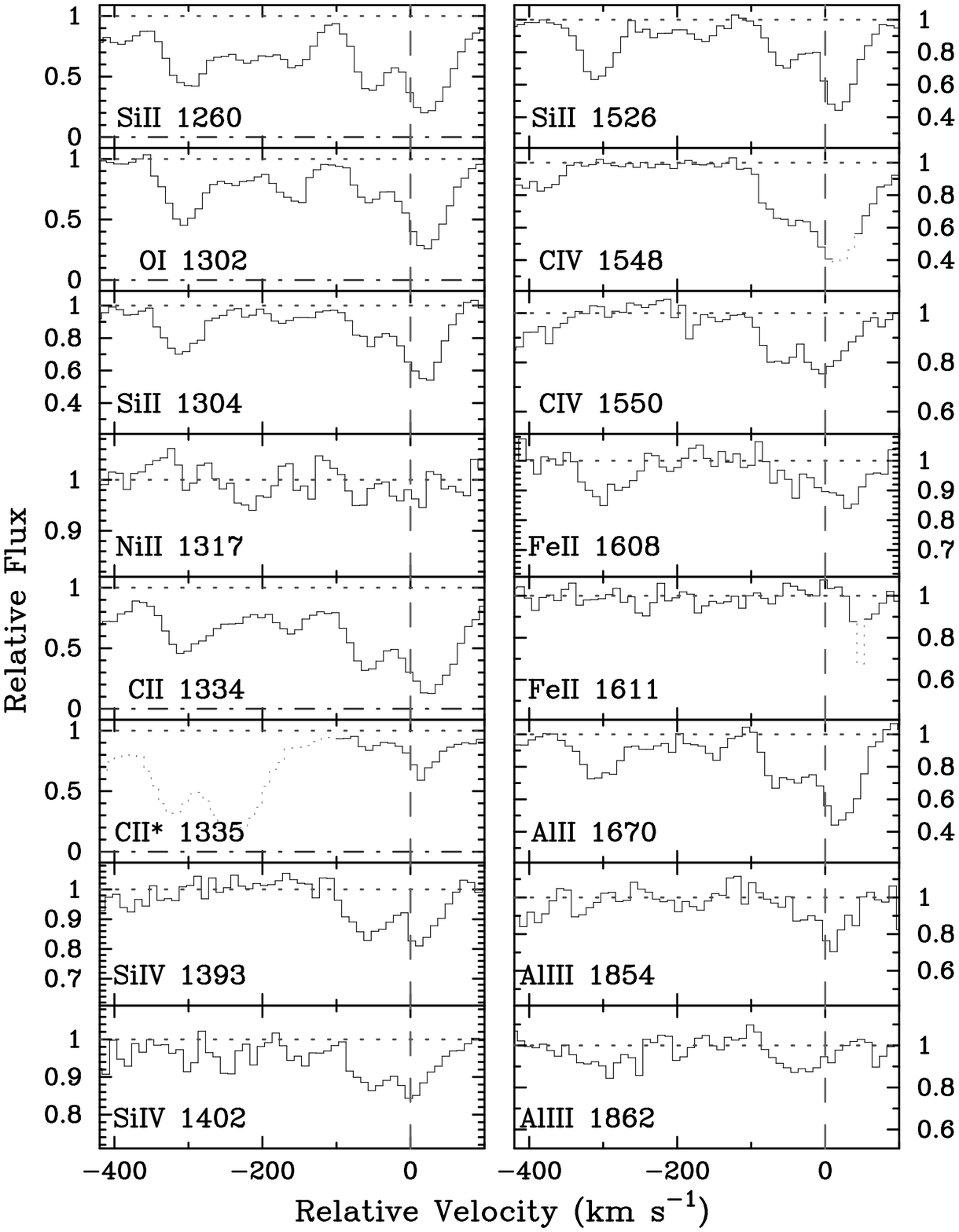}
\figcaption{Velocity plot of the metal-line transitions for the 
damped \lya system at $z = 4.180$ toward PSS0957+33.
The vertical line at $v=0$ corresponds to $z = 4.179825$.  
\label{fig:pss0957B_mtl}}
\end{center}
\end{figure}

\subsection{PSS0957+33, $z = 3.280, 4.180$ \label{subsec:PSS0957+33}}

This sightline was first analysed in PGW01 and is notable for exhibiting
two DLA at $z>3$.  Our re-analysis has revised the H\,I values of the
DLA upward by $\approx 0.1$~dex primarily due to a reassessment of the
continuum (Figures~\ref{fig:pss0957A_lya},\ref{fig:pss0957B_lya}).
The velocity plots for the metal-line transitions are given by 
Figures~\ref{fig:pss0957A_mtl} and \ref{fig:pss0957B_mtl} and the
ionic column densities are listed in Tables~\ref{tab:PSS0957+33_3.280} and
\ref{tab:PSS0957+33_4.180}.  We note in passing that our comparison of 
the ESI data with our HIRES observations \citep{pro01} suggest significant
saturation in most of the ESI profiles of the $z=4.18$ DLA.

\begin{table}[ht]\footnotesize
\begin{center}
\caption{ {\sc
IONIC COLUMN DENSITIES: BQ1021+3001, $z = 2.949$ \label{tab:BQ1021+3001_2.949}}}
\begin{tabular}{lcccc}
\tableline
\tableline
Ion & $\lambda$ & AODM & $N_{\rm adopt}$ & [X/H] \\
\tableline
C  I  &1656.9&$<12.772$\\  
C  II &1334.5&$>14.586$&$>14.586$&$>-2.704$\\  
C  II*&1335.7&$<12.909$\\  
C  IV &1548.2&$13.991 \pm  0.013$\\  
C  IV &1550.8&$13.957 \pm  0.025$\\  
O  I  &1302.2&$>14.848$&$>14.848$&$>-2.592$\\  
Al II &1670.8&$12.693 \pm  0.022$&$12.693 \pm  0.022$&$-2.497 \pm  0.102$\\  
Al III&1854.7&$<12.377$\\  
Si II &1304.4&$14.090 \pm  0.017$&$14.090 \pm  0.017$&$-2.170 \pm  0.101$\\  
Si II &1526.7&$>14.003$\\  
Si II &1808.0&$<14.603$\\  
Si IV &1393.8&$>13.602$\\  
Si IV &1402.8&$13.585 \pm  0.029$\\  
Cr II &2056.3&$<12.833$&$<12.833$&$<-1.537$\\  
Fe II &1608.5&$13.880 \pm  0.026$&$13.880 \pm  0.026$&$-2.320 \pm  0.103$\\  
Fe II &2344.2&$>13.636$\\  
Fe II &2382.8&$>13.561$\\  
Ni II &1370.1&$<13.368$&$<13.368$&$<-1.582$\\  
Ni II &1709.6&$<13.400$\\  
Zn II &2026.1&$<12.189$&$<12.189$&$<-1.181$\\  
\tableline
\end{tabular}
\end{center}
\end{table}

\begin{figure}[ht]
\begin{center}
\includegraphics[height=3.6in, width=2.8in,angle=90]{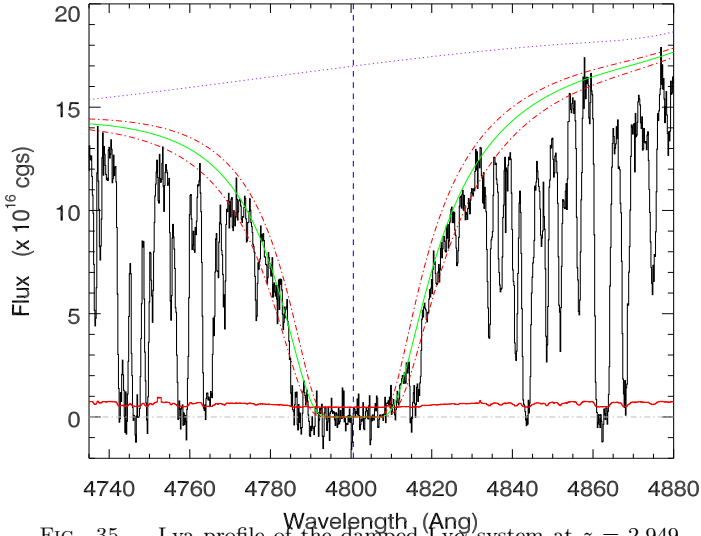}
\figcaption{Lya profile of the damped \lya system at $z=2.949$
toward BQ1021+3001.
The overplotted solid line and accompanying
dash-dot lines trace the best fit solution and the estimated 
bounds corresponding to 
$\log \N{HI} = 20.70^{+0.10}_{-0.10}$.  
Although continuum placement (dotted line) was particularly
challenging across this portion of the spectrum, the core is
rather free from line blending and the resulting fit is well
constrained.
\label{fig:q1021_lya}}
\end{center}
\end{figure}

\begin{figure}[ht]
\begin{center}
\includegraphics[height=6.1in, width=3.9in]{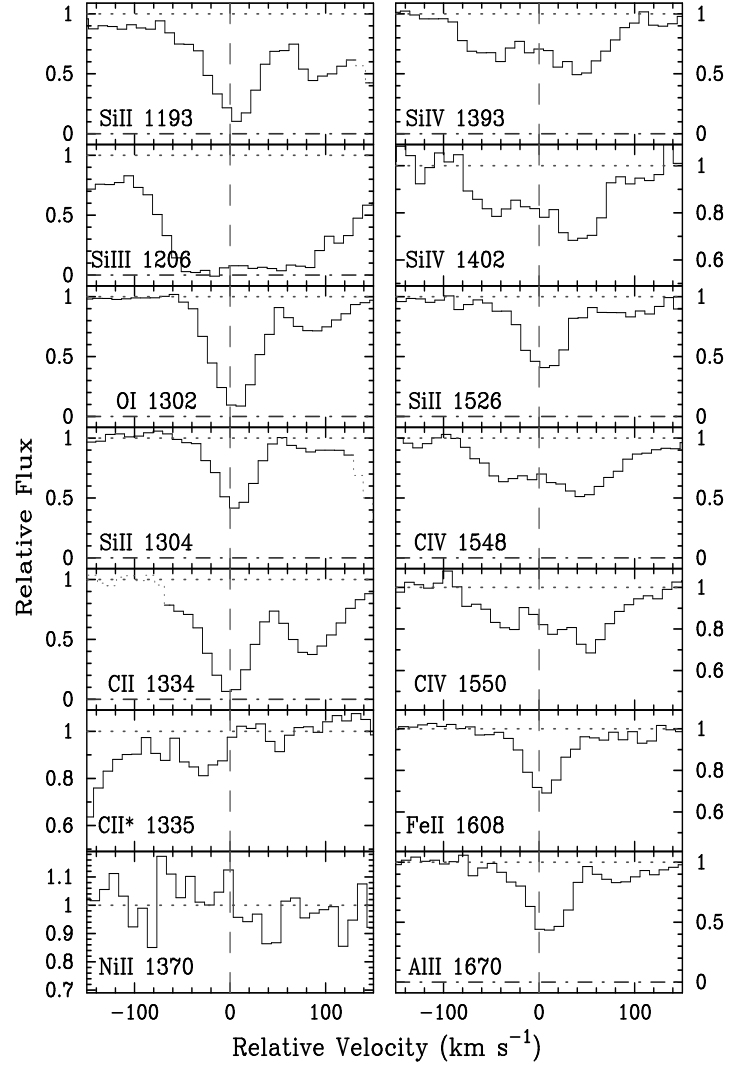}
\figcaption{Velocity plot of the metal-line transitions for the 
damped \lya system at $z = 2.949$ toward BQ1021+3001.
The vertical line at $v=0$ corresponds to $z = 2.9489$.  
\label{fig:q1021_mtl}}
\end{center}
\end{figure}

\begin{figure}[ht]
\begin{center}
\includegraphics[height=6.1in, width=3.9in]{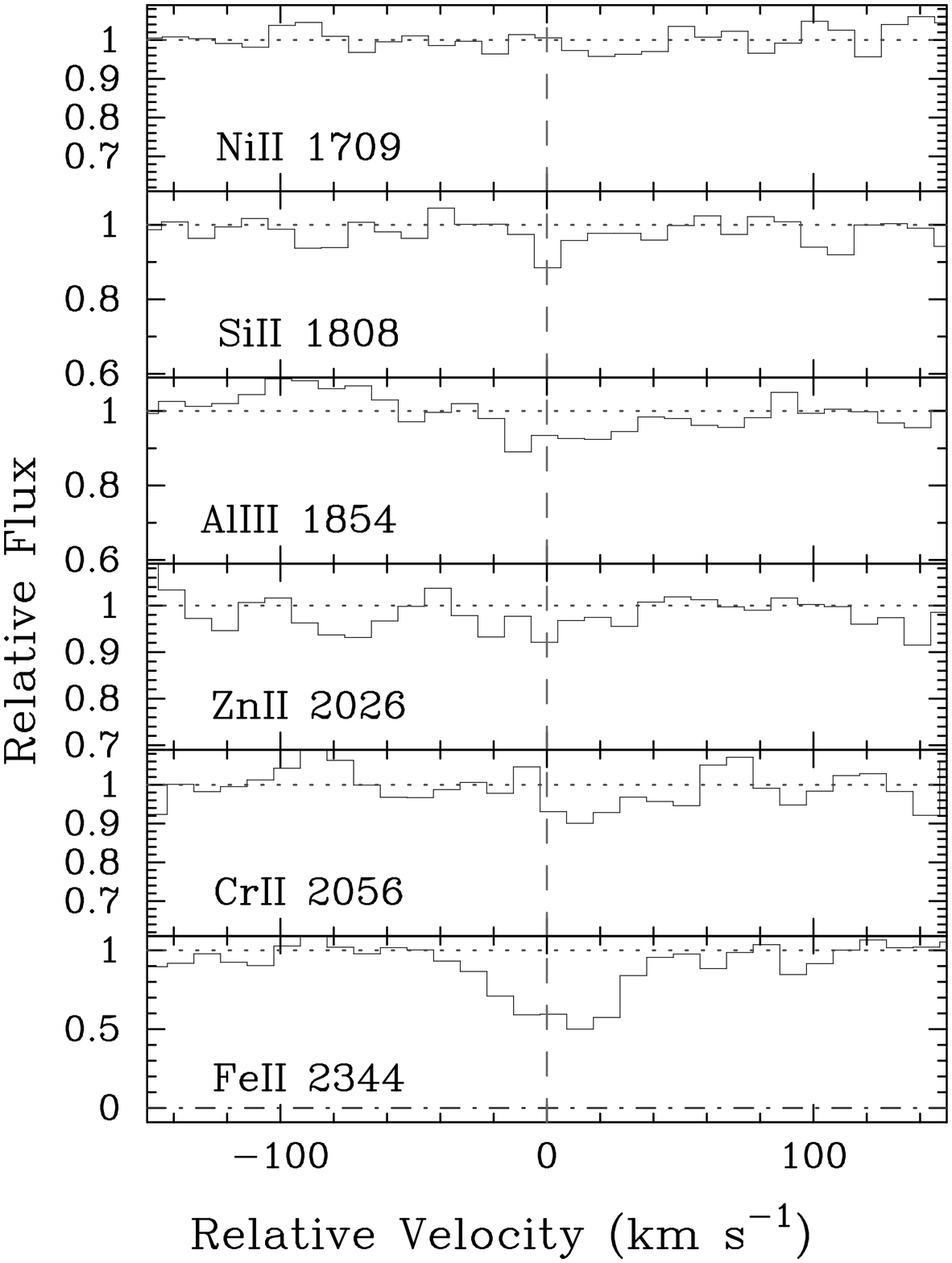}
Fig 36 -- cont
\end{center}
\end{figure}

\subsection{BQ1021+3001, $z = 2.949$ \label{subsec:BQ1021+3001_2.949}}

This quasar was drawn from the FIRST bright quasar survey 
\citep{gregg96} and \cite{white00} identified this DLA as a candidate
in their follow-up observations.
Our fit to the \lya profile is shown in Figure~\ref{fig:q1021_lya},
Figure~\ref{fig:q1021_mtl} presents the velocity profiles, and the
ionic column densities are listed in Table~\ref{tab:BQ1021+3001_2.949}.

\begin{table}[ht]\footnotesize
\begin{center}
\caption{ {\sc
IONIC COLUMN DENSITIES: CTQ460, $z = 2.777$ \label{tab:CTQ460_2.777}}}
\begin{tabular}{lcccc}
\tableline
\tableline
Ion & $\lambda$ & AODM & $N_{\rm adopt}$ & [X/H] \\
\tableline
C  II &1334.5&$>15.151$&$>15.151$&$>-2.439$\\  
C  II*&1335.7&$12.969 \pm  0.051$\\  
C  IV &1548.2&$14.016 \pm  0.011$\\  
C  IV &1550.8&$14.051 \pm  0.017$\\  
O  I  &1302.2&$>15.564$&$>15.564$&$>-2.176$\\  
Mg II &1239.9&$<15.054$&$<15.054$&$<-1.526$\\  
Mg II &1240.4&$<15.292$\\  
Al II &1670.8&$13.425 \pm  0.007$&$13.425 \pm  0.007$&$-2.065 \pm  0.100$\\  
Al III&1854.7&$13.025 \pm  0.017$\\  
Al III&1862.8&$13.057 \pm  0.029$\\  
Si II &1304.4&$>14.970$&$>14.970$&$>-1.590$\\  
Si II &1526.7&$>14.726$\\  
Si IV &1393.8&$13.774 \pm  0.008$\\  
Si IV &1402.8&$13.852 \pm  0.013$\\  
S  II &1259.5&$14.790 \pm  0.022$&$14.791 \pm  0.022$&$-1.409 \pm  0.102$\\  
Cr II &2056.3&$13.008 \pm  0.054$&$13.008 \pm  0.054$&$-1.662 \pm  0.114$\\  
Mn II &2576.9&$12.662 \pm  0.054$&$12.662 \pm  0.054$&$-1.868 \pm  0.114$\\  
Fe II &1608.5&$14.683 \pm  0.008$&$14.684 \pm  0.005$&$-1.816 \pm  0.100$\\  
Fe II &1611.2&$<14.612$\\  
Fe II &2249.9&$<14.758$\\  
Fe II &2260.8&$14.657 \pm  0.056$\\  
Fe II &2344.2&$>14.477$\\  
Fe II &2374.5&$14.686 \pm  0.008$\\  
Fe II &2382.8&$>14.258$\\  
Fe II &2586.7&$>14.503$\\  
Fe II &2600.2&$>14.181$\\  
Ni II &1370.1&$13.572 \pm  0.032$&$13.587 \pm  0.024$&$-1.663 \pm  0.103$\\  
Ni II &1454.8&$13.447 \pm  0.115$\\  
Ni II &1709.6&$13.573 \pm  0.064$\\  
Ni II &1741.6&$13.675 \pm  0.060$\\  
Ni II &1751.9&$13.676 \pm  0.061$\\  
Zn II &2062.7&$<12.359$&$<12.359$&$<-1.311$\\  
\tableline
\end{tabular}
\end{center}
\end{table}

\begin{figure}[ht]
\begin{center}
\includegraphics[height=3.6in, width=2.8in,angle=90]{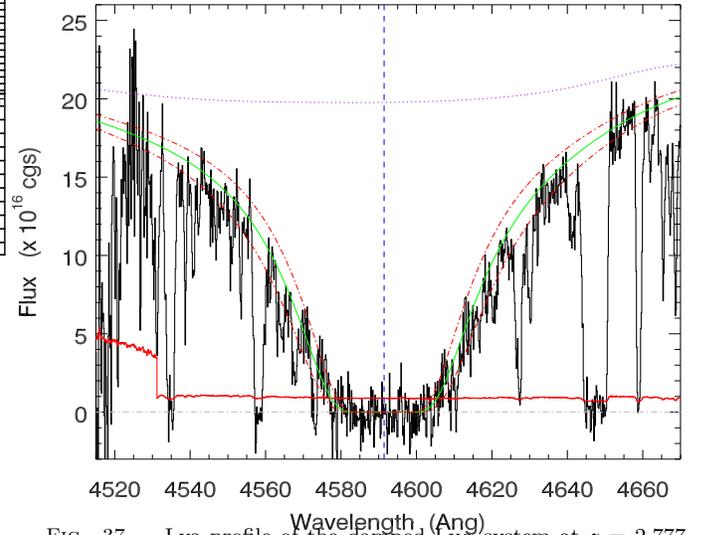}
\figcaption{Lya profile of the damped \lya system at $z=2.777$
toward CTQ460.
The overplotted solid line and accompanying
dash-dot lines trace the best fit solution and the estimated 
bounds corresponding to 
$\log \N{HI} = 21.00^{+0.10}_{-0.10}$.  
Although the continuum is not particularly well constrained across this
DLA and the blue wing approaches the CCD defect, the core is well
sampled and constrains the H\,I column density.
\label{fig:q1036_lya}}
\end{center}
\end{figure}

\begin{figure}[ht]
\begin{center}
\includegraphics[height=6.1in, width=3.9in]{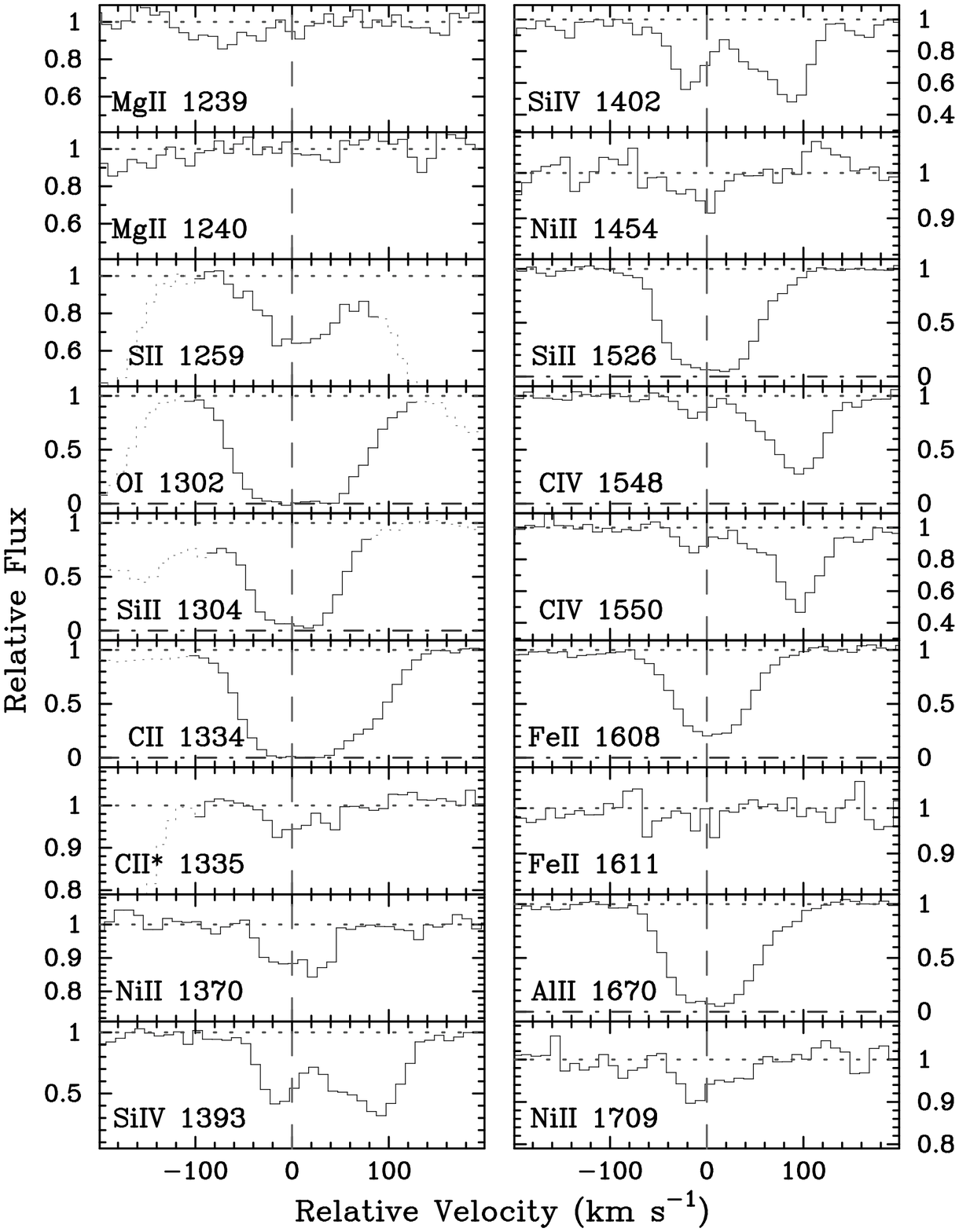}
\figcaption{Velocity plot of the metal-line transitions for the 
damped \lya system at $z = 2.777$ toward CTQ460.
The vertical line at $v=0$ corresponds to $z = 2.7775$.  
\label{fig:q1036_mtl}}
\end{center}
\end{figure}

\begin{figure}[ht]
\begin{center}
\includegraphics[height=6.1in, width=3.9in]{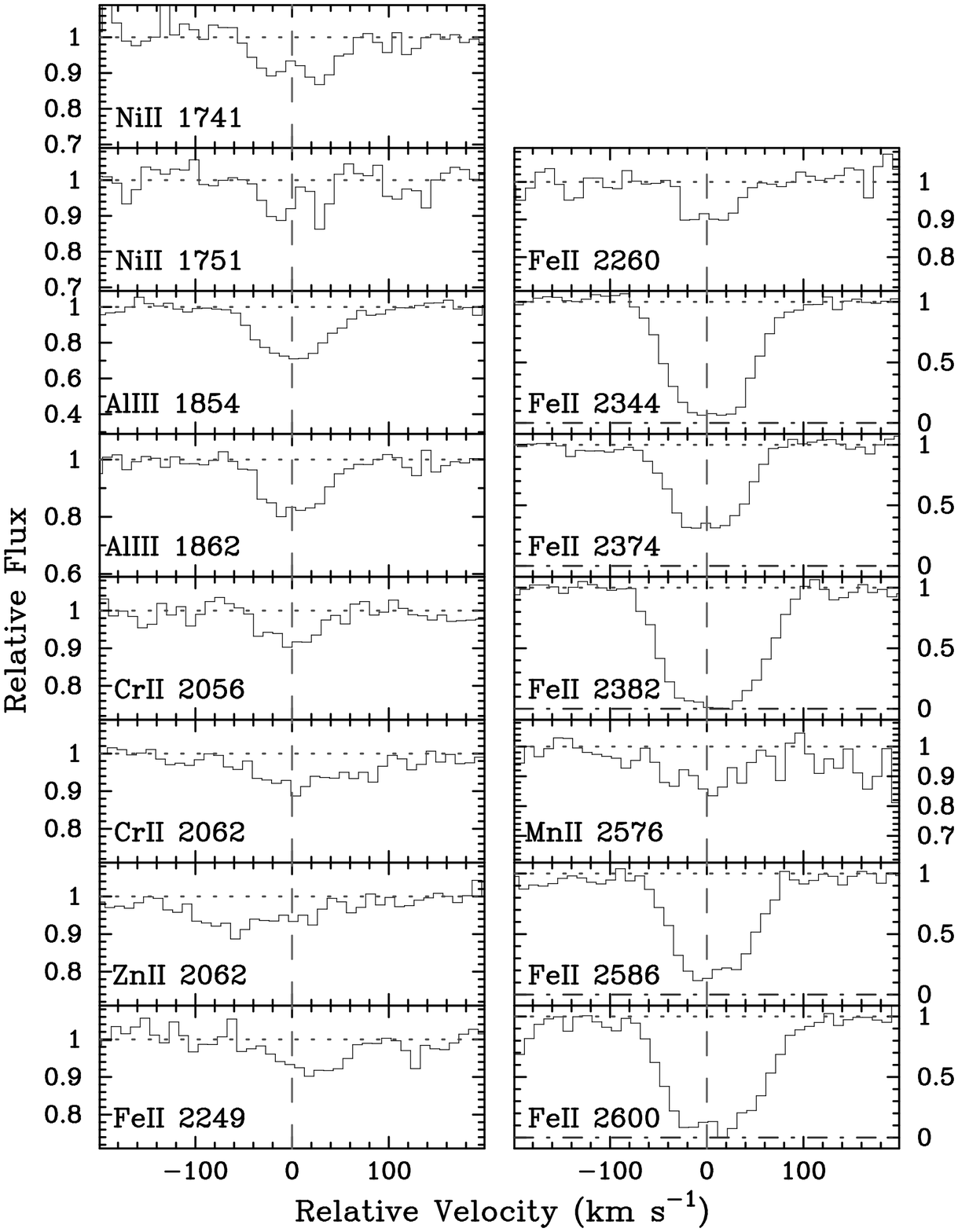}
Fig 38 -- cont
\end{center}
\end{figure}

\subsection{CTQ460, $z = 2.777$ \label{subsec:CTQ460_2.777}}

This DLA was discovered by \cite{lopez01} toward the quasar
CTQ460 which is drawn from the Cal\'an-Tololo survey \citep{maza95}.
The \lya profile presented in Figure~\ref{fig:q1036_lya} has a well
sampled core and a relatively precise H\,I measurement.
Figure~\ref{fig:q1036_mtl} provides an exhaustive list of metal-line profiles
and Table~\ref{tab:CTQ460_2.777} gives the ionic column densities.

\clearpage


\begin{table}[ht]\footnotesize
\begin{center}
\caption{ {\sc
IONIC COLUMN DENSITIES: HS1132+2243, $z = 2.783$ \label{tab:HS1132+2243_2.783}}}
\begin{tabular}{lcccc}
\tableline
\tableline
Ion & $\lambda$ & AODM & $N_{\rm adopt}$ & [X/H] \\
\tableline
C  I  &1656.9&$<12.569$\\  
C  II &1334.5&$>14.580$&$>14.580$&$>-3.010$\\  
C  II*&1335.7&$<12.931$\\  
C  IV &1548.2&$13.370 \pm  0.036$\\  
C  IV &1550.8&$13.365 \pm  0.070$\\  
N  I  &1199.5&$14.127 \pm  0.020$&$14.019 \pm  0.019$&$-2.911 \pm  0.073$\\  
N  I  &1200.2&$13.863 \pm  0.039$\\  
N  II &1084.0&$<13.431$\\  
O  I  &1302.2&$>14.989$&$>14.989$&$>-2.751$\\  
Al II &1670.8&$>12.867$&$>12.867$&$>-2.623$\\  
Al III&1854.7&$12.141 \pm  0.107$\\  
Si II &1193.3&$>14.015$&$14.491 \pm  0.118$&$-2.069 \pm  0.137$\\  
Si II &1260.4&$>13.890$\\  
Si II &1304.4&$>14.303$\\  
Si II &1526.7&$>14.186$\\  
Si II &1808.0&$14.491 \pm  0.118$\\  
Si IV &1393.8&$13.008 \pm  0.039$\\  
Si IV &1402.8&$13.000 \pm  0.081$\\  
S  II &1250.6&$<14.113$&$14.071 \pm  0.060$&$-2.129 \pm  0.092$\\  
S  II &1259.5&$14.071 \pm  0.060$\\  
Cr II &2056.3&$12.834 \pm  0.104$&$12.834 \pm  0.104$&$-1.836 \pm  0.125$\\  
Mn II &2576.9&$<12.457$&$<12.457$&$<-2.073$\\  
Fe II &1144.9&$13.951 \pm  0.044$&$14.024 \pm  0.014$&$-2.476 \pm  0.071$\\  
Fe II &1608.5&$14.084 \pm  0.019$\\  
Fe II &1611.2&$<14.504$\\  
Fe II &2260.8&$<14.386$\\  
Fe II &2344.2&$>13.870$\\  
Fe II &2374.5&$14.082 \pm  0.027$\\  
Fe II &2382.8&$>13.652$\\  
Fe II &2586.7&$13.900 \pm  0.035$\\  
Fe II &2600.2&$>13.596$\\  
Fe III&1122.5&$<13.726$\\  
Ni II &1317.2&$<13.155$&$<13.155$&$<-2.095$\\  
Ni II &1370.1&$<13.164$\\  
Ni II &1741.5&$<13.249$\\  
Ni II &1751.9&$<13.402$\\  
Zn II &2026.1&$<11.986$&$<11.986$&$<-1.684$\\  
\tableline
\end{tabular}
\end{center}
\end{table}

\begin{figure}[ht]
\begin{center}
\includegraphics[height=3.6in, width=2.8in,angle=90]{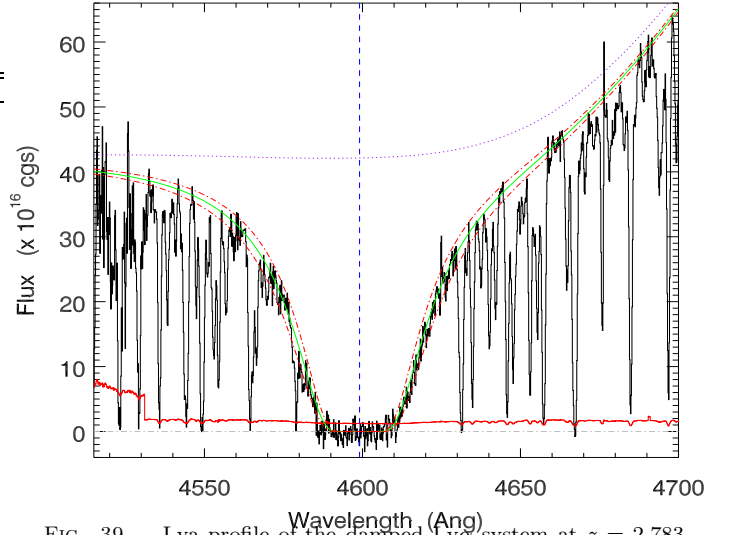}
\figcaption{Lya profile of the damped \lya system at $z=2.783$
toward HS1132+2243.
The overplotted solid line and accompanying
dash-dot lines trace the best fit solution and the estimated 
bounds corresponding to 
$\log \N{HI} = 21.00^{+0.07}_{-0.07}$.  
Although the red wing of the \lya profile occurs along the \lya emission
peak of the quasar, the $\N{HI}$ is very well determined by the core of
the \lya profile.
\label{fig:hs1132_lya}}
\end{center}
\end{figure}

\begin{figure}[ht]
\begin{center}
\includegraphics[height=6.1in, width=3.9in]{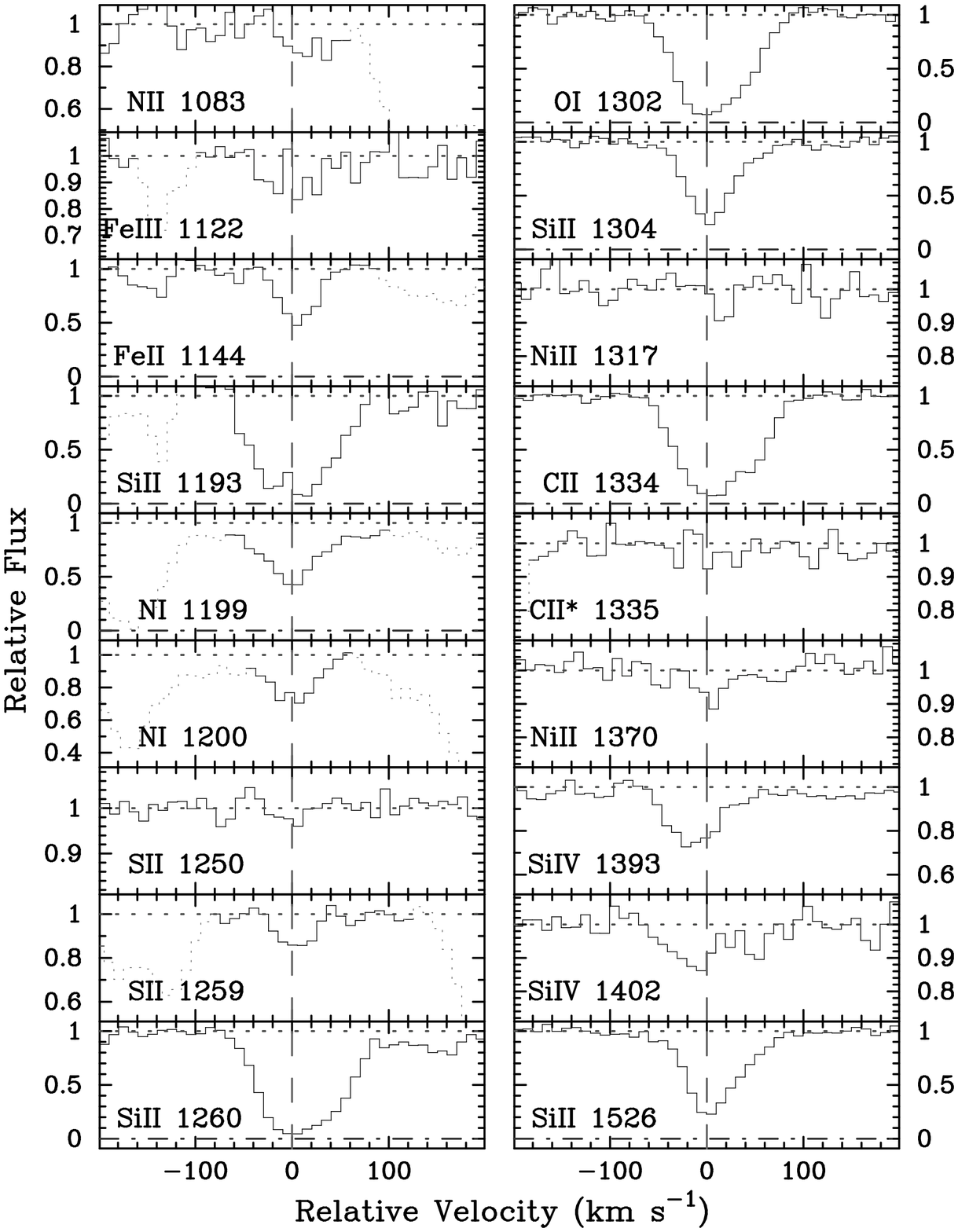}
\figcaption{Velocity plot of the metal-line transitions for the 
damped \lya system at $z = 2.783$ toward HS1132+2243.
The vertical line at $v=0$ corresponds to $z = 2.78347$.  
\label{fig:hs1132_mtl}}
\end{center}
\end{figure}

\begin{figure}[ht]
\begin{center}
\includegraphics[height=6.1in, width=3.9in]{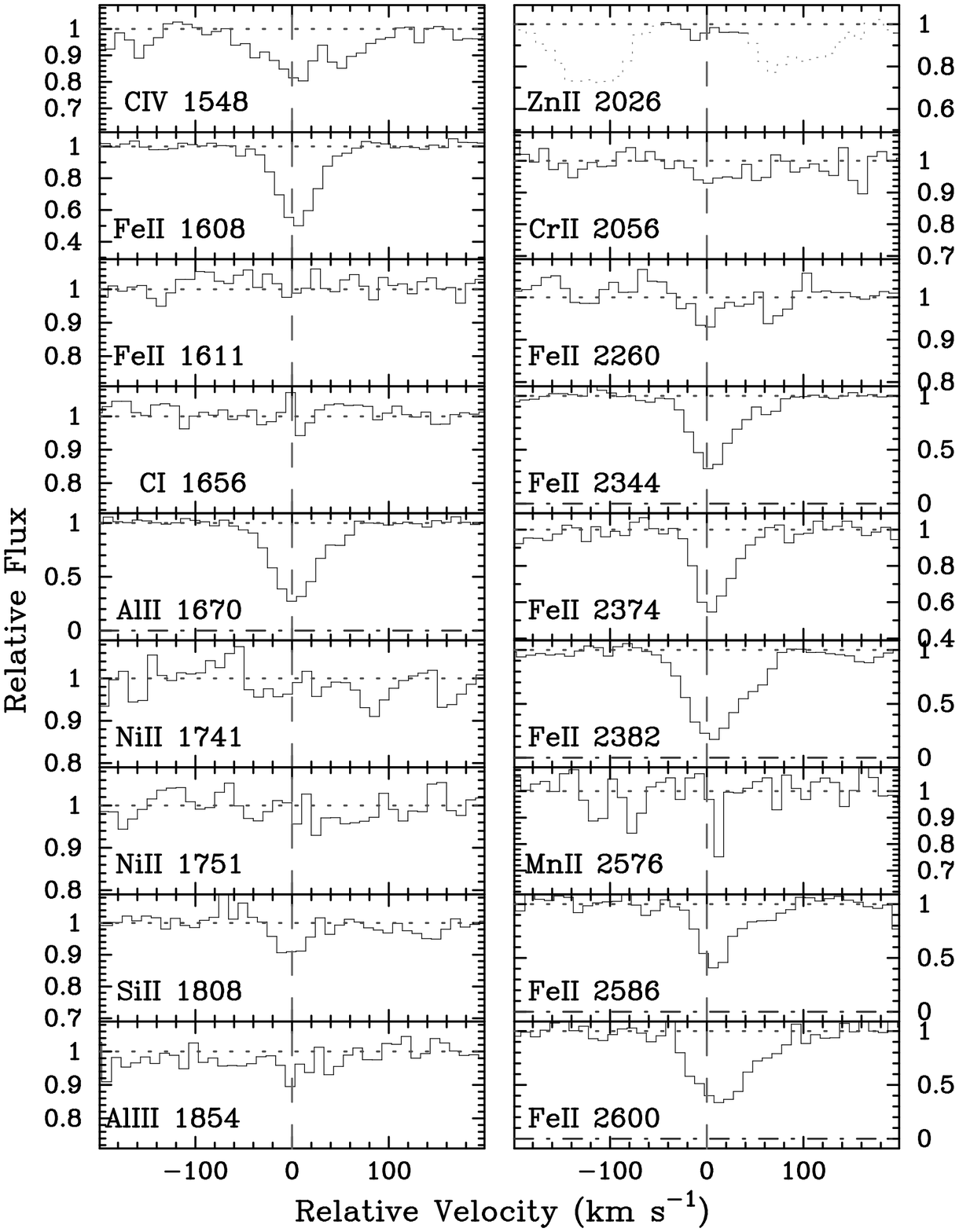}
Fig 40 -- cont
\end{center}
\end{figure}

\subsection{HS1132+2243, $z = 2.783$ \label{subsec:HS1132+2243_2.783}}

This quasar is a member of the Hamburg-ESO Quasar Survey \citep{hagen99}
and one could identify the DLA in their discovery spectrum.
Although the red wing of the \lya profile occurs along the \lya emission
peak of the quasar, the $\N{HI}$ is very well set by the core of
the \lya profile (Figure~\ref{fig:hs1132_lya}).  The extensive set of
metal-line transitions are given in Figure~\ref{fig:hs1132_mtl} and 
Table~\ref{tab:HS1132+2243_2.783} lists the ionic column densities.

\begin{table}[ht]\footnotesize
\begin{center}
\caption{ {\sc
IONIC COLUMN DENSITIES: Q1209+0919, $z = 2.584$ \label{tab:Q1209+0919_2.584}}}
\begin{tabular}{lcccc}
\tableline
\tableline
Ion & $\lambda$ & AODM & $N_{\rm adopt}$ & [X/H] \\
\tableline
C  I  &1560.3&$<13.192$\\  
C  I  &1656.9&$13.276 \pm  0.085$\\  
C  IV &1548.2&$>14.976$\\  
C  IV &1550.8&$>14.973$\\  
Al II &1670.8&$>14.068$&$>14.068$&$>-1.822$\\  
Al III&1854.7&$13.712 \pm  0.013$\\  
Al III&1862.8&$13.826 \pm  0.017$\\  
Si II &1526.7&$>15.295$&$>15.990$&$>-0.970$\\  
Si II &1808.0&$>15.990$\\  
Ti II &1910.6&$<12.652$&$<12.652$&$<-1.688$\\  
Cr II &2056.3&$13.609 \pm  0.053$&$13.581 \pm  0.050$&$-1.489 \pm  0.112$\\  
Cr II &2066.2&$13.468 \pm  0.137$\\  
Mn II &2576.9&$13.134 \pm  0.059$&$13.142 \pm  0.049$&$-1.788 \pm  0.111$\\  
Mn II &2594.5&$13.159 \pm  0.084$\\  
Fe II &1608.5&$>15.004$&$15.219 \pm  0.044$&$-1.681 \pm  0.109$\\  
Fe II &1611.2&$<15.043$\\  
Fe II &2249.9&$<15.283$\\  
Fe II &2260.8&$15.219 \pm  0.044$\\  
Fe II &2344.2&$>14.894$\\  
Fe II &2374.5&$>15.324$\\  
Fe II &2382.8&$>14.730$\\  
Ni II &1703.4&$<14.505$&$<14.218$&$<-1.432$\\  
Ni II &1751.9&$<14.218$\\  
Zn II &2026.1&$12.982 \pm  0.053$&$12.982 \pm  0.053$&$-1.088 \pm  0.113$\\  
\tableline
\end{tabular}
\end{center}
\end{table}

\begin{figure}[ht]
\begin{center}
\includegraphics[height=3.6in, width=2.8in,angle=90]{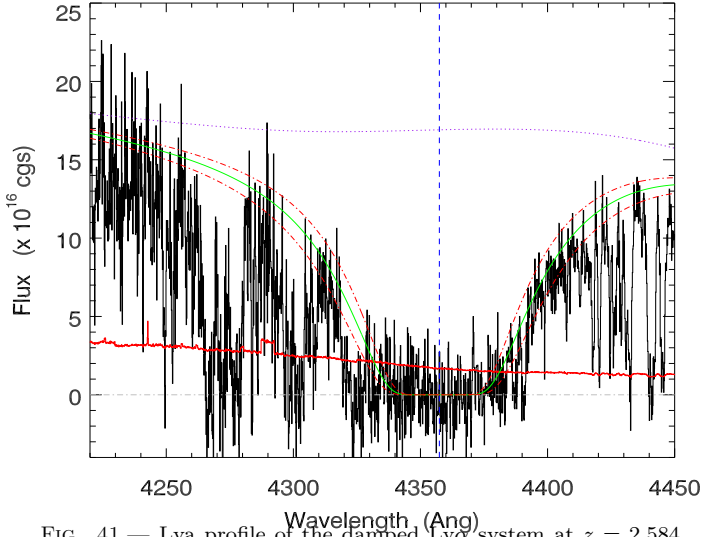}
\figcaption{Lya profile of the damped \lya system at $z=2.584$
toward Q1209+0919.  
The overplotted solid line and accompanying
dash-dot lines trace the best fit solution and the estimated 
bounds corresponding to 
$\log \N{HI} = 21.40^{+0.10}_{-0.10}$.  
The fit is complicated by the poor behavior of the continuum
across the DLA which includes the QSO O\,VI emission feature at
$\lambda \approx 4450$\AA, but the value is relatively well constrained
by the core of the \lya profile.
\label{fig:q1209_lya}}
\end{center}
\end{figure}

\begin{figure}[ht]
\begin{center}
\includegraphics[height=6.1in, width=3.9in]{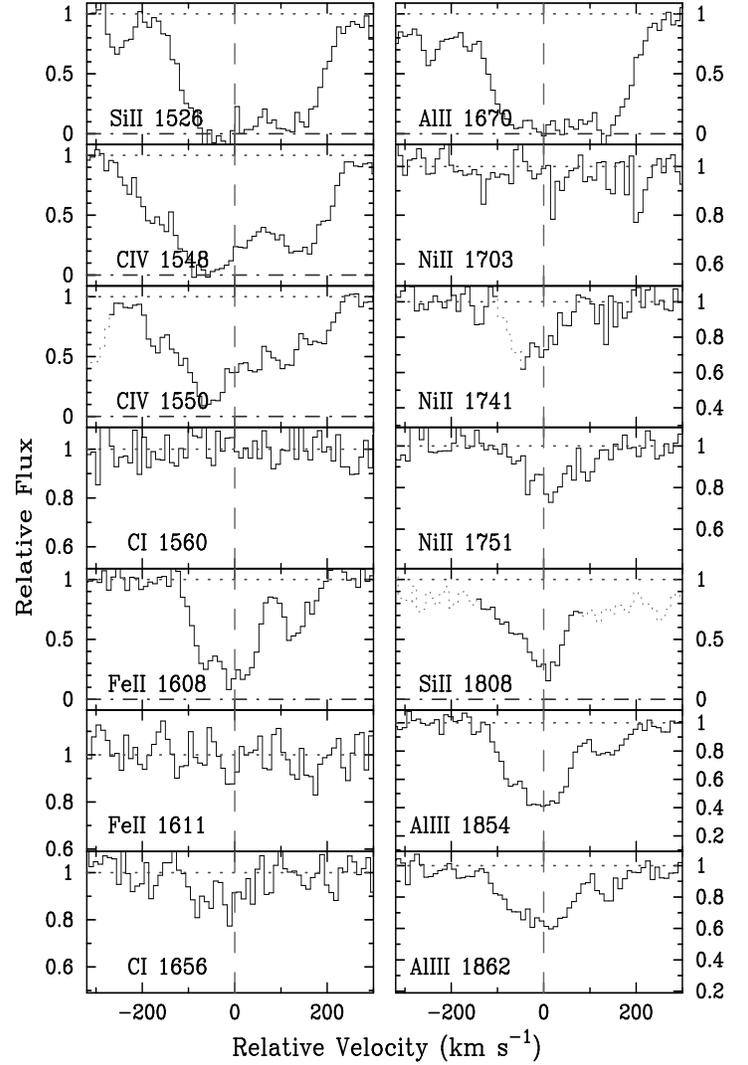}
\figcaption{Velocity plot of the metal-line transitions for the 
damped \lya system at $z = 2.584$ toward Q1209+0919. 
The vertical line at $v=0$ corresponds to $z = 2.5841$.  
\label{fig:q1209_mtl}}
\end{center}
\end{figure}

\begin{figure}[ht]
\begin{center}
\includegraphics[height=6.1in, width=3.9in]{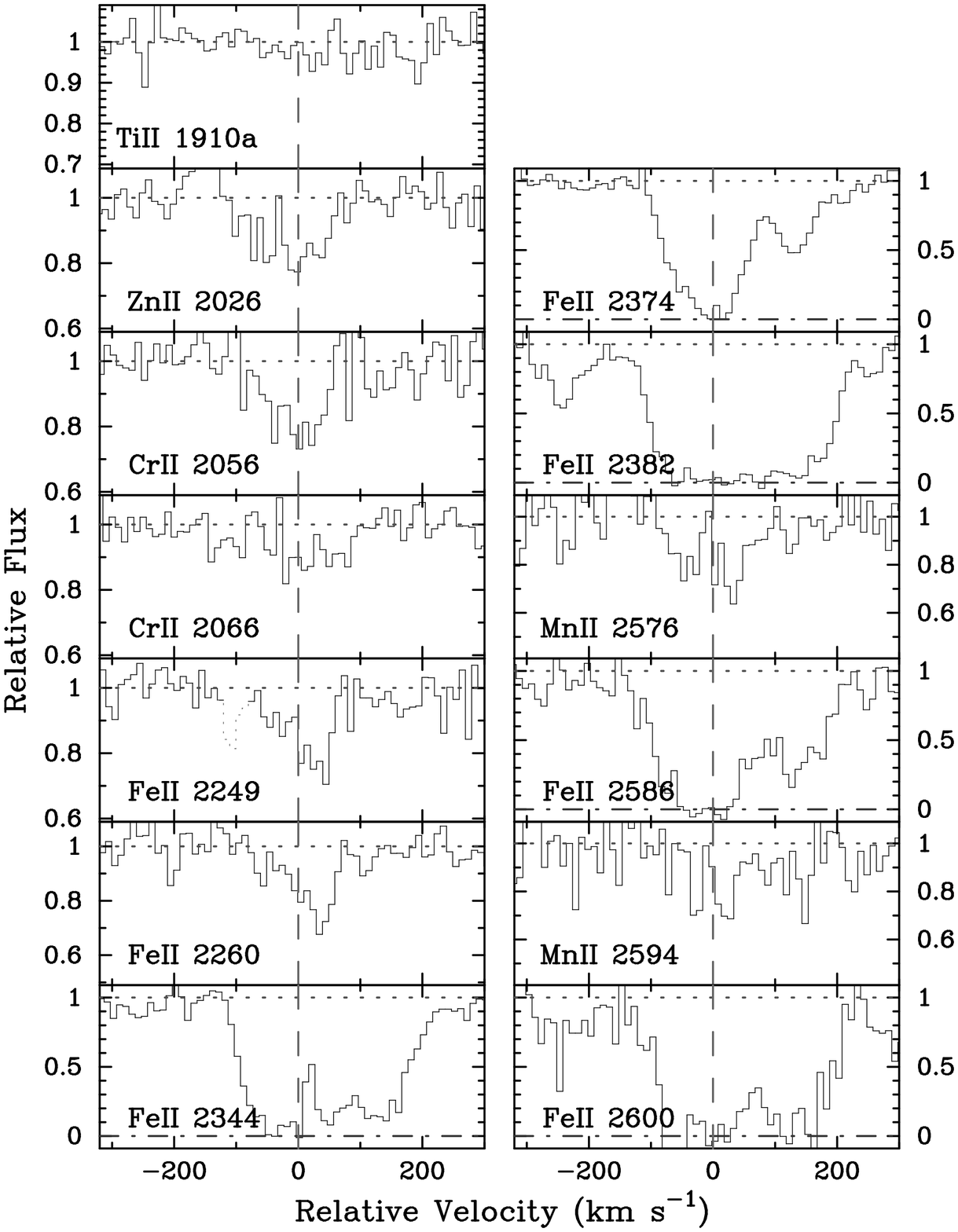}
Fig 42 -- cont
\end{center}
\end{figure}

\subsection{Q1209+0919, $z = 2.584$ \label{subsec:Q1209+0919_2.584}}

This DLA is a member of the LBQS survey of \cite{wolfe95} and exhibits
a relatively large H\,I column density.  It has one of the lowest redshifts 
in this survey and was chosen primarily because of its large H\,I column
density.  Our fit to the \lya profile
is presented in Figure~\ref{fig:q1209_lya}.  Figure~\ref{fig:q1209_mtl}
plots the velocity profiles which include Zn$^+$ and Mn$^+$ detections. 
Finally, the ionic column densities are 
provided in Table~\ref{tab:Q1209+0919_2.584}.

\begin{table}[ht]\footnotesize
\begin{center}
\caption{ {\sc
IONIC COLUMN DENSITIES: PSS1248+31, $z = 3.696$ \label{tab:PSS1248+31_3.696}}}
\begin{tabular}{lcccc}
\tableline
\tableline
Ion & $\lambda$ & AODM & $N_{\rm adopt}$ & [X/H] \\
\tableline
C  IV &1548.2&$13.841 \pm  0.024$\\  
C  IV &1550.8&$13.903 \pm  0.047$\\  
Al II &1670.8&$12.777 \pm  0.020$&$12.777 \pm  0.020$&$-2.343 \pm  0.073$\\  
Al III&1854.7&$<12.191$\\  
Si II &1304.4&$14.393 \pm  0.016$&$14.393 \pm  0.016$&$-1.797 \pm  0.072$\\  
Si II &1526.7&$>14.243$\\  
Si II &1808.0&$<14.610$\\  
Si IV &1393.8&$13.414 \pm  0.017$\\  
Si IV &1402.8&$13.310 \pm  0.045$\\  
Cr II &2056.3&$<13.242$&$<13.242$&$<-1.058$\\  
Fe II &1608.5&$13.890 \pm  0.046$&$13.890 \pm  0.046$&$-2.240 \pm  0.084$\\  
Ni II &1454.8&$<13.622$&$<13.493$&$<-1.387$\\  
Ni II &1741.6&$<13.493$\\  
Zn II &2026.1&$<12.615$&$<12.615$&$<-0.685$\\  
\tableline
\end{tabular}
\end{center}
\end{table}

\begin{figure}[ht]
\begin{center}
\includegraphics[height=3.6in, width=2.8in,angle=90]{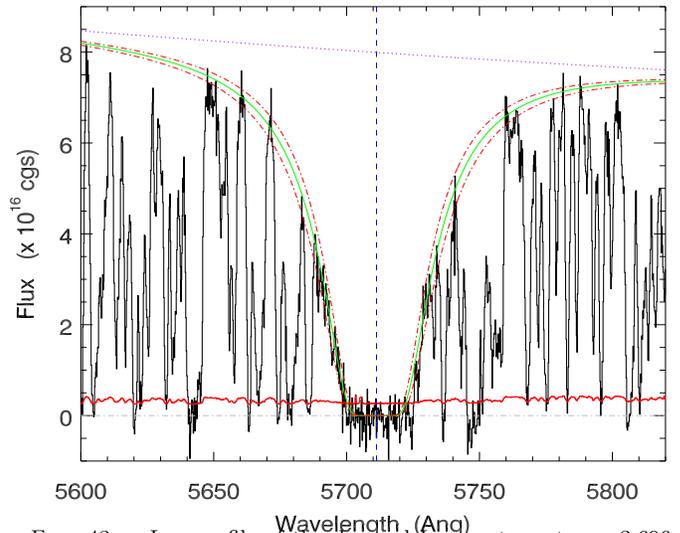}
\figcaption{Lya profile of the damped \lya system at $z=3.696$
toward PSS1248+31.
The overplotted solid line and accompanying
dash-dot lines trace the best fit solution and the estimated 
bounds corresponding to 
$\log \N{HI} = 20.63^{+0.07}_{-0.07}$.  
Because of minimal contamination within the core of the \lya profile
and a well behaved continuum, we are particularly certain of the
H\,I value for this DLA.
\label{fig:pss1248_lya}}
\end{center}
\end{figure}

\begin{figure}[ht]
\begin{center}
\includegraphics[height=6.1in, width=3.9in]{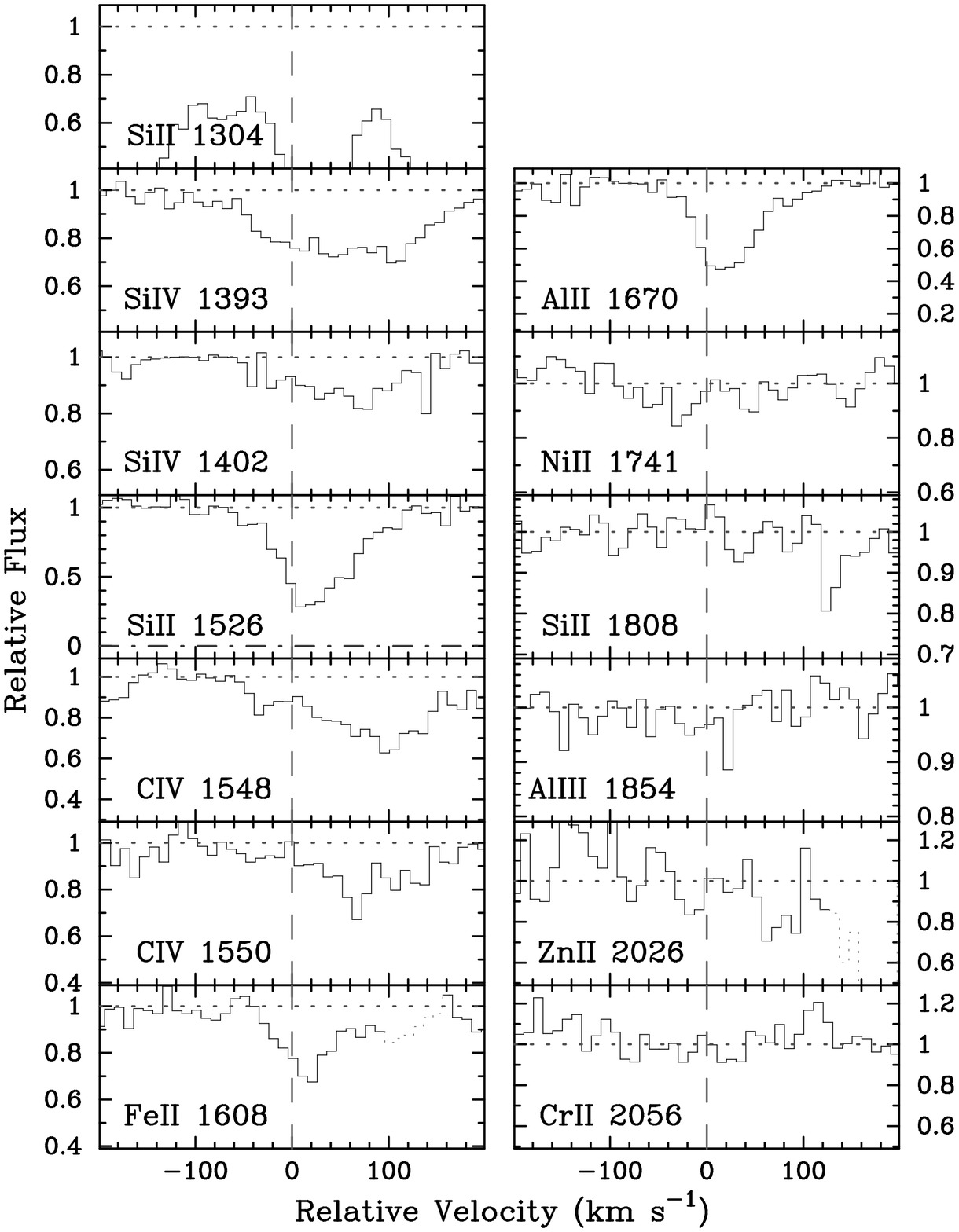}
\figcaption{Velocity plot of the metal-line transitions for the 
damped \lya system at $z = 3.696$ toward PSS1248+31.
The vertical line at $v=0$ corresponds to $z = 3.6960$.  
\label{fig:pss1248_mtl}}
\end{center}
\end{figure}

\subsection{PSS1248+31, $z = 3.696$ \label{subsec:PSS1248+31_3.696}}

This PSS quasar was first analysed in PGW01 and we note that our revised
fit to the \lya profile reveals a significantly larger H\,I column density
(Figure~\ref{fig:pss1248_lya}) owing primarily to a re-evaluation of the quasar
continuum.  In PGW01, we fit the continuum prior to fitting the \lya profile
and in the process significantly underestimated both the 
continuum and the $\N{HI}$ value.
We consider the new solution to be very well constrained by
the observations.  Figure~\ref{fig:pss1248_mtl} gives the metal-line
profiles and Table~\ref{tab:PSS1248+31_3.696} lists the ionic column densities.

\begin{table}[ht]\footnotesize
\begin{center}
\caption{ {\sc
IONIC COLUMN DENSITIES: PSS1253-0228, $z = 2.783$ \label{tab:PSS1253-0228_2.783}}}
\begin{tabular}{lcccc}
\tableline
\tableline
Ion & $\lambda$ & AODM & $N_{\rm adopt}$ & [X/H] \\
\tableline
C  II &1334.5&$>15.012$&$>15.012$&$>-3.428$\\  
O  I  &1302.2&$>15.347$&$>15.347$&$>-3.243$\\  
Al II &1670.8&$>13.399$&$>13.399$&$>-2.941$\\  
Al III&1854.7&$12.537 \pm  0.097$\\  
Al III&1862.8&$<12.664$\\  
Si II &1808.0&$>15.597$&$>15.597$&$>-1.813$\\  
Si IV &1402.8&$13.672 \pm  0.052$\\  
Ti II &1910.6&$<12.841$&$<12.842$&$<-1.948$\\  
Cr II &2056.3&$13.693 \pm  0.029$&$13.632 \pm  0.025$&$-1.888 \pm  0.202$\\  
Cr II &2062.2&$13.537 \pm  0.048$\\  
Mn II &2576.9&$13.420 \pm  0.070$&$13.420 \pm  0.070$&$-1.960 \pm  0.212$\\  
Fe II &1608.5&$>14.627$&$15.358 \pm  0.038$&$-1.992 \pm  0.204$\\  
Fe II &1611.2&$15.340 \pm  0.049$\\  
Fe II &2249.9&$15.393 \pm  0.061$\\  
Fe II &2260.8&$>15.207$\\  
Fe II &2344.2&$>14.331$\\  
Fe II &2374.5&$>14.752$\\  
Fe II &2382.8&$>14.136$\\  
Fe II &2586.7&$>14.388$\\  
Fe II &2600.2&$>14.014$\\  
Ni II &1709.6&$14.359 \pm  0.048$&$14.170 \pm  0.036$&$-1.930 \pm  0.203$\\  
Ni II &1741.5&$14.050 \pm  0.069$\\  
Ni II &1751.9&$14.149 \pm  0.072$\\  
Zn II &2026.1&$<13.068$&$12.768 \pm  0.074$&$-1.752 \pm  0.213$\\  
Zn II &2062.7&$12.768 \pm  0.074$\\  
\tableline
\end{tabular}
\end{center}
\end{table}

\begin{figure}[ht]
\begin{center}
\includegraphics[height=3.6in, width=2.8in,angle=90]{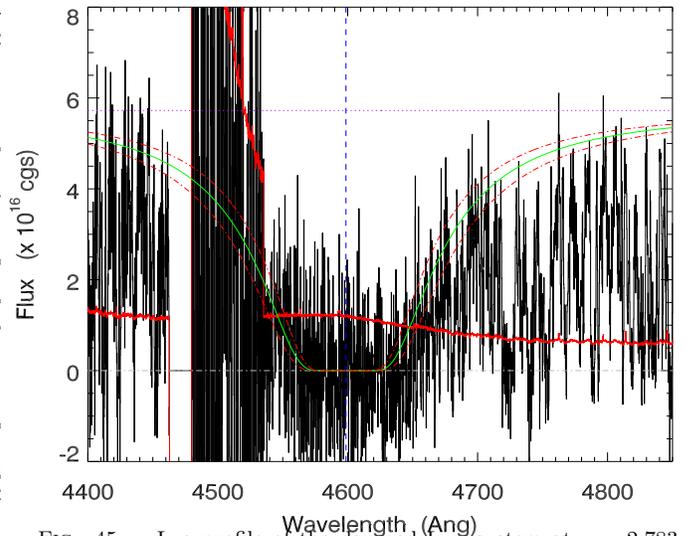}
\figcaption{Lya profile of the damped \lya system at $z=2.783$
toward PSS1253-0228.
The overplotted solid line and accompanying
dash-dot lines trace the best fit solution and the estimated 
bounds corresponding to 
$\log \N{HI} = 21.85^{+0.20}_{-0.20}$.  Although the fit
is complicated by several factors (see the text), we consider this
value to be relatively secure.  In particular, it is difficult 
for the H\,I column density to be significantly lower than 
$10^{21.85} \cm{-2}$.
\label{fig:pss1253_lya}}
\end{center}
\end{figure}

\begin{figure}[ht]
\begin{center}
\includegraphics[height=6.1in, width=3.9in]{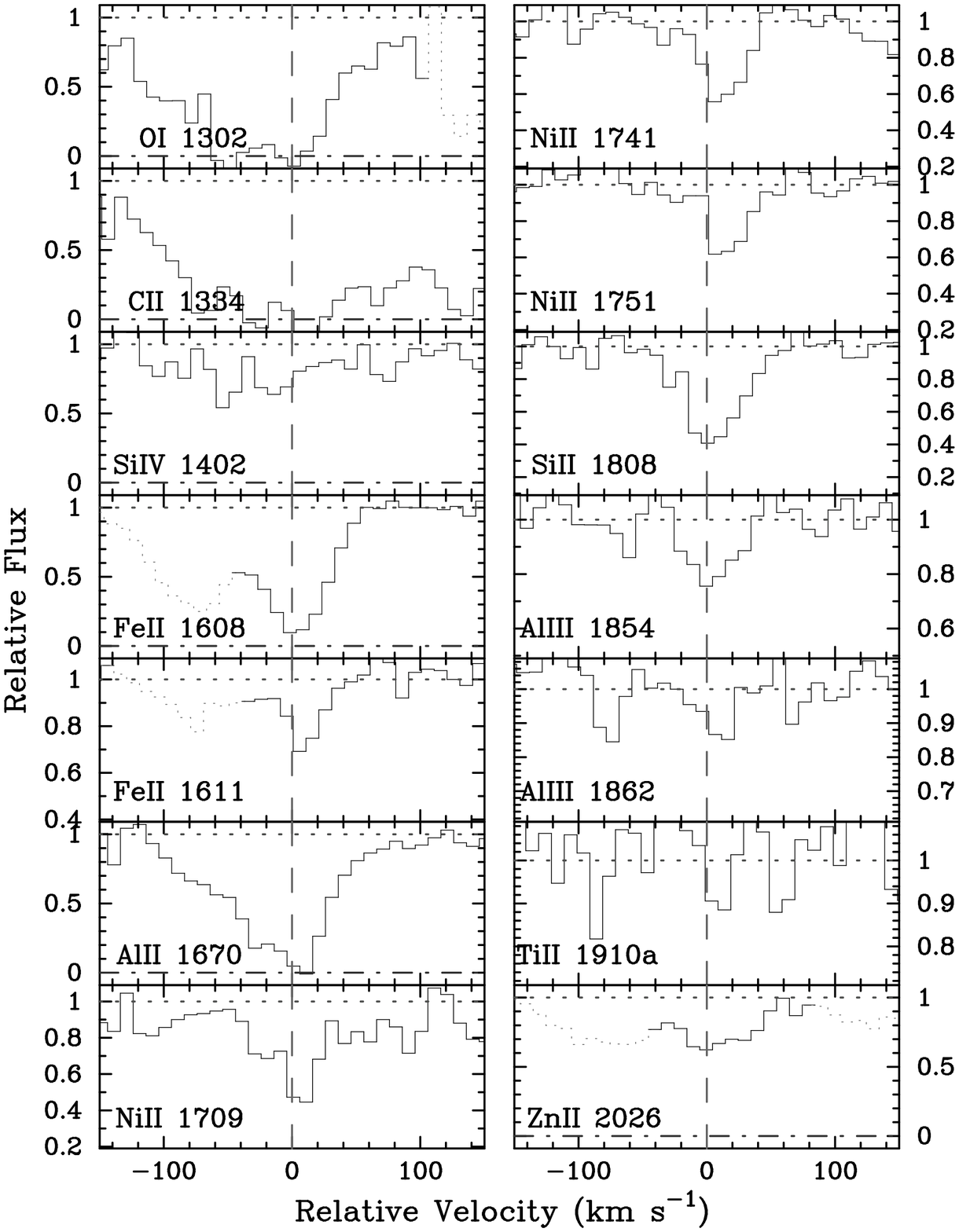}
\figcaption{Velocity plot of the metal-line transitions for the 
damped \lya system at $z = 2.783$ toward PSS1253-0228.
The vertical line at $v=0$ corresponds to $z = 2.78282$.  
\label{fig:pss1253_mtl}}
\end{center}
\end{figure}

\begin{figure}[ht]
\begin{center}
\includegraphics[height=6.1in, width=3.9in]{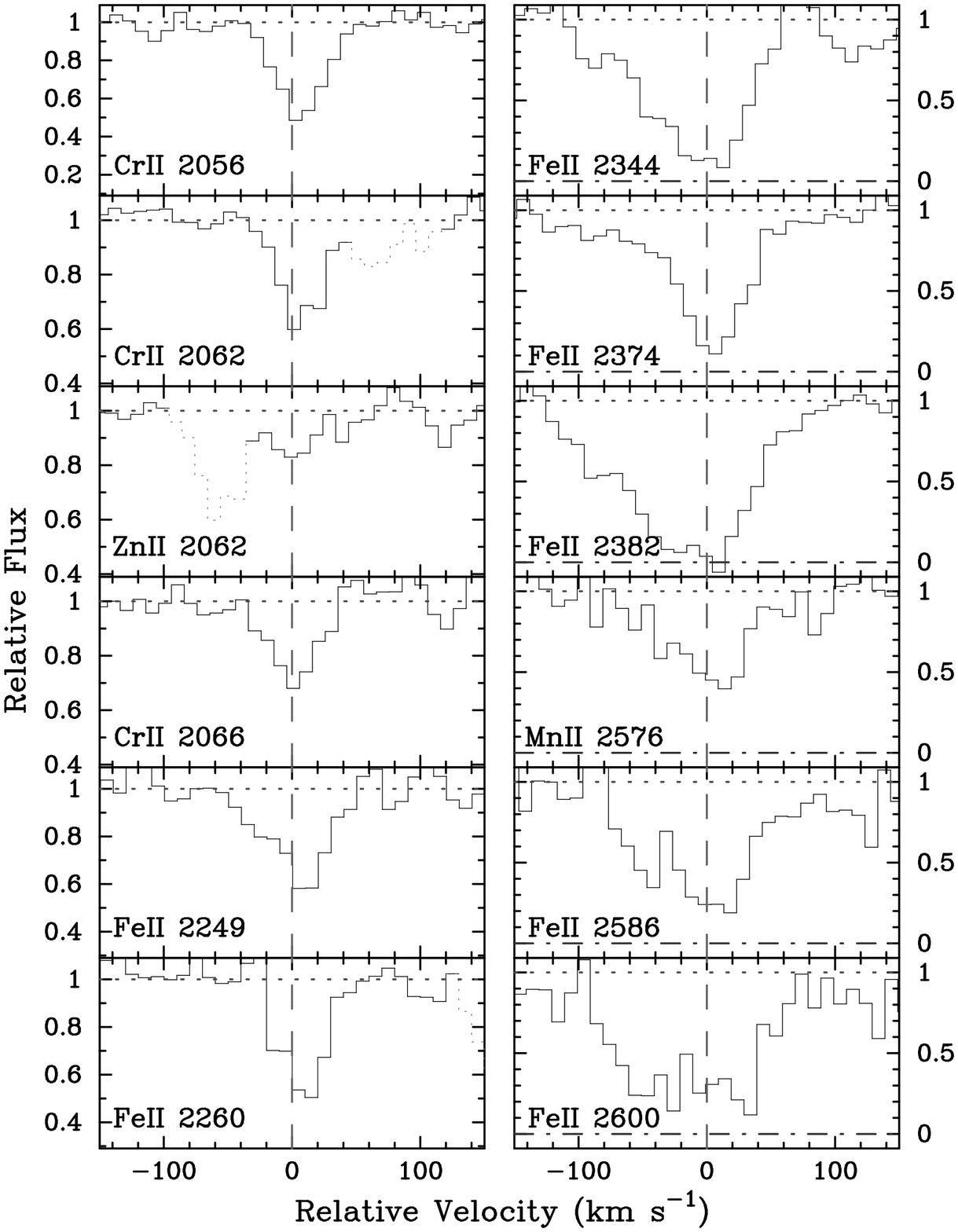}
Fig 46 -- cont
\end{center}
\end{figure}

\subsection{PSS1253-0228, $z = 2.783$ \label{subsec:PSS1253-0228_2.783}}

This quasar sightline is remarkable because it may exhibit the largest $\N{HI}$ 
value of any known high $z$ damped system.  Figure~\ref{fig:pss1253_lya} presents
the \lya profile and our best fit to the data assuming 
$\N{HI} = 10^{21.85} \cm{-2}$.  The fit is complicated by several factors:
(1) the coincidence with the CCD defect of ESI at $\lambda \approx 4500$\AA;
(2) somewhat poor S/N;
(3) a rather uncertain continuum; and
(4) blending with a sub-DLA at $z=3.603$ which is included in our
analysis.  Nevertheless, it is difficult to fit
the core of this profile with an H\,I value significantly less than 
what we have adopted.  Granted the special nature of this DLA, future observations
using an instrument with higher efficiency at $\lambda \approx 4400-4800$\AA\
are warranted.  We stress that \cite{peroux01} reported a significantly
lower value to $\N{HI}$ for this DLA.

Figure~\ref{fig:pss1253_mtl} presents the few metal-line transitions outside
the \lya forest.  Note that Si\,II 1808 is clearly saturated while
Zn\,II 2026 is blended with coincident absorption features.  We do estimate
the Zn$^+$ column density, however, from the unblended portion of the
Zn\,II 2062 profile (apparent as the dotted line in the Cr\,II 2062 profile).
Table~\ref{tab:PSS1253-0228_2.783} provides the column density measurements
for this DLA.

\clearpage

\begin{table}[ht]\footnotesize
\begin{center}
\caption{ {\sc
IONIC COLUMN DENSITIES: Q1337+11, $z = 2.795$ \label{tab:Q1337+11_2.795}}}
\begin{tabular}{lcccc}
\tableline
\tableline
Ion & $\lambda$ & AODM & $N_{\rm adopt}$ & [X/H] \\
\tableline
C  II &1334.5&$>14.483$&$>14.483$&$>-3.057$\\  
C  II*&1335.7&$<13.093$\\  
C  IV &1548.2&$13.476 \pm  0.050$\\  
C  IV &1550.8&$13.508 \pm  0.089$\\  
N  I  &1199.5&$13.751 \pm  0.060$&$13.751 \pm  0.060$&$-3.129 \pm  0.117$\\  
N  II &1084.0&$<13.550$\\  
O  I  &1302.2&$>14.866$&$>14.866$&$>-2.824$\\  
Al II &1670.8&$12.844 \pm  0.021$&$12.844 \pm  0.021$&$-2.596 \pm  0.102$\\  
Al III&1854.7&$12.546 \pm  0.069$\\  
Al III&1862.8&$<12.513$\\  
Si II &1260.4&$>13.568$&$14.718 \pm  0.116$&$-1.792 \pm  0.153$\\  
Si II &1304.4&$>14.274$\\  
Si II &1526.7&$>14.202$\\  
Si II &1808.0&$14.719 \pm  0.116$\\  
Si III&1206.5&$>13.488$\\  
Si IV &1393.8&$12.894 \pm  0.065$\\  
Si IV &1402.8&$<12.861$\\  
S  II &1259.5&$14.270 \pm  0.035$&$14.270 \pm  0.036$&$-1.880 \pm  0.106$\\  
Cr II &2056.3&$<12.900$&$<12.900$&$<-1.720$\\  
Mn II &2576.9&$<12.528$&$<12.528$&$<-1.952$\\  
Fe II &1608.5&$14.051 \pm  0.021$&$14.062 \pm  0.018$&$-2.388 \pm  0.102$\\  
Fe II &1611.2&$<14.767$\\  
Fe II &2260.8&$<14.487$\\  
Fe II &2344.2&$>13.897$\\  
Fe II &2374.5&$14.101 \pm  0.037$\\  
Fe II &2382.8&$>13.634$\\  
Fe II &2586.7&$13.979 \pm  0.052$\\  
Fe II &2600.2&$>13.575$\\  
Ni II &1370.1&$<13.199$&$<13.199$&$<-2.001$\\  
Ni II &1741.6&$<13.424$\\  
Zn II &2026.1&$<12.201$&$<12.201$&$<-1.419$\\  
\tableline
\end{tabular}
\end{center}
\end{table}

\begin{figure}[ht]
\begin{center}
\includegraphics[height=3.6in, width=2.8in,angle=90]{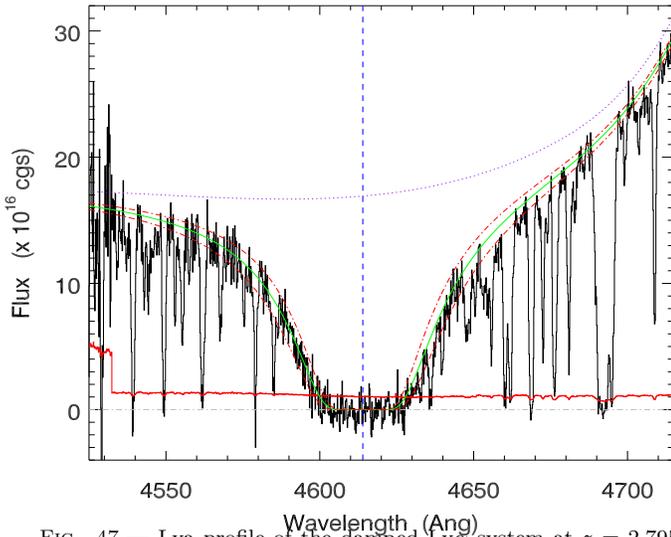}
\figcaption{Lya profile of the damped \lya system at $z=2.795$
toward Q1337+11.
The overplotted solid line and accompanying
dash-dot lines trace the best fit solution and the estimated 
bounds corresponding to 
$\log \N{HI} = 20.95^{+0.10}_{-0.10}$.  
The red wing continuum is complicated by
the \lya emission peak, but the fit is secure.
\label{fig:q1337_lya}}
\end{center}
\end{figure}

\begin{figure}[ht]
\begin{center}
\includegraphics[height=6.1in, width=3.9in]{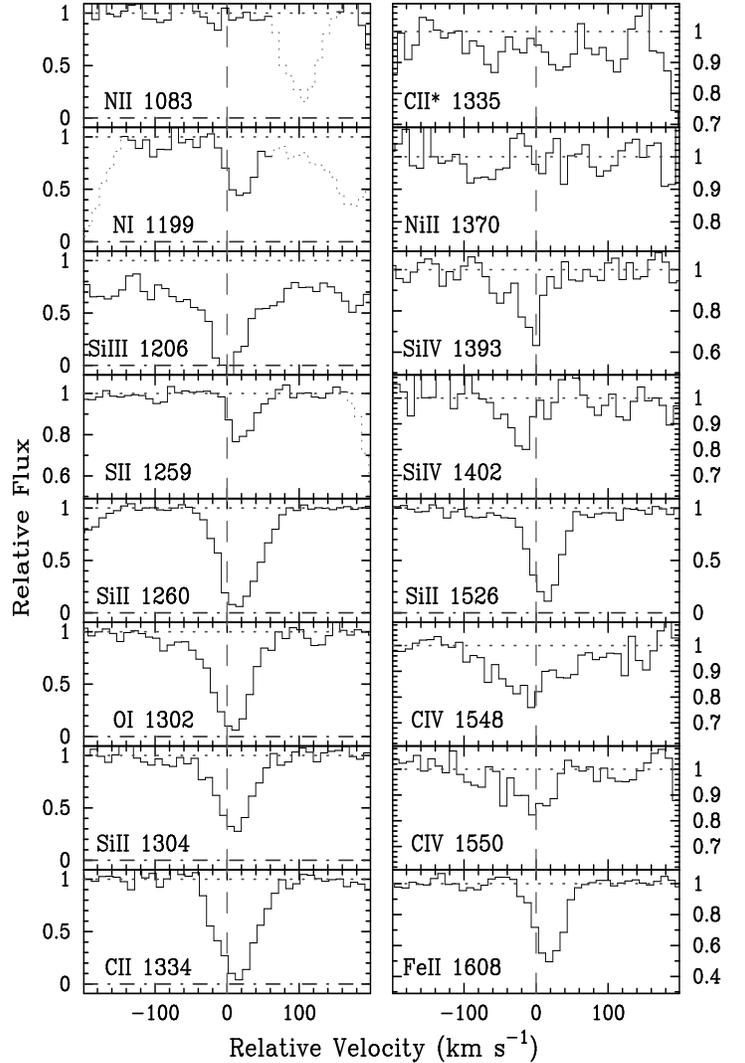}
\figcaption{Velocity plot of the metal-line transitions for the 
damped \lya system at $z = 2.795$ toward Q1337+11.
The vertical line at $v=0$ corresponds to $z = 2.79545$.  
\label{fig:q1337_mtl}}
\end{center}
\end{figure}

\begin{figure}[ht]
\begin{center}
\includegraphics[height=6.1in, width=3.9in]{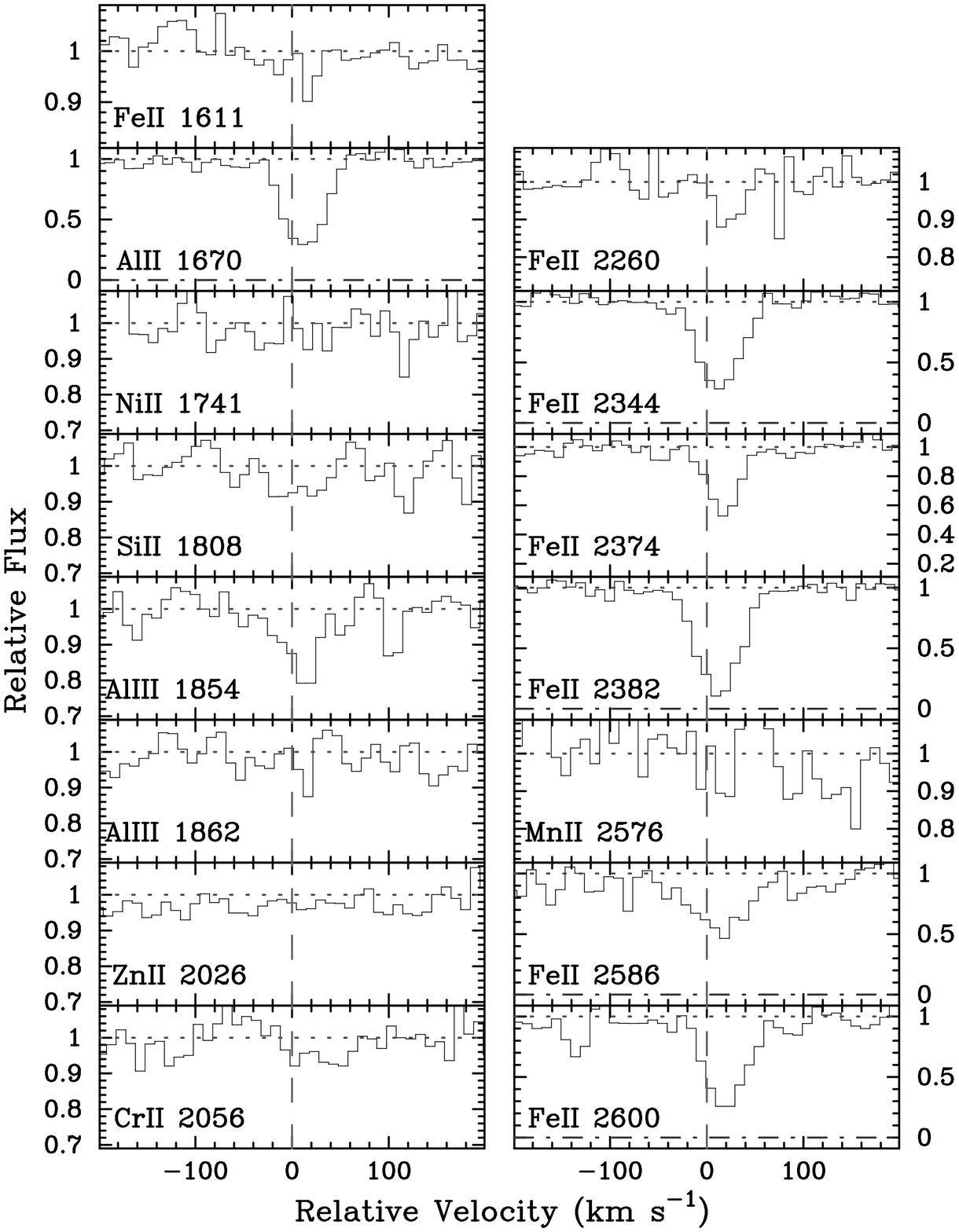}
Fig 48 -- cont
\end{center}
\end{figure}

\subsection{Q1337+11, $z = 2.795$ \label{subsec:Q1337+11_2.795}}

This LBQS quasar \citep{wolfe95} was discovered in the original surveys
of \cite{sargent89} and was chosen principally to fill the gap at
$z \approx 2.8$ of previous metallicity surveys.
The \lya profile of this DLA is shown in Figure~\ref{fig:q1337_lya} along
with the best fit solution.  Although the \lya emission peak complicates
the solution, the fit is reasonably secure.
The metal-line profiles are shown in Figure~\ref{fig:q1337_mtl}
and the ionic column densities are listed in Table~\ref{tab:Q1337+11_2.795}.
Note the Zn\,II 2026 profile is blended with atmospheric absorption and
can only be considered an upper limit to $\N{Zn^+}$.

\begin{table}[ht]\footnotesize
\begin{center}
\caption{ {\sc
IONIC COLUMN DENSITIES: PKS1354-17, $z = 2.780$ \label{tab:PKS1354-17_2.780}}}
\begin{tabular}{lcccc}
\tableline
\tableline
Ion & $\lambda$ & AODM & $N_{\rm adopt}$ & [X/H] \\
\tableline
C  II &1334.5&$>14.142$&$>14.142$&$>-2.748$\\  
C  II*&1335.7&$<12.833$\\  
C  IV &1548.2&$13.670 \pm  0.123$\\  
C  IV &1550.8&$<13.857$\\  
Al II &1670.8&$12.395 \pm  0.095$&$12.395 \pm  0.095$&$-2.395 \pm  0.178$\\  
Si II &1526.7&$13.976 \pm  0.056$&$13.976 \pm  0.056$&$-1.884 \pm  0.160$\\  
Si II &1808.0&$<14.870$\\  
Si IV &1402.8&$<13.347$\\  
Fe II &1608.5&$<13.881$&$13.374 \pm  0.080$&$-2.426 \pm  0.170$\\  
Fe II &2344.2&$13.374 \pm  0.080$\\  
Fe II &2374.5&$<13.761$\\  
\tableline
\end{tabular}
\end{center}
\end{table}
 
\begin{figure}[ht]
\begin{center}
\includegraphics[height=3.6in, width=2.8in,angle=90]{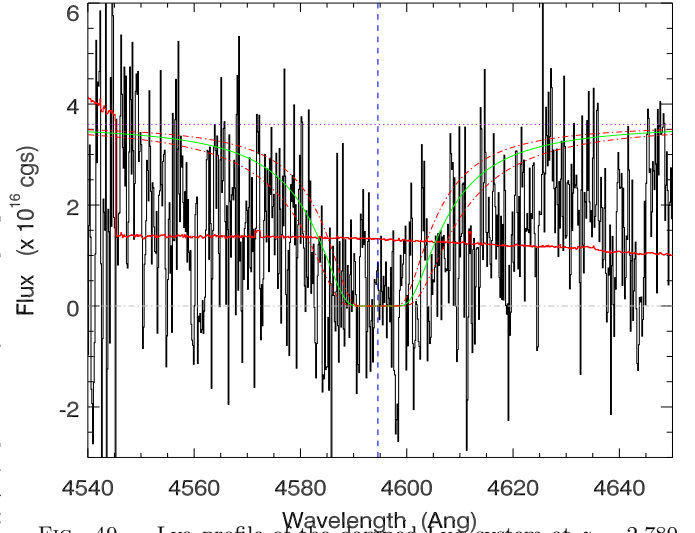}
\figcaption{Lya profile of the damped \lya system at $z=2.780$
toward PKS1354-17.
The overplotted solid line and accompanying
dash-dot lines trace the best fit solution and the estimated 
bounds corresponding to 
$\log \N{HI} = 20.30^{+0.15}_{-0.15}$.  
The signal-to-noise ratio is very poor in this portion of the
spectrum and we report an uncertainty of 0.15~dex in the fit.
\label{fig:q1357_lya}}
\end{center}
\end{figure}

\begin{figure}[ht]
\begin{center}
\includegraphics[height=6.1in, width=3.9in]{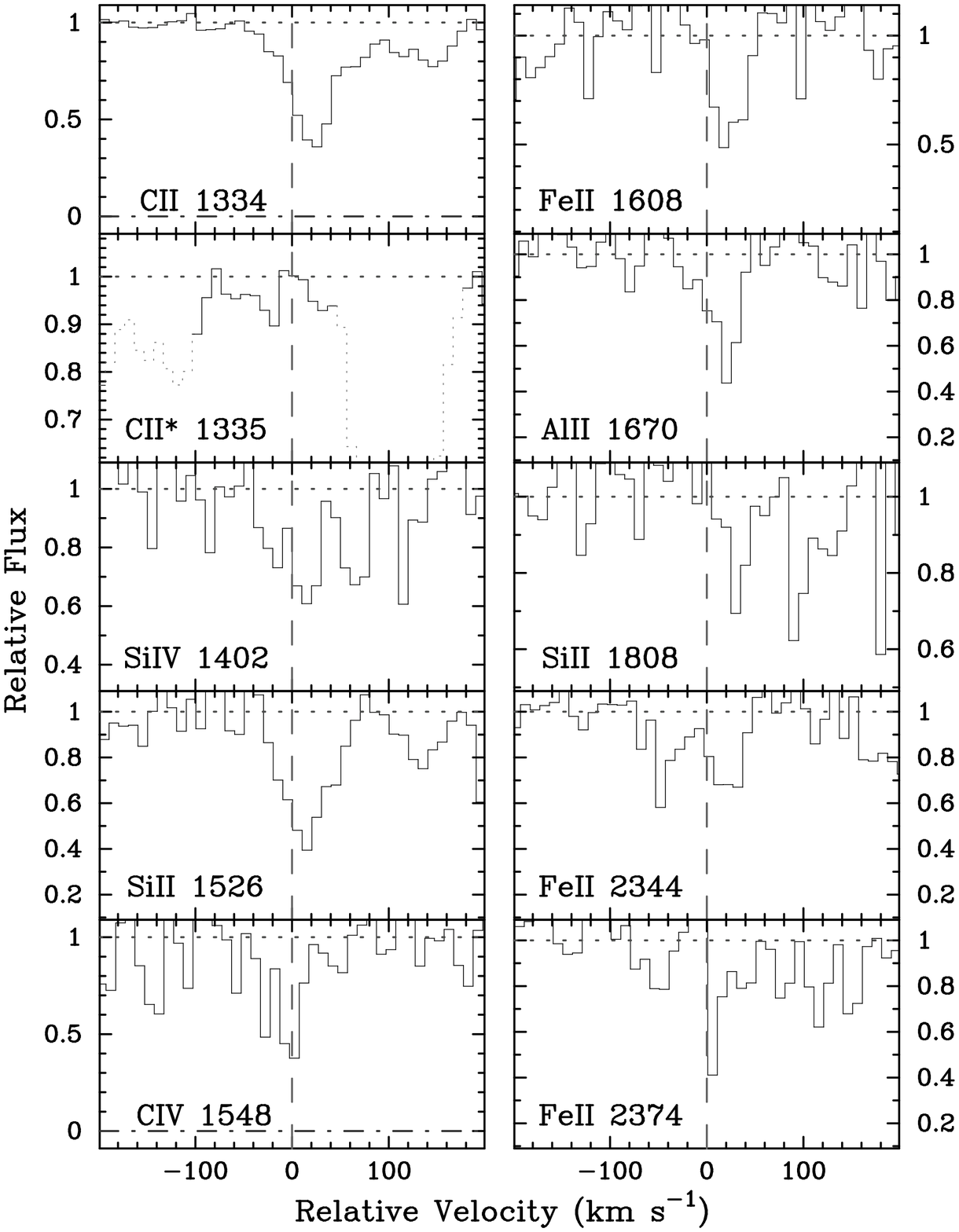}
\figcaption{Velocity plot of the metal-line transitions for the 
damped \lya system at $z = 2.780$ toward PKS1354-17.
The vertical line at $v=0$ corresponds to $z = 2.77993$.  
\label{fig:q1357_mtl}}
\end{center}
\end{figure}

\subsection{PKS1354-17, $z = 2.780$ \label{subsec:PKS1354-17_2.780}}

This radio-bright quasar was first discovered by \cite{schilizzi75}
and we discovered the damped \lya as part of back-up observing plan on the
Magellan telescope.
Figure~\ref{fig:q1357_lya} plots the \lya profile and solution for this
damped system.
Because of the very poor S/N, we report an uncertainty of 0.15~dex and caution
that this system may not meet the DLA criterion of $2 \sci{20} \cm{-2}$ 
when additional observations are made.  Figure~\ref{fig:q1357_mtl} plots
the few metal-line transitions observed and Table~\ref{tab:PKS1354-17_2.780}
provides the measurements of ionic column densities.

\begin{table}[ht]\footnotesize
\begin{center}
\caption{ {\sc
IONIC COLUMN DENSITIES: PSS1432+39, $z = 3.272$ \label{tab:PSS1432+39_3.272}}}
\begin{tabular}{lcccc}
\tableline
\tableline
Ion & $\lambda$ & AODM & $N_{\rm adopt}$ & [X/H] \\
\tableline
C  I  &1656.9&$<13.047$\\  
C  IV &1548.2&$13.751 \pm  0.030$\\  
C  IV &1550.8&$13.637 \pm  0.075$\\  
Al II &1670.8&$>13.749$&$>13.749$&$>-1.991$\\  
Al III&1854.7&$12.920 \pm  0.098$\\  
Si II &1526.7&$>15.062$&$15.666 \pm  0.048$&$-1.144 \pm  0.111$\\  
Si II &1808.0&$15.666 \pm  0.048$\\  
Cr II &2056.3&$<13.160$&$<13.160$&$<-1.760$\\  
Fe II &1608.5&$>14.930$&$>14.930$&$>-1.820$\\  
Fe II &1611.2&$<15.070$\\  
Fe II &2249.9&$<15.312$\\  
Ni II &1709.6&$<13.779$&$13.750 \pm  0.101$&$-1.750 \pm  0.142$\\  
Ni II &1741.6&$13.750 \pm  0.101$\\  
Ni II &1751.9&$<13.834$\\  
Zn II &2026.1&$<12.645$&$<12.645$&$<-1.275$\\  
\tableline
\end{tabular}
\end{center}
\end{table}

\begin{figure}[ht]
\begin{center}
\includegraphics[height=3.6in, width=2.8in,angle=90]{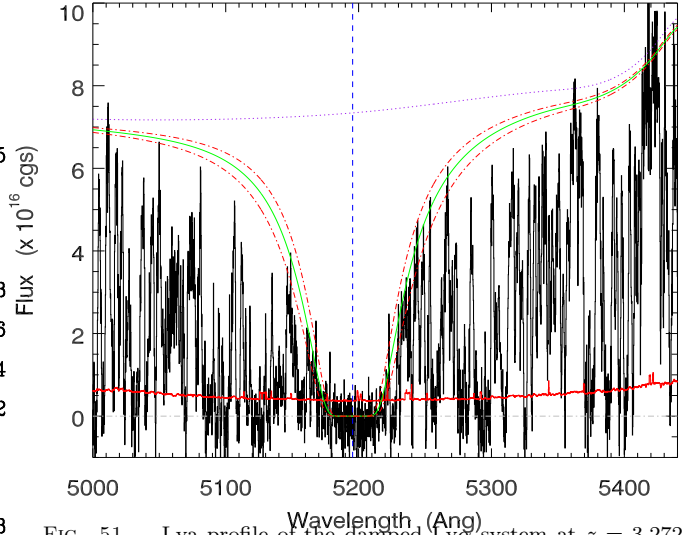}
\figcaption{Lya profile of the damped \lya system at $z=3.272$
toward PSS1432+39.
The overplotted solid line and accompanying
dash-dot lines trace the best fit solution and the estimated 
bounds corresponding to 
$\log \N{HI} = 21.25^{+0.10}_{-0.10}$.  
Note this value is a factor of 0.3~dex higher than the value listed
in PGW01.  
This is primarily due to our decision to fit fluxed
data, i.e.\, including the continuum in the solution when fitting the
profile as described in $\S$~\ref{sec:methods}.
\label{fig:pss1432_lya}}
\end{center}
\end{figure}

\begin{figure}[ht]
\begin{center}
\includegraphics[height=6.1in, width=3.9in]{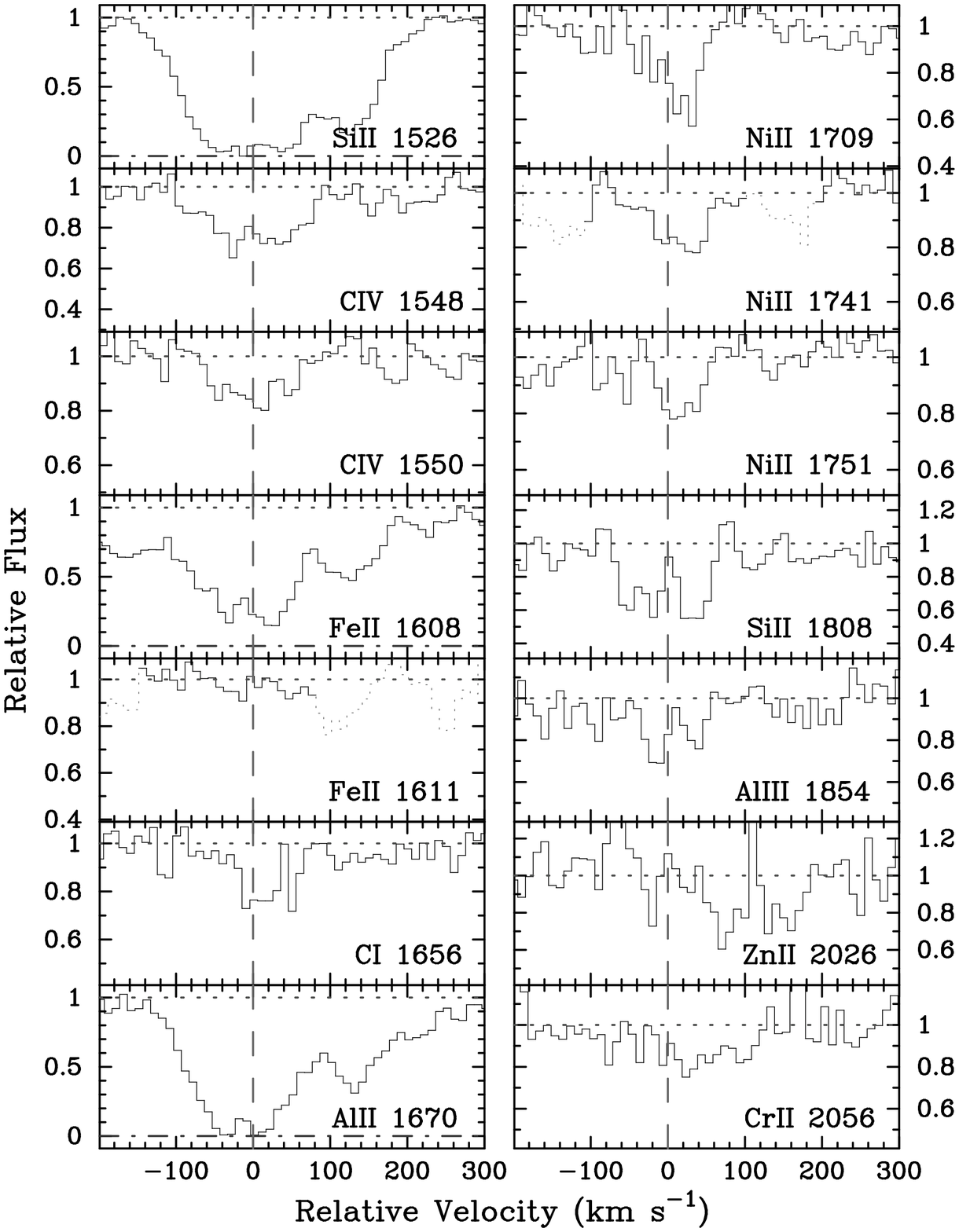}
\figcaption{Velocity plot of the metal-line transitions for the 
damped \lya system at $z = 3.272$ toward PSS1432+39.
The vertical line at $v=0$ corresponds to $z = 3.2720$.  
\label{fig:pss1432_mtl}}
\end{center}
\end{figure}

\subsection{PSS1432+39, $z = 3.272$ \label{subsec:PSS1432+39_3.272}}

We originally discovered and analysed this damped \lya system in PGW01.
Similar to other DLA from the PGW01 sample, we find a somewhat higher
H\,I value in this analysis (Figure~\ref{fig:pss1432_lya}).  Again, this
is due to the improved approach we have taken to measuring H\,I column
densities in the present work.  The metal-line profiles for this DLA
are given in Figure~\ref{fig:pss1432_mtl} and the ionic column densities
are listed in Table~\ref{tab:PSS1432+39_3.272}.  For this DLA, the Ni\,II
1709, 1741, and 1751 profiles are all marginally detected and yield
a Ni$^+$ column density of $10^{13.75} \cm{-2}$ when treated together.

\begin{table}[ht]\footnotesize
\begin{center}
\caption{ {\sc
IONIC COLUMN DENSITIES: Q1502+4837, $z = 2.570$ \label{tab:Q1502+4837_2.570}}}
\begin{tabular}{lcccc}
\tableline
\tableline
Ion & $\lambda$ & AODM & $N_{\rm adopt}$ & [X/H] \\
\tableline
C  II &1334.5&$>15.174$&$>15.174$&$>-1.716$\\  
C  IV &1548.2&$14.240 \pm  0.034$\\  
C  IV &1550.8&$14.149 \pm  0.067$\\  
O  I  &1302.2&$>15.341$&$>15.341$&$>-1.699$\\  
Mg II &2796.4&$>13.961$&$>13.961$&$>-1.919$\\  
Al II &1670.8&$>13.361$&$>13.362$&$>-1.428$\\  
Al III&1854.7&$12.887 \pm  0.144$\\  
Si II &1526.7&$14.241 \pm  0.085$&$14.241 \pm  0.085$&$-1.619 \pm  0.172$\\  
Si IV &1393.8&$>14.434$\\  
Fe II &1608.5&$14.151 \pm  0.121$&$14.151 \pm  0.121$&$-1.649 \pm  0.193$\\  
Fe II &2344.2&$14.103 \pm  0.064$\\  
Fe II &2374.5&$<14.011$\\  
Fe II &2600.2&$>14.081$\\  
\tableline
\end{tabular}
\end{center}
\end{table}

\begin{figure}[ht]
\begin{center}
\includegraphics[height=3.6in, width=2.8in,angle=90]{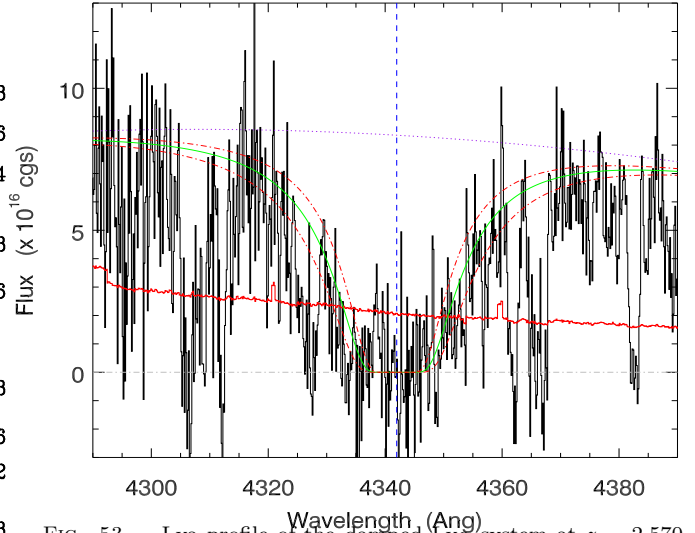}
\figcaption{Lya profile of the damped \lya system at $z=2.570$
toward Q1502+4837.
The overplotted solid line and accompanying
dash-dot lines trace the best fit solution and the estimated 
bounds corresponding to 
$\log \N{HI} = 20.30^{+0.15}_{-0.15}$.  
Although the profile is relatively clean, we report an uncertainty
of 0.15~dex because of the poor S/N.
\label{fig:q1502_lya}}
\end{center}
\end{figure}

\begin{figure}[ht]
\begin{center}
\includegraphics[height=6.1in, width=3.9in]{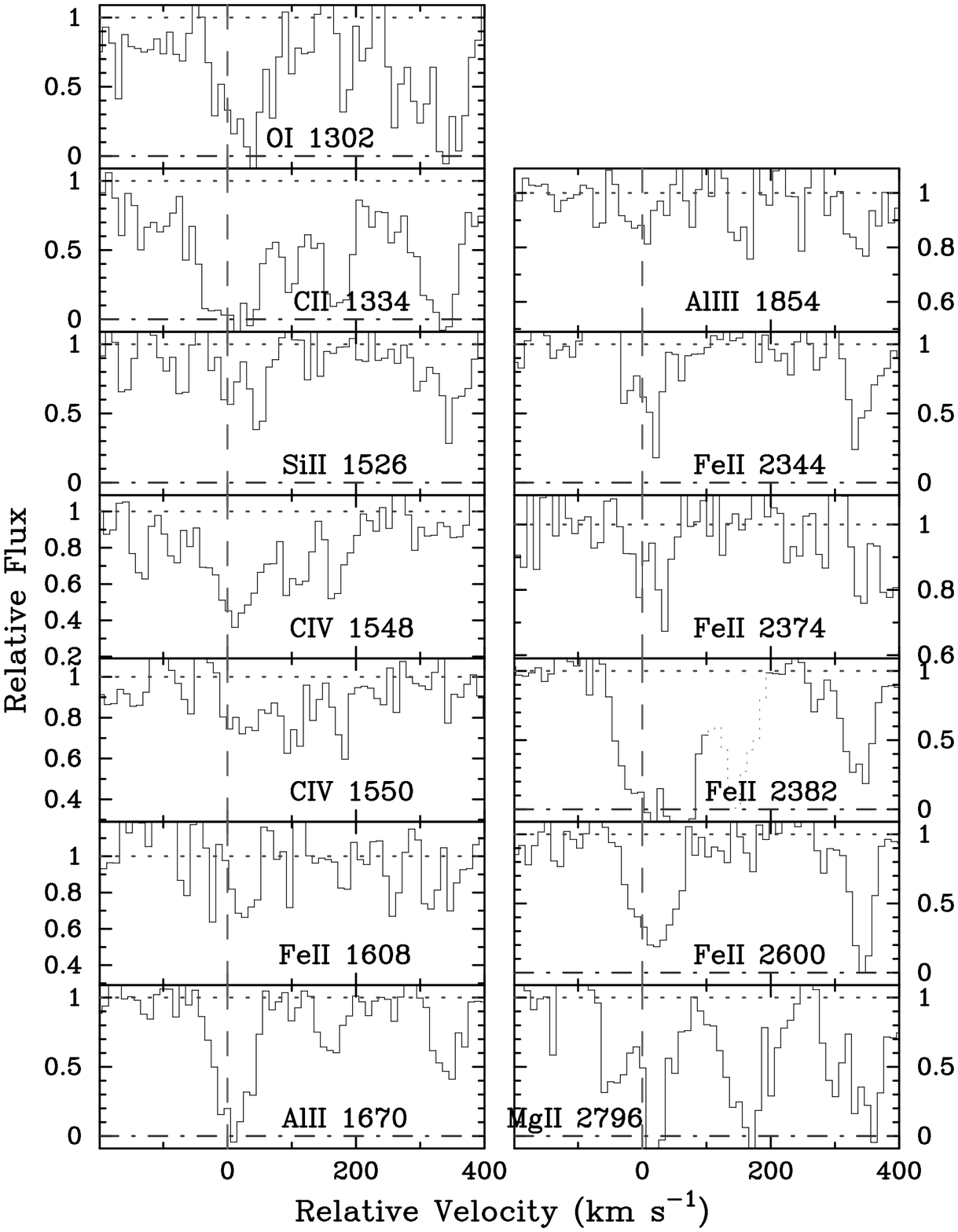}
\figcaption{Velocity plot of the metal-line transitions for the 
damped \lya system at $z = 2.570$ toward Q1502+4837.
The vertical line at $v=0$ corresponds to $z = 2.56956$.  
\label{fig:q1502_mtl}}
\end{center}
\end{figure}

\subsection{Q1502+4837, $z = 2.570$ \label{subsec:Q1502+4837_2.570}}

This quasar is a member of the Cosmic Lens All Sky Survey (CLASS)
\citep{snellen01} and the DLA at $z=2.57$ was identified in their
discovery spectrum.
The fit to the \lya profile (Figure~\ref{fig:q1502_lya}) is complicated
by the poor efficiency of ESI at $\lambda < 4400$\AA, but the fit is 
reasonable.  We note, however, that future observations may favor a 
central value below the DLA criterion.
Figure~\ref{fig:q1502_mtl} provides the metal-line transitions and
Table~\ref{tab:Q1502+4837_2.570} gives the ionic column densities.
The S/N of this spectrum is poor and only a few profiles provide
meaningful measurements.  We caution using this DLA in ones analysis.
Finally, we note in passing that the velocity profiles span an impressive 
400~km/s.

\begin{table}[ht]\footnotesize
\begin{center}
\caption{ {\sc
IONIC COLUMN DENSITIES: PSS1506+5220, $z = 3.224$ \label{tab:PSS1506+5220_3.224}}}
\begin{tabular}{lcccc}
\tableline
\tableline
Ion & $\lambda$ & AODM & $N_{\rm adopt}$ & [X/H] \\
\tableline
C  I  &1656.9&$<12.787$\\  
C  II &1334.5&$>15.210$&$>15.210$&$>-2.050$\\  
C  II*&1335.7&$<12.890$\\  
C  IV &1548.2&$13.581 \pm  0.033$\\  
C  IV &1550.8&$13.746 \pm  0.048$\\  
O  I  &1302.2&$>14.656$&$>14.657$&$>-2.753$\\  
Al II &1670.8&$>12.504$&$>12.503$&$>-2.657$\\  
Al III&1854.7&$<12.080$\\  
Si II &1260.4&$>13.536$&$13.882 \pm  0.018$&$-2.348 \pm  0.072$\\  
Si II &1526.7&$13.882 \pm  0.018$\\  
Si III&1206.5&$>13.379$\\  
Cr II &2056.3&$<12.784$&$<12.784$&$<-1.556$\\  
Fe II &1608.5&$13.896 \pm  0.036$&$13.708 \pm  0.034$&$-2.462 \pm  0.078$\\  
Fe II &2344.2&$13.546 \pm  0.062$\\  
Ni II &1709.6&$<13.413$&$<13.413$&$<-1.507$\\  
Zn II &2026.1&$<12.105$&$<12.105$&$<-1.235$\\  
\tableline
\end{tabular}
\end{center}
\end{table}

\begin{figure}[ht]
\begin{center}
\includegraphics[height=3.6in, width=2.8in,angle=90]{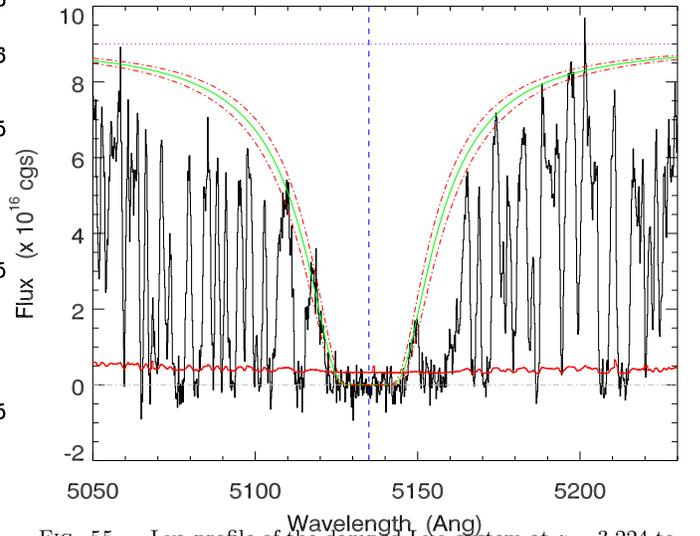}
\figcaption{Lya profile of the damped \lya system at $z=3.224$
toward PSS1506+5220.
The overplotted solid line and accompanying
dash-dot lines trace the best fit solution and the estimated 
bounds corresponding to 
$\log \N{HI} = 20.67^{+0.07}_{-0.07}$.  
The \lya profile has a poorly determined blue wing but 
sufficient sampling in its core and red wing to provide a
rather precise fit.
\label{fig:pss1506_lya}}
\end{center}
\end{figure}

\begin{figure}[ht]
\begin{center}
\includegraphics[height=6.1in, width=3.9in]{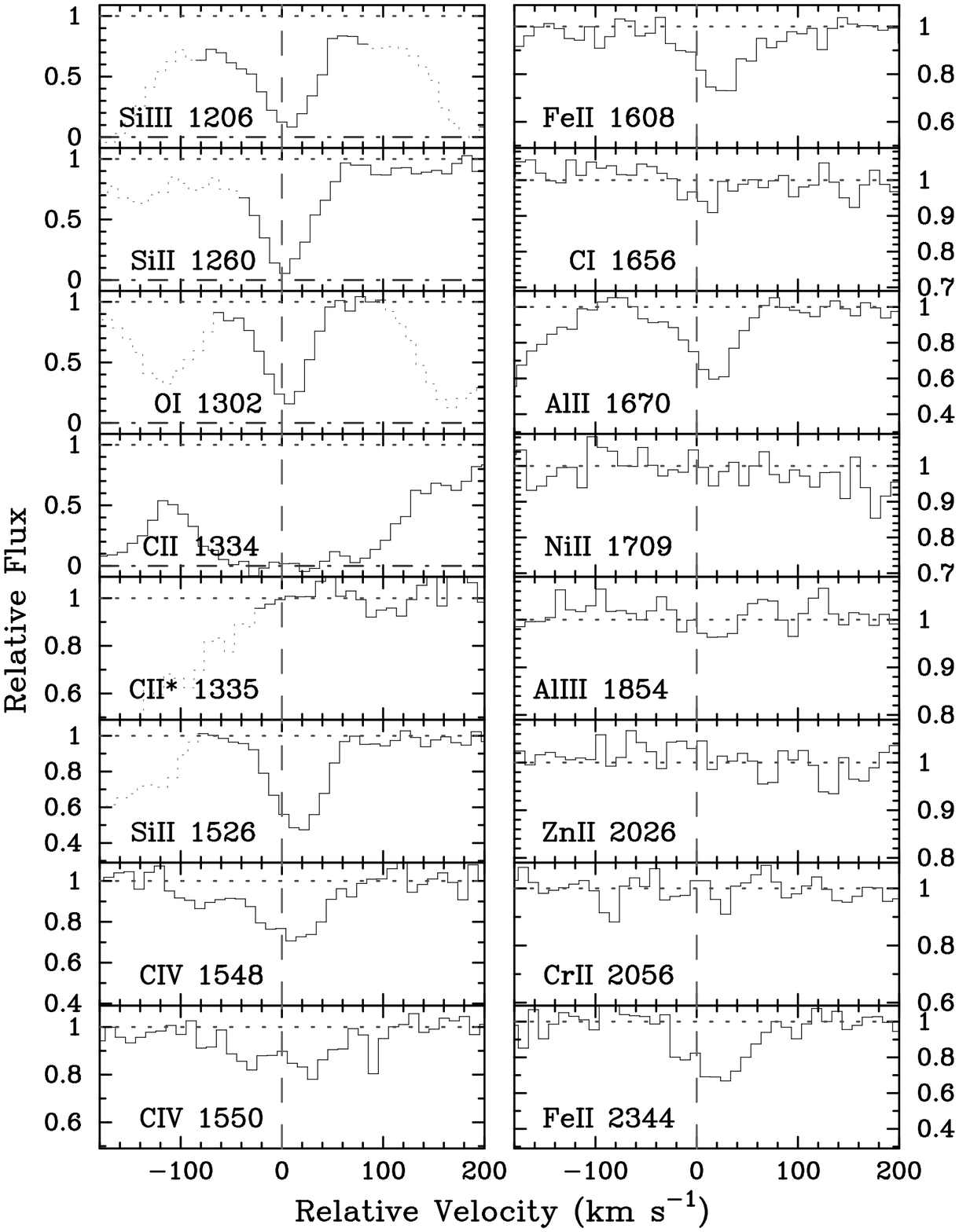}
\figcaption{Velocity plot of the metal-line transitions for the 
damped \lya system at $z = 3.224$ toward PSS1506+5220.
The vertical line at $v=0$ corresponds to $z = 3.22442$.  
\label{fig:pss1506_mtl}}
\end{center}
\end{figure}

\subsection{PSS1506+5220, $z = 3.224$ \label{subsec:PSS1506+5220_3.224}}

This PSS quasar exhibits a damped system at $z=3.224$ with a reasonably
simple \lya profile (Figure~\ref{fig:pss1506_lya}).
Because the continuum is also well behaved, we place an uncertainty on the
$\N{HI}$ value of only 0.07~dex.
Figure~\ref{fig:pss1506_mtl} plots the velocity profiles of this low 
metallicity, simple DLA and the ionic column densities are given in 
Table~\ref{tab:PSS1506+5220_3.224}.

\begin{table}[ht]\footnotesize
\begin{center}
\caption{ {\sc
IONIC COLUMN DENSITIES: PSS1723+2243, $z = 3.695$ \label{tab:PSS1723+2243_3.695}}}
\begin{tabular}{lcccc}
\tableline
\tableline
Ion & $\lambda$ & AODM & $N_{\rm adopt}$ & [X/H] \\
\tableline
C  IV &1548.2&$>14.475$\\  
C  IV &1550.8&$14.761 \pm  0.007$\\  
Al II &1670.8&$>13.564$&$>13.564$&$>-1.426$\\  
Al III&1862.8&$<12.603$\\  
Si II &1526.7&$>14.861$&$>14.861$&$>-1.199$\\  
Si II &1808.0&$<15.497$\\  
Cr II &2056.3&$<13.214$&$<13.214$&$<-0.956$\\  
Fe II &1608.5&$>14.575$&$>14.575$&$>-1.425$\\  
Ni II &1741.6&$<13.950$&$<13.950$&$<-0.800$\\  
Zn II &2026.1&$>12.508$&$>12.508$&$>-0.662$\\  
\tableline
\end{tabular}
\end{center}
\end{table}

\begin{figure}[ht]
\begin{center}
\includegraphics[height=3.6in, width=2.8in,angle=90]{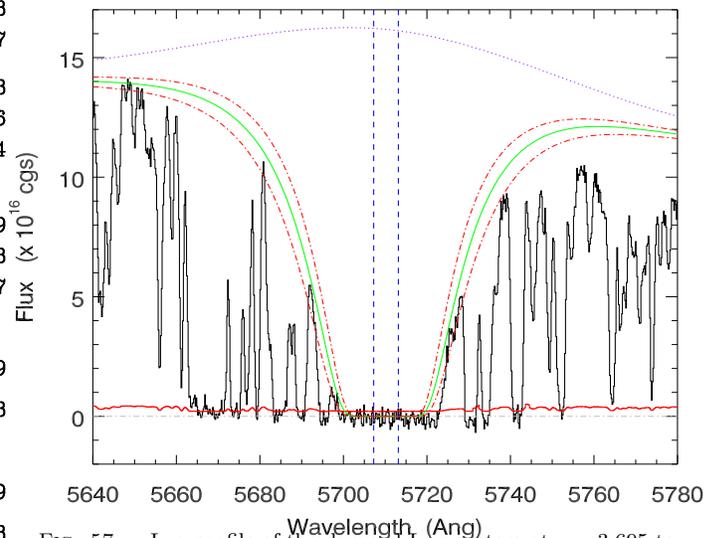}
\figcaption{Lya profile of the damped \lya system at $z=3.695$
toward PSS1723+2243.
The overplotted solid line and accompanying
dash-dot lines trace the best fit solution and the estimated 
bounds corresponding to 
$\log \N{HI} = 20.50^{+0.15}_{-0.15}$.  
The continuum is difficult to predict for the region surrounding the DLA
because it includes the O\,VI emission peak of the quasar.  Furthermore,
the profile is best fit by two components.  The individual $\N{HI}$ values
are not well determined but we feel the total column density is accurate
to 0.15~dex.
\label{fig:pss1723_lya}}
\end{center}
\end{figure}

\begin{figure}[ht]
\begin{center}
\includegraphics[height=6.1in, width=3.9in]{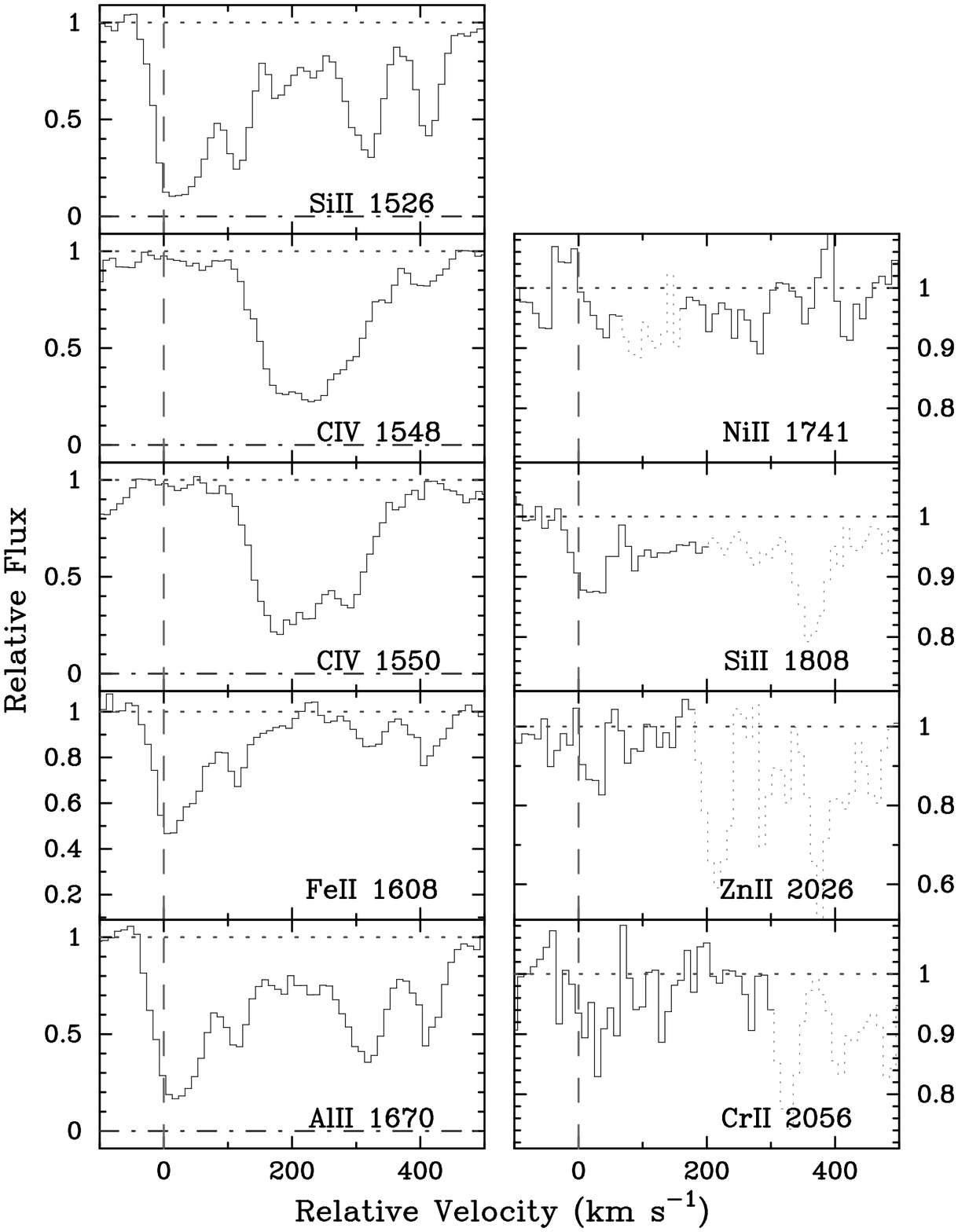}
\figcaption{Velocity plot of the metal-line transitions for the 
damped \lya system at $z = 3.695$ toward PSS1723+2243.
The vertical line at $v=0$ corresponds to $z = 3.69474$.  
\label{fig:pss1723_mtl}}
\end{center}
\end{figure}

\subsection{PSS1723+2243, $z = 3.695$ \label{subsec:PSS1723+2243_3.695}}

The damped \lya system toward this Palomar Sky Survey quasar is remarkable
for exhibiting a particularly large metallicity at $z>3.5$ as well as metal-line
profiles which span over 400 km/s.   The H\,I column density is best fit
with two components separated by $\approx 300 \mkms$ as shown in 
Figure~\ref{fig:pss1723_lya}.   The $\N{HI}$ values for the individual
components is not very well constrained and if one divides the total
H\,I column density evenly, then neither absorber satisfies the DLA
criterion of $10^{20.3} \cm{-2}$.  In this case, one might refer to this
damped system as a pair of sub-DLA.  To further complicate matters, if
the \lya profile were observed at lower resolution, one may have been
likely compose a fit with a single component with 
$\N{HI} \approx 10^{20.5} \cm{-2}$.  
Unfortunately, a higher redshift Lyman limit system precludes an 
analysis of the higher order Lyman series for the system at $z=3.695$.
For now, we fit the profile with two 
components and report the total $\N{HI}$ value.   Similarly, we have
chosen to integrate the metal-line profiles over the full velocity range.
Any future studies which focus on the relative abundances of this system
must take caution in interpreting the metallicity of the gas.

The metal-line profiles for this DLA are shown in Figure~\ref{fig:pss1723_mtl}
which includes only a few species because of bad line blending both
within and outside the \lya forest.  The ionic column densities are 
provided in Table~\ref{tab:PSS1723+2243_3.695}.  We have cautiously listed
most of the values as lower limits because our experience with ESI suggests
several of the line profiles are saturated.

\begin{table}[ht]\footnotesize
\begin{center}
\caption{ {\sc
IONIC COLUMN DENSITIES: PSS2155+1358, $z = 3.316$ \label{tab:PSS2155+1358_3.316}}}
\begin{tabular}{lcccc}
\tableline
\tableline
Ion & $\lambda$ & AODM & $N_{\rm adopt}$ & [X/H] \\
\tableline
C  IV &1548.2&$>14.221$\\  
C  IV &1550.8&$14.209 \pm  0.018$\\  
Al II &1670.8&$>13.316$&$>13.316$&$>-1.724$\\  
Si II &1526.7&$>14.649$&$14.847 \pm  0.072$&$-1.263 \pm  0.166$\\  
Si II &1808.0&$14.848 \pm  0.072$\\  
Cr II &2056.3&$<13.150$&$<13.151$&$<-1.069$\\  
Fe II &1608.5&$>14.373$&$>14.373$&$>-1.677$\\  
Fe II &2249.9&$<15.124$\\  
Fe II &2344.2&$>14.284$\\  
Ni II &1709.6&$<13.437$&$<13.287$&$<-1.513$\\  
Ni II &1741.6&$<13.287$\\  
Ni II &1751.9&$<13.494$\\  
Zn II &2026.1&$<12.053$&$<12.053$&$<-1.167$\\  
\tableline
\end{tabular}
\end{center}
\end{table}

\begin{figure}[ht]
\begin{center}
\includegraphics[height=3.6in, width=2.8in,angle=90]{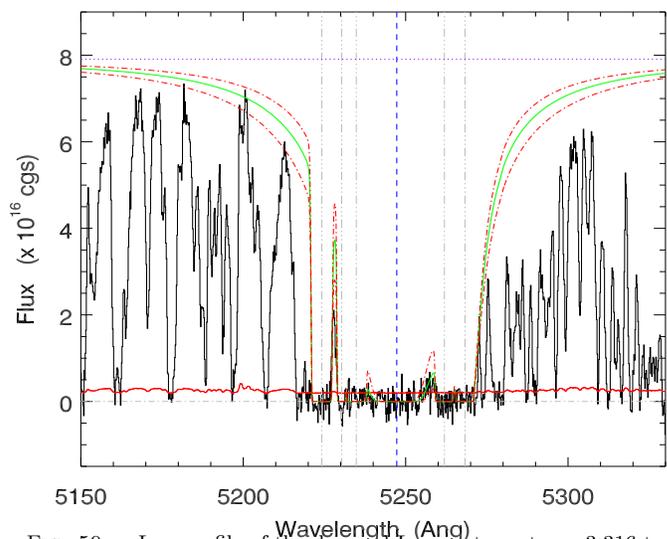}
\figcaption{Lya profile of the damped \lya system at $z=3.316$
toward PSS2155+1358.
The overplotted solid line and accompanying
dash-dot lines trace the best fit solution and the estimated 
bounds corresponding to 
$\log \N{HI} = 20.55^{+0.15}_{-0.15}$.  
One notes the presence of several coincident \lya clouds which 
absorb the core and red wing of the damped \lya profile.  In this case,
we have fit the trace of flux at $\lambda \approx 5258$\AA\ and
the few regions of the blue wing which appear uncontaminated.
The fit is further complicated by the uncertainty of the continuum
and we estimate an error of 0.15~dex to our $\N{HI}$ measurement.
\label{fig:pss2155_lya}}
\end{center}
\end{figure}

\clearpage

\begin{figure}[ht]
\begin{center}
\includegraphics[height=6.1in, width=3.9in]{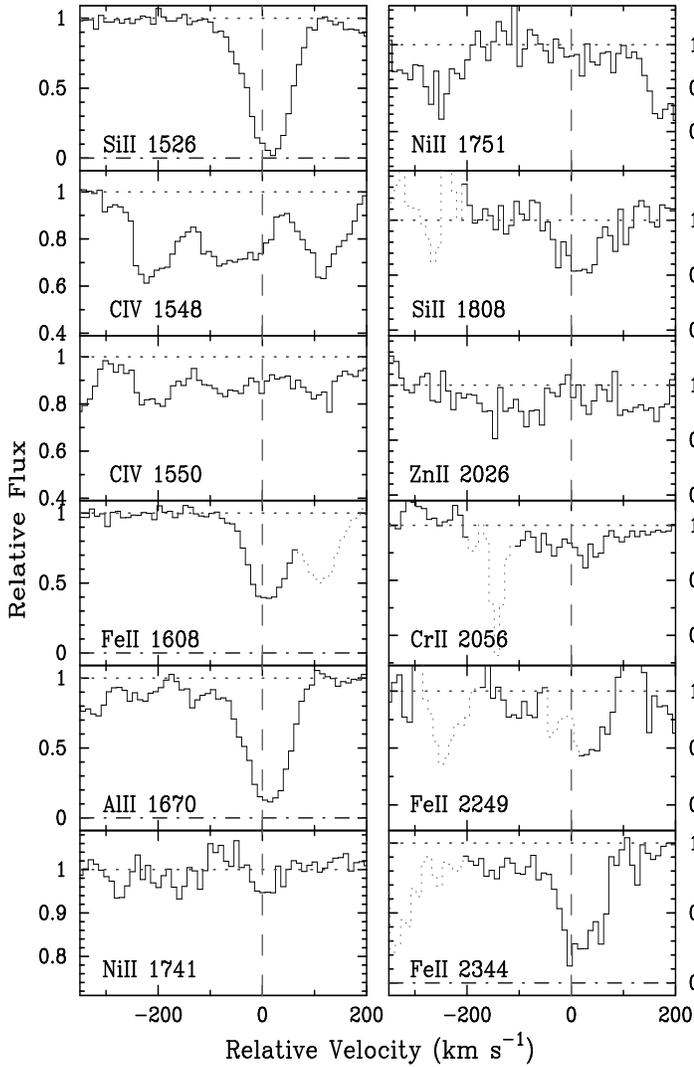}
\figcaption{Velocity plot of the metal-line transitions for the 
damped \lya system at $z = 3.316$ toward PSS2155+1358.
The vertical line at $v=0$ corresponds to $z = 3.31625$.  
\label{fig:pss2155_mtl}}
\end{center}
\end{figure}

\subsection{PSS2155+1358, $z = 3.316$ \label{subsec:PSS2155+1358_3.316}}

This PSS quasar shows a damped \lya system with a complicated \lya profile
(Figure~\ref{fig:pss2155_lya}).  We note that the quasar was observed 
in poor conditions and has poor S/N for its total exposure time.
There is severe blending in the core and
red wing of the profile from coincident \lya clouds which limits the 
precision of the $\N{HI}$ measurement.  Our adopted value 
$\N{HI} = 10^{20.55} \cm{-2}$ is significantly lower than the value 
reported by \cite{peroux01}.  We suspect the difference can be attributed
to their inability to identify the coincident \lya features in their spectra.
Figure~\ref{fig:pss2155_mtl} shows the few
metal-line profiles observed for this DLA and Table~\ref{tab:PSS2155+1358_3.316}
lists the column density measurements.

\begin{table}[ht]\footnotesize
\begin{center}
\caption{ {\sc
IONIC COLUMN DENSITIES: Q2223+20, $z = 3.119$ \label{tab:Q2223+20_3.119}}}
\begin{tabular}{lcccc}
\tableline
\tableline
Ion & $\lambda$ & AODM & $N_{\rm adopt}$ & [X/H] \\
\tableline
C  IV &1548.2&$13.572 \pm  0.024$\\  
C  IV &1550.8&$13.678 \pm  0.034$\\  
N  I  &1134.4&$<13.619$&$<13.619$&$<-2.611$\\  
O  I  &1302.2&$>14.842$&$>14.842$&$>-2.198$\\  
Al II &1670.8&$<12.329$&$<12.329$&$<-2.461$\\  
Si II &1260.4&$>13.787$&$13.640 \pm  0.036$&$-2.220 \pm  0.106$\\  
Si II &1526.7&$13.640 \pm  0.036$\\  
Si IV &1393.8&$13.236 \pm  0.019$\\  
Si IV &1402.8&$13.124 \pm  0.043$\\  
Fe II &1608.5&$13.376 \pm  0.090$&$13.324 \pm  0.061$&$-2.476 \pm  0.117$\\  
Fe II &2344.2&$13.292 \pm  0.081$\\  
Fe II &2382.8&$>13.438$\\  
Fe III&1122.5&$<13.600$\\  
\tableline
\end{tabular}
\end{center}
\end{table}

\begin{figure}[ht]
\begin{center}
\includegraphics[height=3.6in, width=2.8in,angle=90]{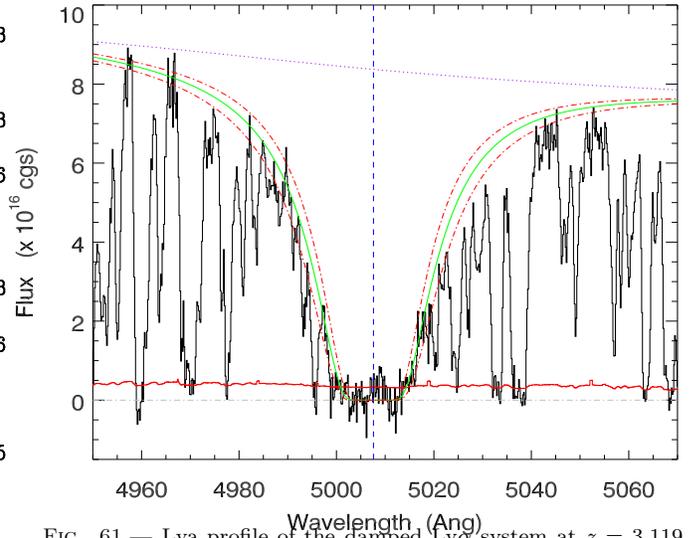}
\figcaption{Lya profile of the damped \lya system at $z=3.119$
toward Q2223+20.
The overplotted solid line and accompanying
dash-dot lines trace the best fit solution and the estimated 
bounds corresponding to 
$\log \N{HI} = 20.30^{+0.10}_{-0.10}$.  
This fit was straightforward except for the determination of the
quasar continuum.  We note that this DLA only just meets the DLA
criterion of $2 \sci{20} \cm{-2}$ and warn that future observations
may remove it from the statistical sample.
\label{fig:q2223_lya}}
\end{center}
\end{figure}

\begin{figure}[ht]
\begin{center}
\includegraphics[height=6.1in, width=3.9in]{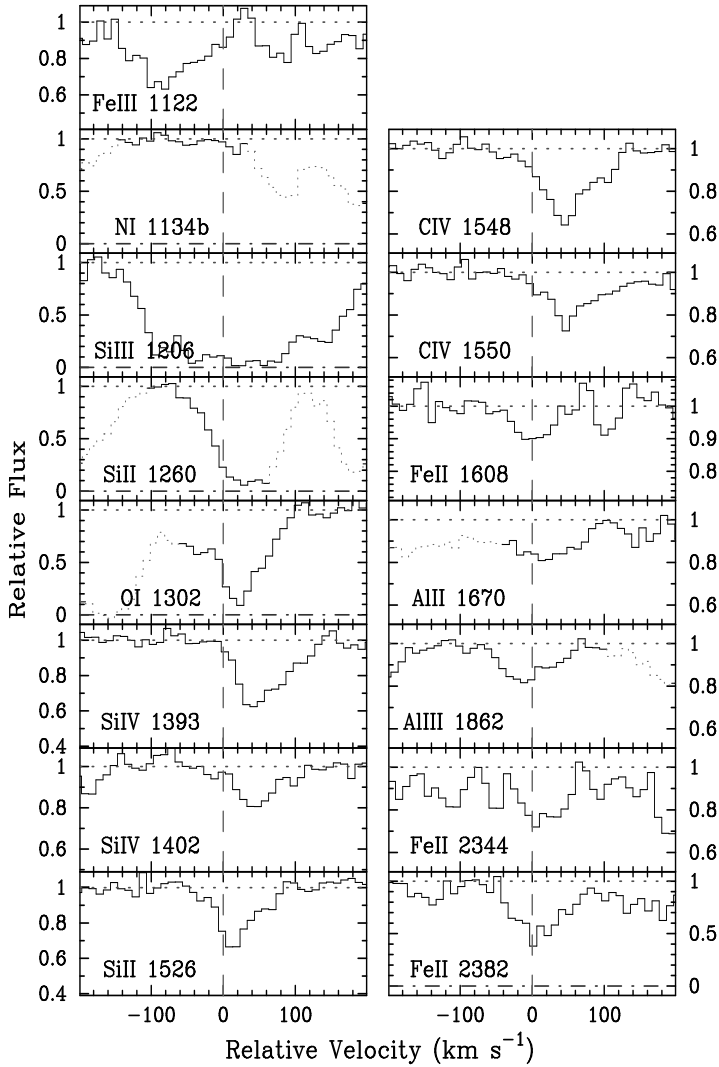}
\figcaption{Velocity plot of the metal-line transitions for the 
damped \lya system at $z = 3.119$ toward Q2223+20.
The vertical line at $v=0$ corresponds to $z = 3.11921$.  
\label{fig:q2223_mtl}}
\end{center}
\end{figure}

\subsection{Q2223+20, $z = 3.119$ \label{subsec:Q2223+20_3.119}}

This radio-loud quasar was discovered as part of the Third MIT-Green
Bank 5GHz Survey \citep{griffith90}.  Although \cite{storrie00} did
not identify the $z=3.119$ absorption system as a DLA, our best fit
(Figure~\ref{fig:q2223_lya})
just exceeds the DLA criterion of statistical samples. 
Its metal-line profiles are shown in Figure~\ref{fig:q2223_mtl} and
Table~\ref{tab:Q2223+20_3.119} lists the ionic column densities.

\begin{table}[ht]\footnotesize
\begin{center}
\caption{ {\sc
IONIC COLUMN DENSITIES: PSS2241+1352, $z = 4.282$ \label{tab:PSS2241+1352_4.282}}}
\begin{tabular}{lcccc}
\tableline
\tableline
Ion & $\lambda$ & AODM & $N_{\rm adopt}$ & [X/H] \\
\tableline
C  II &1334.5&$>15.015$&$>15.015$&$>-2.725$\\  
C  II*&1335.7&$13.467 \pm  0.086$\\  
C  IV &1550.8&$14.055 \pm  0.050$\\  
N  I  &1134.4&$<14.651$&$>14.808$&$>-2.272$\\  
N  I  &1200.2&$>14.809$\\  
Al II &1670.8&$>13.520$&$>13.519$&$>-2.121$\\  
Si II &1260.4&$>14.162$&$>15.011$&$>-1.699$\\  
Si II &1304.4&$>15.011$\\  
Si II &1526.7&$>14.727$\\  
Si II &1808.0&$<15.103$\\  
Si IV &1393.8&$13.376 \pm  0.054$\\  
S  II &1259.5&$14.575 \pm  0.033$&$14.575 \pm  0.033$&$-1.775 \pm  0.105$\\  
Fe II &1608.5&$>14.646$&$>14.646$&$>-2.004$\\  
Fe II &1611.2&$<14.864$\\  
Ni II &1317.2&$<13.223$&$<13.223$&$<-2.177$\\  
Ni II &1709.6&$<13.588$\\  
Ni II &1741.5&$<13.507$\\  
\tableline
\end{tabular}
\end{center}
\end{table}

\begin{figure}[ht]
\begin{center}
\includegraphics[height=3.6in, width=2.8in,angle=90]{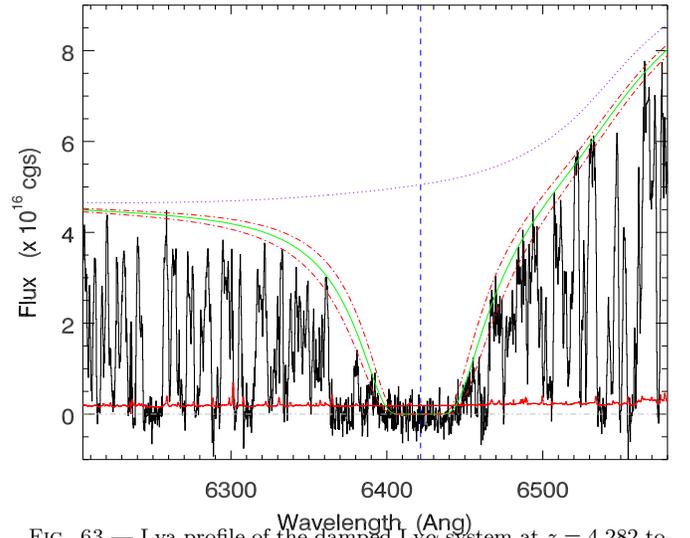}
\figcaption{Lya profile of the damped \lya system at $z=4.282$
toward PSS2241+1352.
The overplotted solid line and accompanying
dash-dot lines trace the best fit solution and the estimated 
bounds corresponding to 
$\log \N{HI} = 21.15^{+0.10}_{-0.10}$.  
Although the quasar continuum (dotted line) for this portion of the
spectrum is complicated, the core is well isolated from contamination
and we are reasonably confident in the best fit value.  
\label{fig:pss2241_lya}}
\end{center}
\end{figure}

\begin{figure}[ht]
\begin{center}
\includegraphics[height=6.1in, width=3.9in]{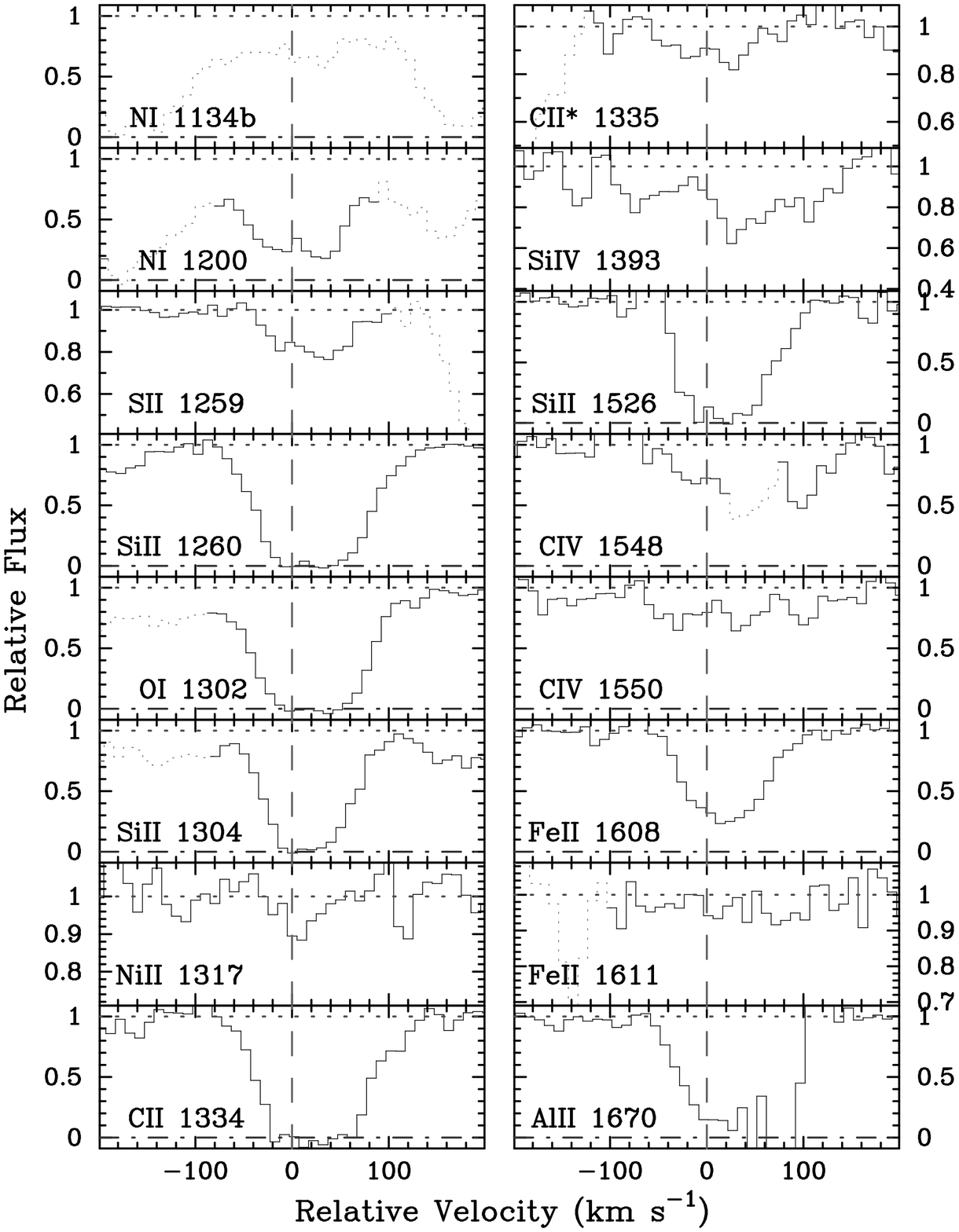}
\figcaption{Velocity plot of the metal-line transitions for the 
damped \lya system at $z = 4.282$ toward PSS2241+1352.
The vertical line at $v=0$ corresponds to $z = 4.28239$.  
\label{fig:pss2241_mtl}}
\end{center}
\end{figure}

\subsection{PSS2241+1352, $z = 4.282$ \label{subsec:PSS2241+1352_4.282}}

This damped \lya system \citep{peroux01} has one of the highest redshifts in our sample.
While the determination of its $\N{HI}$ value is complicated by the \lya
emission peak (Figure~\ref{fig:pss2241_lya}), its H\,I value is 
constrained by the higher order Lyman lines and 
our best fit value is $\N{HI} = 10^{21.15} \cm{-2}$.  At present,
this is the largest H\,I value for a damped system at $z>4$.
The metal-line profiles for the DLA are shown in Figure~\ref{fig:pss2241_mtl}
which includes a probable detection of C\,II$^*$ 1335.
The column densities are given by Table~\ref{tab:PSS2241+1352_4.282}.

\begin{table}[ht]\footnotesize
\begin{center}
\caption{ {\sc
IONIC COLUMN DENSITIES: PSS2323+2758, $z = 3.684$ \label{tab:PSS2323+2758_3.684}}}
\begin{tabular}{lcccc}
\tableline
\tableline
Ion & $\lambda$ & AODM & $N_{\rm adopt}$ & [X/H] \\
\tableline
C  IV &1548.2&$13.417 \pm  0.066$\\  
O  I  &1039.2&$>15.607$&$>15.607$&$>-2.083$\\  
O  I  &1302.2&$>14.824$\\  
Al II &1670.8&$12.354 \pm  0.034$&$12.354 \pm  0.033$&$-3.086 \pm  0.105$\\  
Al III&1854.7&$12.500 \pm  0.127$\\  
Si II &1526.7&$13.917 \pm  0.030$&$13.917 \pm  0.030$&$-2.593 \pm  0.104$\\  
Si II &1808.0&$<14.667$\\  
Si IV &1393.8&$<12.376$\\  
Fe II &1608.5&$13.321 \pm  0.128$&$13.321 \pm  0.128$&$-3.129 \pm  0.162$\\  
\tableline
\end{tabular}
\end{center}
\end{table}

\begin{figure}[ht]
\begin{center}
\includegraphics[height=3.6in, width=2.8in,angle=90]{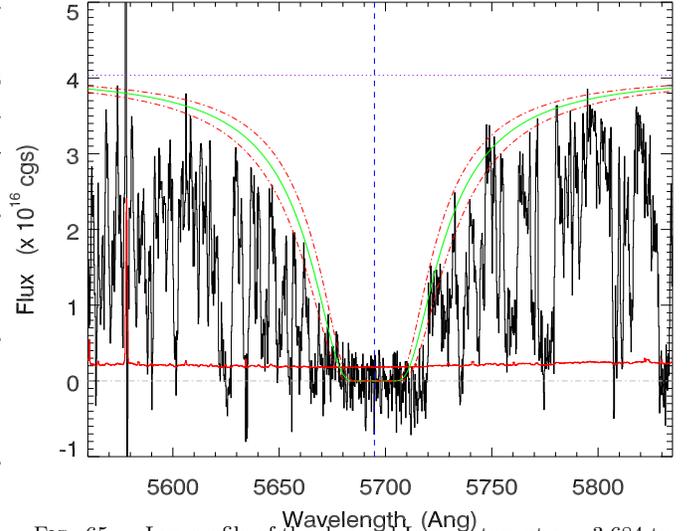}
\figcaption{Lya profile of the damped \lya system at $z=3.684$
toward PSS2323+2758.
The overplotted solid line and accompanying
dash-dot lines trace the best fit solution and the estimated 
bounds corresponding to 
$\log \N{HI} = 20.95^{+0.10}_{-0.10}$.  
This profile is reasonably well fit even though the blue wing
is largely unconstrained by the observations.  
\label{fig:pss2323_lya}}
\end{center}
\end{figure}

\begin{figure}[ht]
\begin{center}
\includegraphics[height=6.1in, width=3.9in]{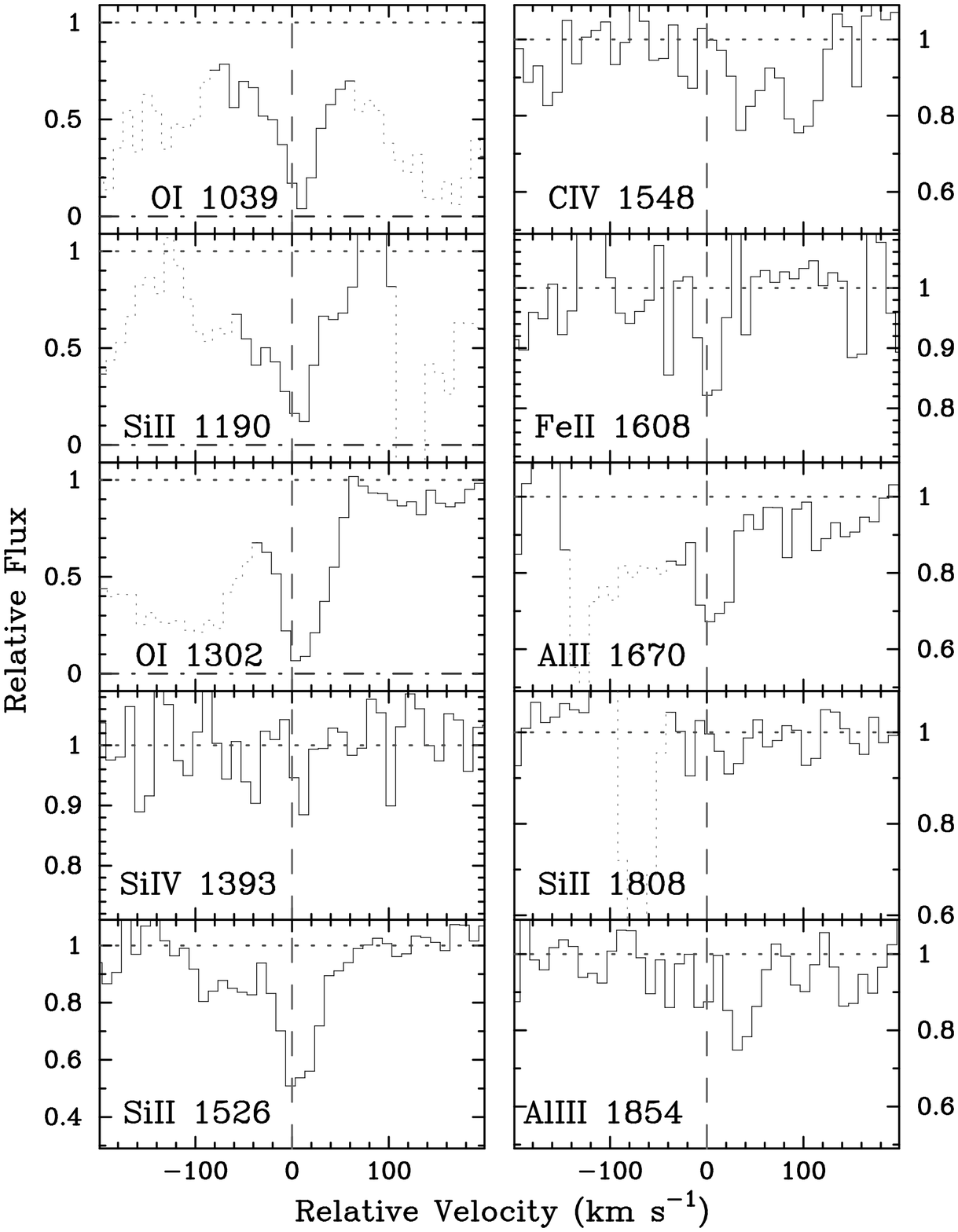}
\figcaption{Velocity plot of the metal-line transitions for the 
damped \lya system at $z = 3.684$ toward PSS2323+2758.
The vertical line at $v=0$ corresponds to $z = 3.6845$.  
\label{fig:pss2323_mtl}}
\end{center}
\end{figure}

\subsection{PSS2323+2758, $z = 3.684$ \label{subsec:PSS2323+2758_3.684}}

This PSS quasar exhibits a damped system at $z=3.684$ with a considerable
H\,I column density $\N{HI} \approx 10^{21} \cm{-2}$ (Figure~\ref{fig:pss2323_lya}).
Furthermore, it has one of the lowest metallicities of any DLA, with an
Fe abundance below $10^{-3}$.  The few valuable metal-line profiles are
shown in Figure~\ref{fig:pss2323_mtl} and Table~\ref{tab:PSS2323+2758_3.684}
provides the ionic column densities.
In passing, we note that a measurement of the O\,I~1039 transition reveals
$\N{O^+} > 15.6$ if one assumes that this profile is unblended.  We 
expect that this is a false assumption but suggest the system warrants
further analysis.

\begin{table}[ht]\footnotesize
\begin{center}
\caption{ {\sc
IONIC COLUMN DENSITIES: FJ2334-09, $z = 3.057$ \label{tab:FJ2334-09_3.057}}}
\begin{tabular}{lcccc}
\tableline
\tableline
Ion & $\lambda$ & AODM & $N_{\rm adopt}$ & [X/H] \\
\tableline
C  II &1334.5&$>15.120$&$>15.120$&$>-1.920$\\  
C  II*&1335.7&$<12.934$\\  
C  IV &1548.2&$>14.389$\\  
C  IV &1550.8&$>14.485$\\  
O  I  &1302.2&$>15.497$&$>15.497$&$>-1.693$\\  
Al II &1670.8&$>13.317$&$>13.317$&$>-1.623$\\  
Al III&1854.7&$12.683 \pm  0.047$\\  
Al III&1862.8&$12.630 \pm  0.082$\\  
Si II &1526.7&$>14.560$&$14.860 \pm  0.070$&$-1.150 \pm  0.122$\\  
Si II &1808.0&$14.860 \pm  0.070$\\  
Si IV &1393.8&$>13.913$\\  
Si IV &1402.8&$>13.924$\\  
Fe II &1608.5&$14.334 \pm  0.015$&$14.323 \pm  0.014$&$-1.627 \pm  0.101$\\  
Fe II &1611.2&$<14.893$\\  
Fe II &2374.5&$14.258 \pm  0.041$\\  
Ni II &1317.2&$<13.568$&$<13.125$&$<-1.575$\\  
Ni II &1741.6&$<13.125$\\  
Zn II &2026.1&$<12.168$&$<12.167$&$<-0.953$\\  
\tableline
\end{tabular}
\end{center}
\end{table}

\begin{figure}[ht]
\begin{center}
\includegraphics[height=3.6in, width=2.8in,angle=90]{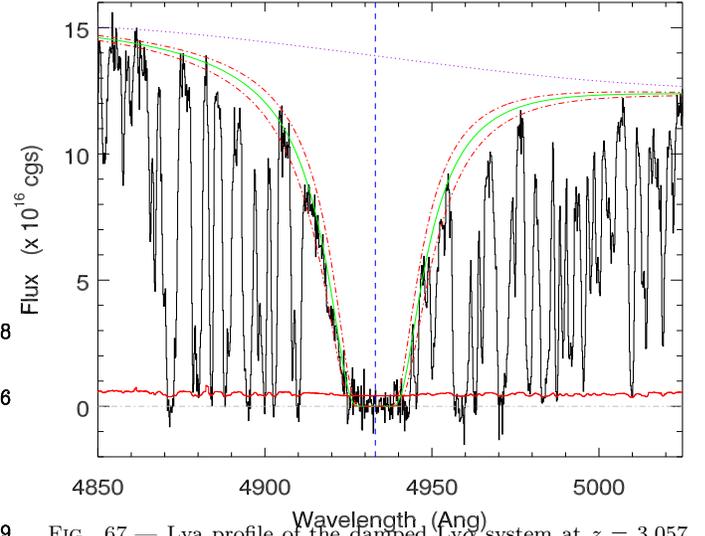}
\figcaption{Lya profile of the damped \lya system at $z=3.057$
toward FJ2334-09.
The overplotted solid line and accompanying
dash-dot lines trace the best fit solution and the estimated 
bounds corresponding to 
$\log \N{HI} = 20.45^{+0.10}_{-0.10}$.  
The continuum varies
significantly across the \lya profile, but the core constrains the H\,I
column density to reasonable precision.
\label{fig:fj2334_lya}}
\end{center}
\end{figure}

\begin{figure}[ht]
\begin{center}
\includegraphics[height=6.1in, width=3.9in]{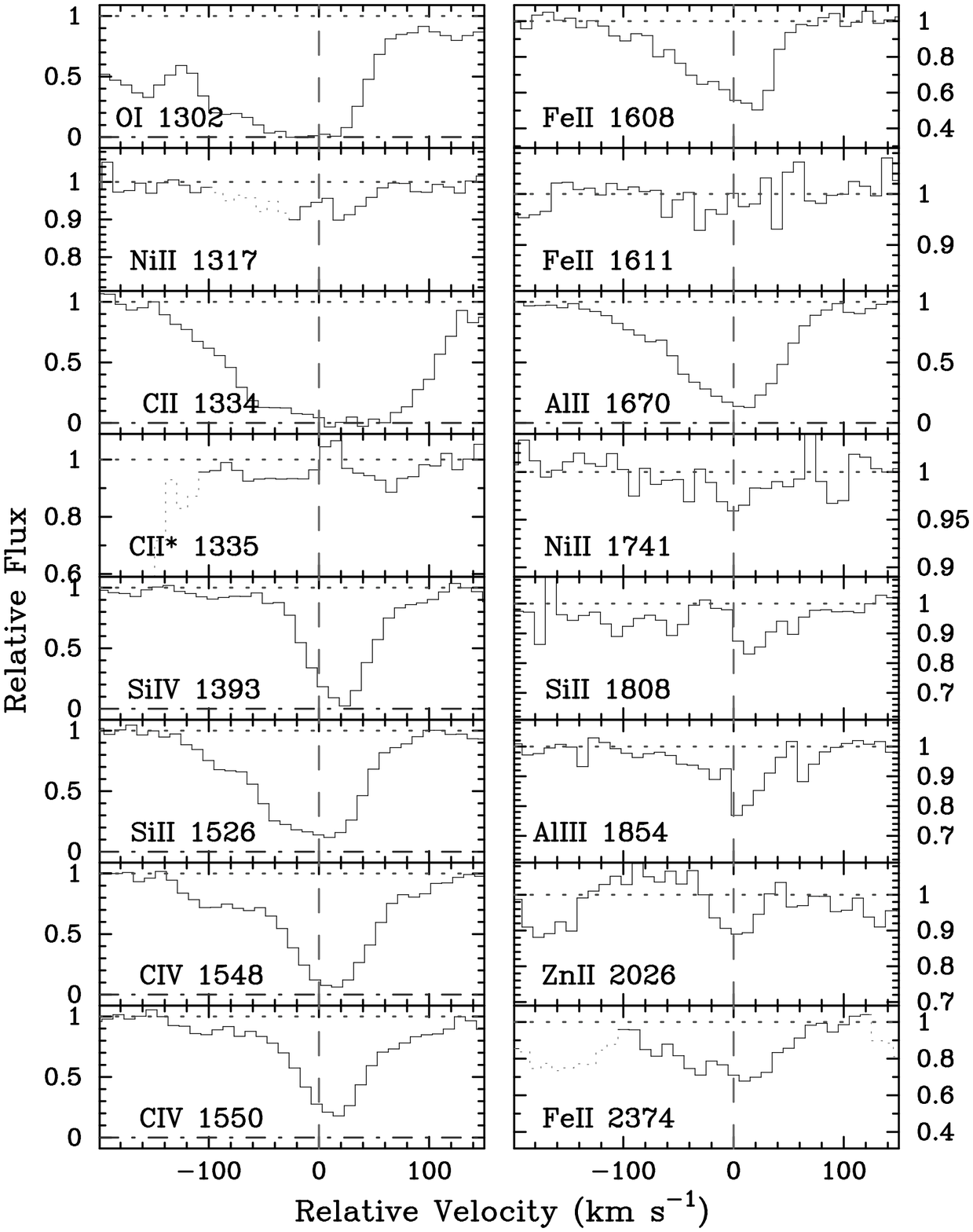}
\figcaption{Velocity plot of the metal-line transitions for the 
damped \lya system at $z = 3.057$ toward FJ2334-09.
The vertical line at $v=0$ corresponds to $z = 3.057195$.  
\label{fig:fj2334_mtl}}
\end{center}
\end{figure}

\subsection{FJ2334-09, $z = 3.057$ \label{subsec:FJ2334-09_3.057}}

This FIRST quasar \citep{white00} exhibits a modest damped \lya system
at $z=3.057$ (Figure~\ref{fig:fj2334_lya}).  
The continuum varies
significantly across the \lya profile, but the core constrains the H\,I
column density to reasonable precision.  We present the metal-line
profiles in Figure~\ref{fig:fj2334_mtl} and the ionic column densities
in Table~\ref{tab:FJ2334-09_3.057}.

\begin{table}[ht]\footnotesize
\begin{center}
\caption{ {\sc
IONIC COLUMN DENSITIES: Q2342+34, $z = 2.908$ \label{tab:Q2342+34_2.908}}}
\begin{tabular}{lcccc}
\tableline
\tableline
Ion & $\lambda$ & AODM & $N_{\rm adopt}$ & [X/H] \\
\tableline
C  I  &1656.9&$<12.909$\\  
C  II &1334.5&$>15.188$&$>15.188$&$>-2.502$\\  
C  IV &1548.2&$>14.403$\\  
C  IV &1550.8&$14.403 \pm  0.016$\\  
N  I  &1135.0&$14.887 \pm  0.045$&$14.887 \pm  0.045$&$-2.143 \pm  0.110$\\  
O  I  &1302.2&$>15.526$&$>15.526$&$>-2.314$\\  
O  I  &1355.6&$<18.223$\\  
Mg II &1240.4&$<15.612$&$<15.612$&$<-1.068$\\  
Al II &1670.8&$>13.615$&$>13.615$&$>-1.975$\\  
Al III&1854.7&$13.112 \pm  0.035$\\  
Si II &1304.4&$>15.047$&$15.467 \pm  0.033$&$-1.193 \pm  0.105$\\  
Si II &1526.7&$>14.845$\\  
Si II &1808.0&$15.467 \pm  0.033$\\  
Si IV &1393.8&$>13.999$\\  
Si IV &1402.8&$14.004 \pm  0.022$\\  
S  II &1259.5&$>15.090$&$>15.090$&$>-1.210$\\  
Cr II &2056.3&$13.361 \pm  0.074$&$13.361 \pm  0.074$&$-1.409 \pm  0.124$\\  
Fe II &1122.0&$14.982 \pm  0.064$&$14.982 \pm  0.064$&$-1.618 \pm  0.119$\\  
Fe II &1608.5&$>14.896$\\  
Fe II &1611.2&$<14.942$\\  
Fe II &2249.9&$<15.219$\\  
Fe II &2344.2&$>14.672$\\  
Fe III&1122.5&$<13.940$\\  
Ni II &1709.6&$13.852 \pm  0.085$&$13.827 \pm  0.045$&$-1.523 \pm  0.110$\\  
Ni II &1741.6&$13.771 \pm  0.069$\\  
Ni II &1751.9&$13.922 \pm  0.078$\\  
Zn II &2026.1&$12.501 \pm  0.108$&$12.501 \pm  0.108$&$-1.269 \pm  0.147$\\  
\tableline
\end{tabular}
\end{center}
\end{table}

\begin{figure}[ht]
\begin{center}
\includegraphics[height=3.6in, width=2.8in,angle=90]{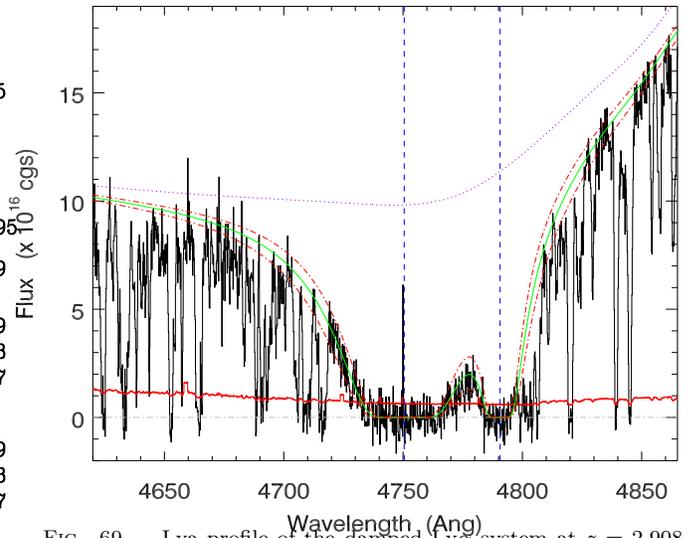}
\figcaption{Lya profile of the damped \lya system at $z=2.908$
toward Q2342+34.
The overplotted solid line and accompanying
dash-dot lines trace the best fit solution and the estimated 
bounds corresponding to 
$\log \N{HI} = 21.10^{+0.10}_{-0.10}$.  
The solution for this \lya profile is complicated by a
neighboring sub-DLA and the \lya emission peak 
but the blue wing and the flux at $\lambda \approx 4780$\AA\
do nicely constrain the $\N{HI}$ value.  Note the 1 pixel
spike at $\lambda \approx 4750$\AA\ is a cosmic ray.
\label{fig:q2342_lya}}
\end{center}
\end{figure}

\begin{figure}[ht]
\begin{center}
\includegraphics[height=6.1in, width=3.9in]{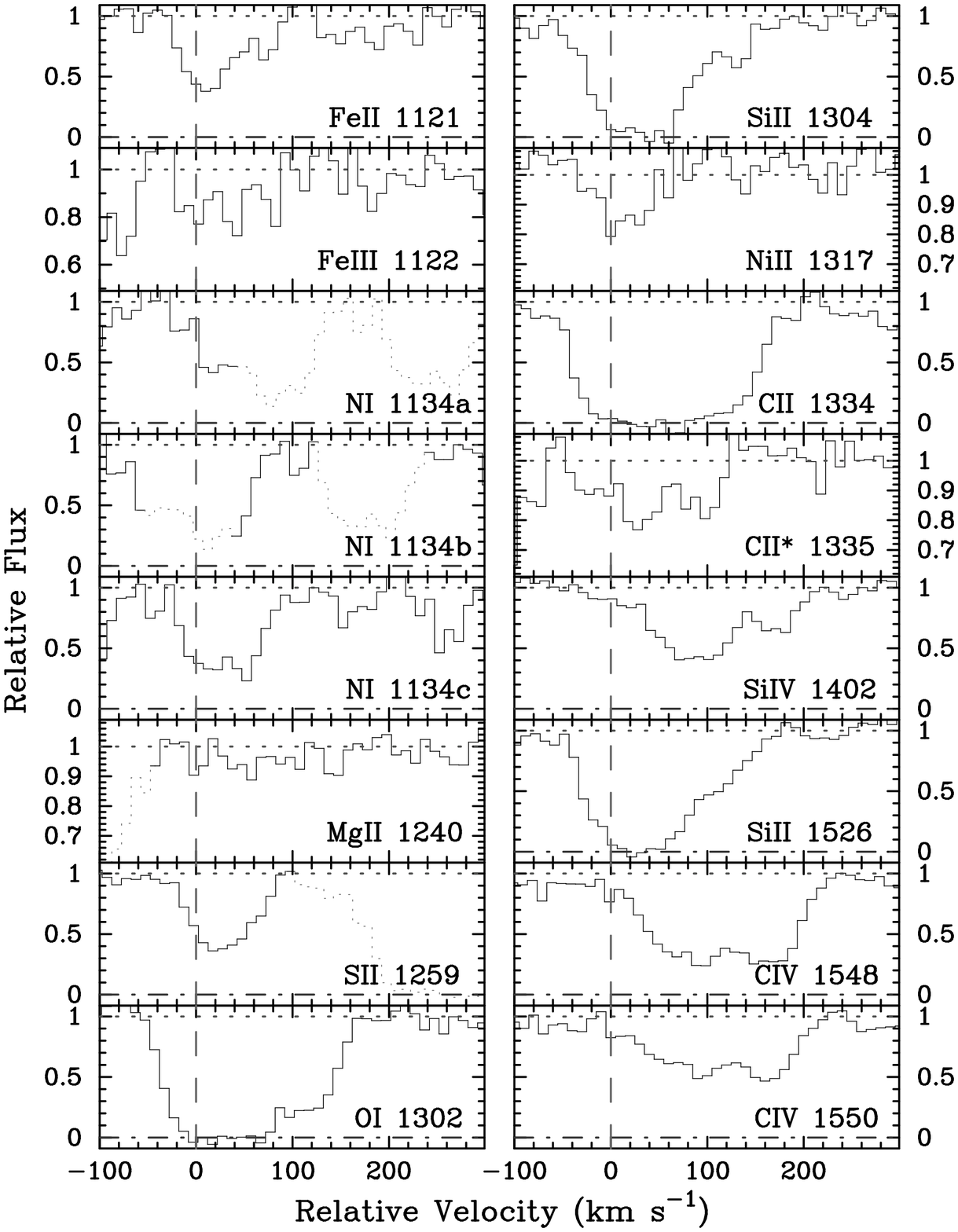}
\figcaption{Velocity plot of the metal-line transitions for the 
damped \lya system at $z = 2.908$ toward Q2342+34.
The vertical line at $v=0$ corresponds to $z = 2.90823$.  
\label{fig:q2342_mtl}}
\end{center}
\end{figure}

\begin{figure}[ht]
\begin{center}
\includegraphics[height=6.1in, width=3.9in]{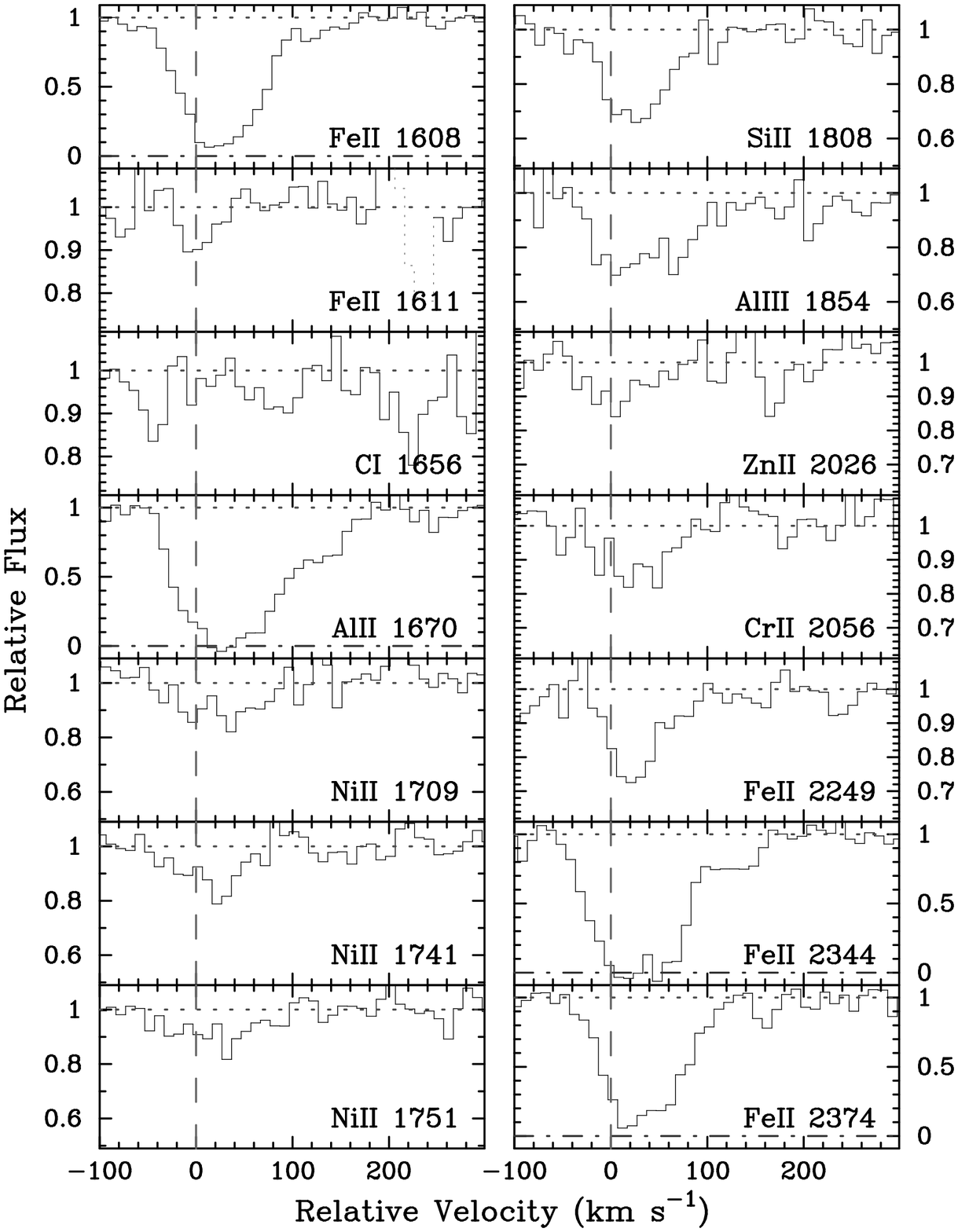}
Fig 70 -- cont
\end{center}
\end{figure}
\subsection{Q2342+34, $z = 2.908$ \label{subsec:Q2342+34_2.908}}

This quasar is another member of the 
Third MIT-Green Bank 5GHz Survey  \citep{griffith90} while
the damped \lya system was identified by \cite{white93}.
The \lya profile of this damped system (Figure~\ref{fig:q2342_lya}
is blended with a neighboring
sub-DLA and the red wing is dominated by the \lya emission peak.  
Nevertheless, the observations nicely constrain the 
$\N{HI}$ value.  Figure~\ref{fig:q2342_mtl} plots the comprehensive
set of velocity profiles
of this DLA and Table~\ref{tab:Q2342+34_2.908} gives the ionic column
densities.

\clearpage

\begin{table}[ht]\footnotesize
\begin{center}
\caption{ {\sc
IONIC COLUMN DENSITIES: PSS2344+0342, $z = 3.219$ \label{tab:PSS2344+0342_3.219}}}
\begin{tabular}{lcccc}
\tableline
\tableline
Ion & $\lambda$ & AODM & $N_{\rm adopt}$ & [X/H] \\
\tableline
C  IV &1548.2&$>14.568$\\  
C  IV &1550.8&$>14.726$\\  
N  I  &1135.0&$>14.648$&$>14.648$&$>-2.632$\\  
Al II &1670.8&$>13.535$&$>13.535$&$>-2.305$\\  
Al III&1854.7&$12.845 \pm  0.046$\\  
Si II &1526.7&$>14.866$&$>14.865$&$>-2.045$\\  
Cr II &2056.3&$<13.112$&$13.306 \pm  0.088$&$-1.714 \pm  0.112$\\  
Cr II &2062.2&$13.306 \pm  0.088$\\  
Fe II &1608.5&$>14.707$&$15.234 \pm  0.075$&$-1.616 \pm  0.103$\\  
Fe II &1611.2&$15.234 \pm  0.075$\\  
Fe II &2374.5&$>14.517$\\  
Ni II &1709.6&$13.546 \pm  0.144$&$13.545 \pm  0.144$&$-2.055 \pm  0.160$\\  
Ni II &1751.9&$<13.599$\\  
Zn II &2026.1&$<12.267$&$<12.267$&$<-1.753$\\  
Zn II &2062.7&$<12.601$\\  
\tableline
\end{tabular}
\end{center}
\end{table}

\begin{figure}[ht]
\begin{center}
\includegraphics[height=3.6in, width=2.8in,angle=90]{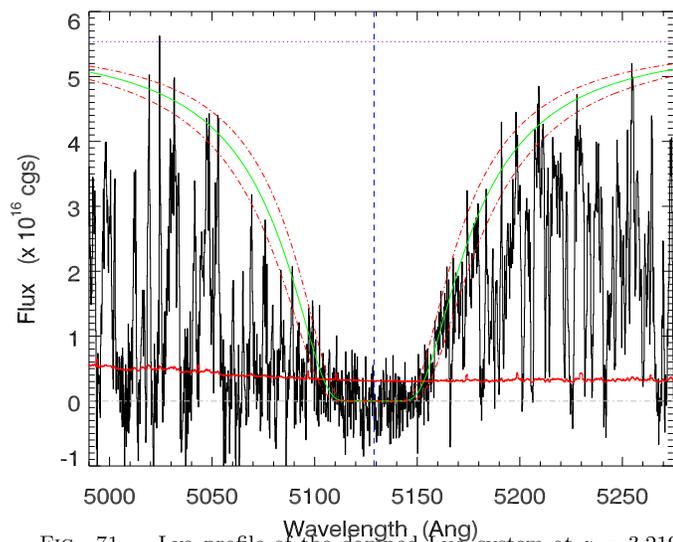}
\figcaption{Lya profile of the damped \lya system at $z=3.219$
toward PSS2344+0342.
The overplotted solid line and accompanying
dash-dot lines trace the best fit solution and the estimated 
bounds corresponding to 
$\log \N{HI} = 21.35^{+0.07}_{-0.07}$.  
Because of the well behaved continuum and a relatively clean
profile, we assess an uncertainty of only 0.07~dex to this best fit 
value.
\label{fig:pss2344_lya}}
\end{center}
\end{figure}

\begin{figure}[ht]
\begin{center}
\includegraphics[height=6.1in, width=3.9in]{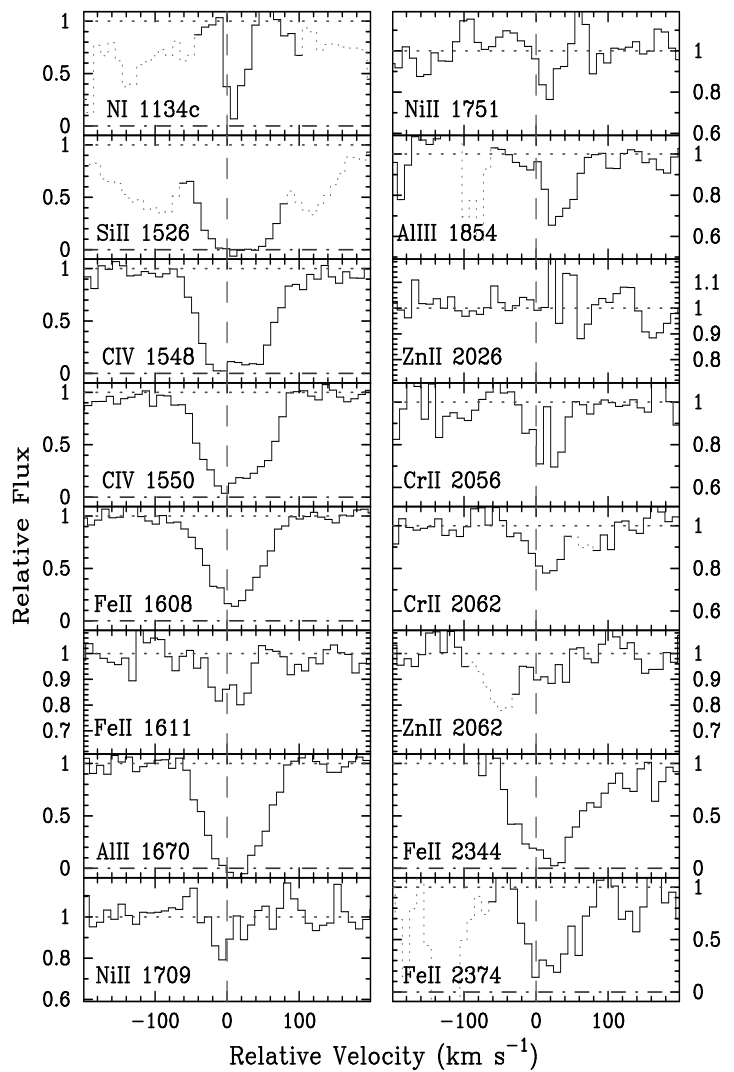}
\figcaption{Velocity plot of the metal-line transitions for the 
damped \lya system at $z = 3.219$ toward PSS2344+0342.
The vertical line at $v=0$ corresponds to $z = 3.2194$.  
\label{fig:pss2344_mtl}}
\end{center}
\end{figure}

\subsection{PSS2344+0342, $z = 3.219$ \label{subsec:PSS2344+0342_3.219}}

This damped \lya system was discovered by \cite{peroux01} toward the
PSS quasar PSS2344+0342 \citep{pssweb}.
Its \lya profile is displayed in Figure~\ref{fig:pss2344_lya} which is 
somewhat blended with coincident \lya lines.  The continuum is well
behaved across the profile, though, and the $\N{HI}$ value is tightly
constrained $(10^{21.35 \pm 0.07} \cm{-2}$). This H\,I value is considerable;
this DLA has one of the largest column densities at $z>3$.
Note that our value is considerably larger (0.45~dex) than the one reported
by \cite{peroux01}.  Given the quality of our fit, we suspect a systematic
error in their continuum placement.
We also note that \cite{peroux01} reported a second DLA along this sightline
at $z=2.68$.  Unfortunately, this profile lands at the defect in the ESI CCD
which prevents us from confirming its existence.  We do worry, however, that
the higher order Lyman series and Lyman limit of a $z=3.93$ system do cut into
the spectrum at the position where they reported the \lya profile.  Also, 
we have been unable to identify and Fe\,II, Zn\,II, or Cr\,II transitions
with the claimed DLA.  For these reasons, we suspect the DLA was a mistaken
identification.

An extensive set of metal-line profiles of the $z=3.2194$ DLA
are presented in Figure~\ref{fig:pss2344_mtl}
and the column densities are listed in Table~\ref{tab:PSS2344+0342_3.219}.
One notes that the system appears to exhibit [Zn/Fe]~$< 0$,
implying a very low dust content.


\begin{sidewaystable*}\footnotesize
\begin{center}
\caption{
{\sc ABUNDANCE SUMMARY \label{tab:XHsum}}}
\begin{tabular}{lccrrrrrrrrrrrrrr}
\tableline
\tableline \tskip
Name & $z_{abs}$ & $\N{HI}$
& [C/H] & [N/H] & [O/H] & [Al/H] & [Si/H] & [S/H] & [Ti/H] & 
[Cr/H] & [Fe/H] & [Ni/H] & [Zn/H] \\
\tableline
SDSS0127-00&3.727&21.15&$>-3.068$&$-3.052$&$>-2.705$&$>-2.897$&$>-2.404$&&&&&$<-1.709$&\\  
PSS0133+0400&3.693&20.70&$>-3.130$&&&$-2.818$&$>-2.004$&&&&$-2.689$&&\\  
PSS0133+0400&3.774&20.55&$>-1.831$&&&$>-1.248$&$-0.644$&&$-0.449$&$<-0.980$&$>-1.182$&$-0.822$&$<-0.121$\\  
PSS0134+3317&3.761&20.85&&&&$-2.728$&$>-2.521$&&&&&&\\  
PSS0209+0517&3.667&20.45&&$<-2.588$&$>-2.416$&$-2.420$&$>-1.903$&&&$<-1.041$&$-2.315$&$<-1.450$&\\  
PSS0209+0517&3.864&20.55&$>-3.120$&&$>-2.922$&$-2.894$&$-2.647$&&&&$<-2.710$&$<-1.677$&\\  
BRJ0426-2202&2.983&21.50&&&&$>-3.059$&$<-2.064$&&&$<-2.269$&$-2.848$&$<-2.377$&$<-2.005$\\  
FJ0747+2739&3.423&20.85&$>-2.344$&&&$>-2.136$&$>-1.871$&&&&$>-1.920$&$<-1.828$&\\  
FJ0747+2739&3.900&20.50&$>-2.587$&&$>-2.434$&$-2.529$&$-2.029$&$<-1.339$&&&$<-2.201$&$<-1.644$&$<-0.772$\\  
PSS0808+52&3.113&20.65&&$<-1.356$&$>-2.445$&&$-1.611$&$<-0.929$&&$<-1.585$&$-1.980$&$<-1.662$&$<-1.186$\\  
FJ0812+32&2.626&21.35&$>-2.800$&$< 0.326$&$>-2.602$&$>-2.181$&$>-1.128$&$>-1.112$&$-1.921$&$-1.700$&$-1.753$&$-1.754$&$-0.985$\\  
Q0930+28&3.235&20.35&$>-2.767$&&$>-2.894$&$-2.471$&$-2.117$&&&&$-2.360$&&\\  
PC0953+47&3.404&21.15&&&&$>-2.280$&$>-2.093$&&&$<-1.259$&$>-2.178$&$<-1.504$&\\  
PC0953+47&3.891&21.20&&&&$>-1.859$&$>-1.599$&&&&$>-1.712$&$-1.698$&\\  
PC0953+47&4.244&20.90&$>-2.771$&&$>-2.478$&$<-2.517$&$-2.225$&$<-1.382$&&&$-2.500$&$<-1.537$&\\  
PSS0957+33&3.280&20.45&&$<-1.936$&&$>-1.812$&$-1.334$&$<-1.073$&&$<-1.229$&$>-1.862$&$<-1.548$&$<-0.952$\\  
PSS0957+33&4.180&20.65&$>-2.441$&&$>-2.418$&$-2.102$&$>-1.834$&&&&$>-2.253$&$<-1.952$&\\  
BQ1021+3001&2.949&20.70&$>-2.704$&&$>-2.592$&$-2.497$&$-2.170$&&&$<-1.537$&$-2.320$&$<-1.582$&$<-1.181$\\  
CTQ460&2.777&21.00&$>-2.439$&&$>-2.176$&$-2.065$&$>-1.590$&$-1.409$&&$-1.662$&$-1.816$&$-1.663$&$<-1.311$\\  
HS1132+2243&2.783&21.00&$>-3.010$&$-2.911$&$>-2.751$&$>-2.623$&$-2.069$&$-2.129$&&$-1.836$&$-2.476$&$<-2.095$&$<-1.684$\\  
Q1209+0919&2.584&21.40&&&&$>-1.822$&$>-0.970$&&$<-1.688$&$-1.489$&$-1.681$&$<-1.432$&$-1.088$\\  
PSS1248+31&3.696&20.63&&&&$-2.343$&$-1.797$&&&$<-1.058$&$-2.240$&$<-1.387$&$<-0.685$\\  
PSS1253-0228&2.783&21.85&$>-3.428$&&$>-3.243$&$>-2.941$&$>-1.813$&&$<-1.948$&$-1.888$&$-1.992$&$-1.930$&$-1.752$\\  
Q1337+11&2.795&20.95&$>-3.057$&$-3.129$&$>-2.824$&$-2.596$&$-1.792$&$-1.880$&&$<-1.720$&$-2.388$&$<-2.001$&$<-1.419$\\  
PKS1354-17&2.780&20.30&$>-2.748$&&&$-2.395$&$-1.884$&&&&$-2.426$&&\\  
PSS1432+39&3.272&21.25&&&&$>-1.991$&$-1.144$&&&$<-1.760$&$>-1.820$&$-1.750$&$<-1.275$\\  
Q1502+4837&2.570&20.30&$>-1.716$&&$>-1.699$&$>-1.428$&$-1.619$&&&&$-1.649$&&\\  
PSS1506+5220&3.224&20.67&$>-2.050$&&$>-2.753$&$>-2.657$&$-2.348$&&&$<-1.556$&$-2.462$&$<-1.507$&$<-1.235$\\  
PSS1723+2243&3.695&20.50&&&&$>-1.426$&$>-1.199$&&&$<-0.956$&$>-1.425$&$<-0.800$&$>-0.662$\\  
PSS2155+1358&3.316&20.55&&&&$>-1.724$&$-1.263$&&&$<-1.069$&$>-1.677$&$<-1.513$&$<-1.167$\\  
Q2223+20&3.119&20.30&&$<-2.611$&$>-2.198$&$<-2.461$&$-2.220$&&&&$-2.476$&&\\  
PSS2241+1352&4.282&21.15&$>-2.725$&$>-2.272$&&$>-2.121$&$>-1.699$&$-1.775$&&&$>-2.004$&$<-2.177$&\\  
PSS2323+2758&3.684&20.95&&&$>-2.083$&$-3.086$&$-2.593$&&&&$-3.129$&&\\  
FJ2334-09&3.057&20.45&$>-1.920$&&$>-1.693$&$>-1.623$&$-1.150$&&&&$-1.627$&$<-1.575$&$<-0.953$\\  
Q2342+34&2.908&21.10&$>-2.502$&$-2.143$&$>-2.314$&$>-1.975$&$-1.193$&$>-1.210$&&$-1.409$&$-1.618$&$-1.523$&$-1.269$\\  
PSS2344+0342&3.219&21.35&&$>-2.632$&&$>-2.305$&$>-2.045$&&&$-1.714$&$-1.616$&$-2.055$&$<-1.753$\\  
\tskip \tableline
\end{tabular}
\end{center}
\end{sidewaystable*}

\begin{sidewaystable*}\footnotesize
\begin{center}
\caption{
{\sc RELATIVE ABUNDANCE SUMMARY \label{tab:XFesum}}}
\begin{tabular}{lccrrrrrrrrrrrrrr}
\tableline
\tableline \tskip
Name & $z_{abs}$ & $\N{HI}$ & [C/Fe] & [N/Fe] & [O/Fe] & [Al/Fe] 
& [Si/Fe] & [S/Fe] & [Ti/Fe] & [Cr/Fe] & [Ni/Fe] & [Zn/Fe] \\
\tableline
SDSS0127-00&3.727&21.15\\  
PSS0133+0400&3.693&20.70&$>-0.441$&&&$-0.129$&$>+ 0.685$&&&&&\\  
PSS0133+0400$^a$&3.774&20.55&$>-1.009$&&&$>-0.426$&$+ 0.178$&&$+ 0.373$&$<-0.158$&&$<+ 0.701$\\  
PSS0134+3317$^b$&3.761&20.85&&&&&$>+ 0.207$&&&&&\\  
PSS0209+0517&3.667&20.45&&$<-0.273$&$>-0.101$&$-0.105$&$>+ 0.412$&&&$<+ 1.274$&$<+ 0.865$&\\  
PSS0209+0517$^b$&3.864&20.55&$>-0.226$&&$>-0.028$&&$+ 0.247$&&&&$<+ 1.217$&\\  
BRJ0426-2202&2.983&21.50&&&&$>-0.211$&$<+ 0.784$&&&$<+ 0.579$&$<+ 0.471$&$<+ 0.843$\\  
FJ0747+2739&3.423&20.85\\  
FJ0747+2739$^b$&3.900&20.50&$>-0.058$&&$>+ 0.095$&&$+ 0.500$&$<+ 1.190$&&&$<+ 0.885$&$<+ 1.757$\\  
PSS0808+52&3.113&20.65&&$<+ 0.624$&$>-0.465$&&$+ 0.369$&$<+ 1.051$&&$<+ 0.395$&$<+ 0.318$&$<+ 0.794$\\  
FJ0812+32&2.626&21.35&$>-1.047$&$<+ 2.079$&$>-0.849$&$>-0.428$&$>+ 0.625$&$>+ 0.641$&$-0.168$&$+ 0.053$&$-0.001$&$+ 0.768$\\  
Q0930+28&3.235&20.35&$>-0.407$&&$>-0.534$&$-0.111$&$+ 0.243$&&&&&\\  
PC0953+47$^b$&3.404&21.15&&&&&$>+ 0.187$&&&$<+ 1.021$&$<+ 0.776$&\\  
PC0953+47$^a$&3.891&21.20&&&&$>-0.161$&$>+ 0.099$&&&&&\\  
PC0953+47&4.244&20.90&$>-0.271$&&$>+ 0.022$&$<-0.017$&$+ 0.275$&$<+ 1.118$&&&$<+ 0.963$&\\  
PSS0957+33&3.280&20.45\\  
PSS0957+33$^b$&4.180&20.65&$>-0.339$&&$>-0.316$&&$>+ 0.268$&&&&$<+ 0.150$&\\  
BQ1021+3001&2.949&20.70&$>-0.384$&&$>-0.272$&$-0.177$&$+ 0.150$&&&$<+ 0.783$&$<+ 0.738$&$<+ 1.139$\\  
CTQ460&2.777&21.00&$>-0.623$&&$>-0.360$&$-0.249$&$>+ 0.226$&$+ 0.407$&&$+ 0.154$&$+ 0.153$&$<+ 0.505$\\  
HS1132+2243&2.783&21.00&$>-0.534$&$-0.435$&$>-0.275$&$>-0.147$&$+ 0.407$&$+ 0.347$&&$+ 0.640$&$<+ 0.381$&$<+ 0.792$\\  
Q1209+0919&2.584&21.40&&&&$>-0.141$&$>+ 0.711$&&$<-0.007$&$+ 0.192$&$<+ 0.249$&$+ 0.593$\\  
PSS1248+31&3.696&20.63&&&&$-0.103$&$+ 0.443$&&&$<+ 1.182$&$<+ 0.853$&$<+ 1.555$\\  
PSS1253-0228&2.783&21.85&$>-1.436$&&$>-1.251$&$>-0.949$&$>+ 0.179$&&$<+ 0.044$&$+ 0.104$&$+ 0.062$&$+ 0.240$\\  
Q1337+11&2.795&20.95&$>-0.669$&$-0.741$&$>-0.436$&$-0.208$&$+ 0.596$&$+ 0.508$&&$<+ 0.668$&$<+ 0.387$&$<+ 0.969$\\  
PKS1354-17&2.780&20.30&$>-0.322$&&&$+ 0.031$&$+ 0.542$&&&&&\\  
PSS1432+39$^a$&3.272&21.25&&&&$>-0.241$&$+ 0.606$&&&$<-0.010$&&$<+ 0.475$\\  
Q1502+4837&2.570&20.30&$>-0.067$&&$>-0.050$&$>+ 0.221$&$+ 0.030$&&&&&\\  
PSS1506+5220&3.224&20.67&$>+ 0.412$&&$>-0.291$&$>-0.195$&$+ 0.114$&&&$<+ 0.906$&$<+ 0.955$&$<+ 1.227$\\  
PSS1723+2243$^b$&3.695&20.50&&&&&$>+ 0.227$&&&$<+ 0.470$&$<+ 0.626$&$>+ 0.764$\\  
PSS2155+1358$^b$&3.316&20.55&&&&&$+ 0.461$&&&$<+ 0.655$&$<+ 0.211$&$<+ 0.557$\\  
Q2223+20&3.119&20.30&&$<-0.135$&$>+ 0.278$&$<+ 0.015$&$+ 0.256$&&&&&\\  
PSS2241+1352&4.282&21.15\\  
PSS2323+2758&3.684&20.95&&&$>+ 1.046$&$+ 0.043$&$+ 0.536$&&&&&\\  
FJ2334-09&3.057&20.45&$>-0.293$&&$>-0.066$&$>+ 0.004$&$+ 0.477$&&&&$<+ 0.052$&$<+ 0.674$\\  
Q2342+34&2.908&21.10&$>-0.884$&$-0.525$&$>-0.696$&$>-0.357$&$+ 0.425$&$>+ 0.408$&&$+ 0.209$&$+ 0.095$&$+ 0.349$\\  
PSS2344+0342&3.219&21.35&&$>-1.016$&&$>-0.689$&$>-0.429$&&&$-0.098$&$-0.439$&$<-0.137$\\  
\tskip \tableline
\end{tabular}
\end{center}
$^a$Ni is serving as a proxy for Fe\\
$^b$Al is serving as a proxy for Fe
\end{sidewaystable*}

\section{SUMMARY}

Tables~\ref{tab:XHsum} and \ref{tab:XFesum} present a summary of the
absolute and relative abundances of the \ndla\ damped \lya systems
in our complete database.  In Table~\ref{tab:XFesum}, where
we present abundances relative to Fe, we have 
considered Ni or Al as a proxy for Fe in a few cases as noted.

We have presented ionic column density measurements for our complete
sample of damped \lya systems.  
We have measured
metal-line column densities with the apparent optical depth method and
$\N{HI}$ values through qualitative fits to the \lya profiles.  
Therefore, all of the data has been reduced and analysed with an identical
approach.  We
have used the most up to date atomic data and will continue to update
the database as new information becomes available.  Visit 
http://www.ucolick.org/$\sim$xavier/ESI/index.html for tables, figures and updated
measurements as well as http://kingpin.ucsd.edu/$\sim$hiresdla/ for the
HIRES abundance database.  Future papers will
present new scientific results based on the ESI database.
It is our intention to release all of the reduced spectra within 3~years
of publication.  Monitor the ESI data website for details.

\acknowledgments

The authors wish to recognize and acknowledge the very significant cultural
role and reverence that the summit of Mauna Kea has always had within the
indigenous Hawaiian community.  We are most fortunate to have the
opportunity to conduct observations from this mountain.
We also acknowledge the Keck support staff for their efforts
in performing these observations.  
We thank the referee Sara Ellison for her extensive and extremely
valuable comments and criticism.
E.G. is supported by Fundaci\'{o}n Andes and by an
NSF Astronomy and Astrophysics Postdoctoral Fellowship under
award AST-0201667.

\clearpage

\end{document}